\documentclass[a4paper,12pt,twoside,openright]{report}
\usepackage[english]{babel}

\usepackage{epsfig}
\usepackage{amsmath,amssymb}
\usepackage{eufrak}
\usepackage{mathrsfs}

\newcommand{\slh}[1]{#1\!\!\!/}

\begin{document}

\thispagestyle{empty}
\centerline{\large{\bf{University of Parma, Italy}}}
\centerline{Faculty of Mathematical, Physical and Natural Sciences}
\centerline{\large{\textbf{Research Doctorate in Physics - Cycle XVII}}}
\vspace{1.0cm}
\begin{figure}[h]
\centering\epsfig{file=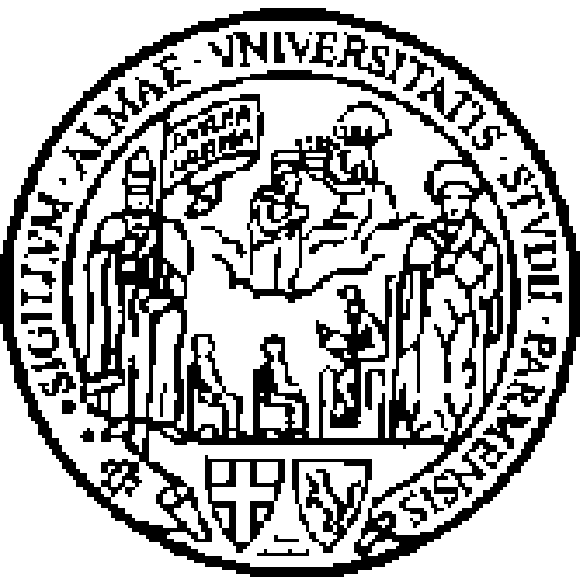,width=3.0cm,height=3.0cm}
\end{figure}
\vspace{1cm}
{\Large
\boldmath
\centerline{\textbf {Semi-Exclusive Heavy Quark Production}}
\boldmath
\centerline{\textbf {in Charged-Current Deep Inelastic Scattering}}
\vspace{2.5cm} 
\begin{flushleft}
\small \rm Ph.D. Thesis by \\ 
\large \textbf {Nicola Pessina}
\end{flushleft}
\begin{flushright}
\small \rm Supervisor \\ 
\large \rm \textbf {Dr. Matteo Cacciari}  \\
\end{flushright}
\begin{flushright}
\small \rm Coordinator  \\
\large \rm \textbf {Prof. Antonio Deriu}
\end{flushright} 
\bigskip
\bigskip
\bigskip
\bigskip 
\bigskip
\centerline{\textbf {February 2005}}
}

\rm
\thispagestyle{empty}
{\it To Cristina and Augusto, my parents}\\
\\
\\
{
Theory is when you know everything but nothing works.\\
Practice is when everything works but no one knows why.\\
If practice and theory are combined, nothing works and no one knows why.
}
{\flushright{{\bf Albert Einstein} }}\\
\\
\\
{
An expert is a person who has made\\
all the mistakes that can be made\\
in a very narrow field.
}
{\flushright{{\bf Niels Bohr} }}

{\pagenumbering{Roman}
\noindent
\centerline{\textbf {Semi-Exclusive Heavy Quark Production}} \\
\centerline{\textbf {in Charged-Current Deep Inelastic Scattering}}\\
\\
\centerline{\bf Abstract}\\
\noindent
The main aim of this research consists in the description of 
Deep Inelastic Scattering (DIS) processes mediated by 
Charged Currents (CC), producing a massive quark in the final state. \\
First of all, we have carried out the semi-exclusive hadronic cross section
parametrised by multi-differential structure functions,
including in the treatise also the mass of the charged leptons.\\
Structure functions are reconstructed by means of the parton model,
introducing Partonic Distribution Functions (PDF) and Fragmentation Functions (FF); 
in the context of Quantum Chromodynamics (QCD), we have obtained
the functional form of the semi-exclusive coefficient functions
for heavy quark production up to Next-to-Leading Order (NLO).\\
The obtained results have been therefore numerically implemented
in order to carry out a quantitative analysis, in view of 
a more complete phenomenological investigation.\\
In this way we have pointed out that the semi-exclusive result obtained 
with techniques at fixed order (NLO) is quantitatively
inadequate in some kinematic regions, highlighting the need 
of contributions coming from higher orders and therefore
showing the necessity of an all-order resummed calculation. \\
In order to make easier the inclusion of resummation, we have considered
the possibility to perform convolutions of parton densities, coefficient 
functions and fragmentation functions directly inside the Mellin space.
For this purpose we have written a C/C++ code allowing to calculate
evolved Mellin transforms of some parametrisation of modern PDF, 
as CTEQ6 and MRST2001.\\
The only missing element is the extension of the analytical results 
for semi-exclusive coefficient functions to the whole complex plane:
in this case the numerical approach appears to be difficult and 
at the moment an implementation has not yet carried out. \\
The first application of our results is the description of
Strange quark inside nucleons: the CC DIS process 
with production of a hadron containing Charm is very useful
for this purpose because quark and antiquark distributions
can be independently studied.
The means offered by our semi-exclusive results allow to 
accurately reproduce experimentally accessible observables
and therefore it is possible a precise measurement for both
Strange partonic distribution and $s - \bar s$ asymmetry.

\tableofcontents
\listoffigures}

\pagenumbering{arabic}

\rm
\chapter*{Introduction}
\addcontentsline{toc}{chapter}{Introduction}

In processes where the transferred momentum is high,
the structure of nucleons is characterized by mean of
Parton Distribution Functions which describe neutrons and 
protons in terms of quarks and gluons, according to the parton model 
at first, and -more in general- to the QCD Factorization 
Theorem \cite{THE_FACTOR}.\\
These distributions are a fundamental requirement to formulate predictions,
based on a perturbative approach, for observable quantities of 
high energy physics and particles experiments.\\
In Deep Inelastic Scattering (DIS) processes the interaction of 
leptons with hadrons is under investigation: it is then possible to
probe the inner structure of the latter.
In particular when Charged Currents are exchanged ($W^{\pm}$ bosons),
interactions involve (at Leading Order of the perturbative theory)
only quarks with positive electrical charge
($u$, $c$, $t$, $\bar{d}$, $\bar{s}$, $\bar{b}$)
or negative ($d$, $s$, $b$, $\bar{u}$, $\bar{c}$, $\bar{t}$);
it is then possible to select interaction channels 
sensitive to specific partonic distributions, simply imposing further 
constraints to final state configurations and considering the suppression 
originated by CKM matrix elements.
For example, for Charm quark production in DIS processes with 
Charged Currents and with neutrinos in initial state
\begin{equation*}
\nu + N \to \ell + H_c + X
\end{equation*}
the only partonic distributions contributing - at Leading Order -
are the ones of Strange and Down (CKM suppressed) quarks.
Consequently this class of events is suitable for the study of
Strange quark distribution inside nucleons;
the same holds about the Strange antiquark distribution
taking into account the analogous case where in the initial state there is an antineutrino,
producing Charm antiquark in the final state.\\
Through DIS processes mediated by $W^{\pm}$ it is then possible 
to carry out information on quark and antiquark {\it independently};
this is favourable to investigate the asymmetry between 
$s$ and $\bar{s}$, suggested by recent experimental measurements
\cite{S_ASYMM_3},\cite{S_ASYMM_4}.
Among light quarks, the Strange distribution inside nucleons is
the less accurate: following the previously illustrated approach,
it is possible to study it in a dedicated and profitable manner.\\
\\
\noindent
In this thesis we have examined in depth the formalism of DIS processes
mediated by Charged Currents (CC), including moreover the mass contributions
of charged lepton participating in the interaction (first chapter).
In the second chapter we have carried out the semi-exclusive partonic 
coefficient functions for heavy quark production up to Next-to-Leading Order, 
in the context of perturbative QCD; it has been the first check of the only
result present in literature and we have confirmed it.
The property of exclusiveness is important in order to carry out predictions on 
realistic observable and not to restrict only to inclusive quantities,
so that the four-momentum of Charm in the final state can be investigated.
In the third chapter we have shown the outcomes of a numerical implementation 
of the obtained results, analysing also its theoretical uncertainties.
Finally in the fourth chapter we have described the Mellin Transform approach,
with a formalism that could facilitate the numerical treatment of convolutions
found and allowing a more natural inclusion of resummation effects.

\rm
\chapter{Charged-Current Deep Inelastic Scattering}
\label{chapter_DIS}

{\it
The main aim of this chapter is to introduce to cross section
calculation for Deep Inelastic Scattering (DIS) process mediated 
by Charged Currents.
Only fundamental aspects have been reported here, whereas 
more complete and accurate passages are contained in appendix
\ref{DIS Formalism}; in this way the explanation
should be more fluent and essential.
Such an appendix is not simply a mere list of formulae
but is integral part of the treatise.\\
\\
In paragraph \ref{standard_approach} the generic semi-exclusive cross section
of DIS processes mediated by Charged Currents is carried out
{\small
\begin{equation*}
\frac{d^{3}\sigma}{dxdydz}
=\frac{G_{F}^{2}ME}{\pi}\frac{M_{W}^{4}}{\left(Q^{2}+M_{W}^{2}\right)^{2}}
\left[
\sum_{k=1}^{5}c_{k}(x,y)\frac{dF_{k}}{dz}(x,y,z)
\right]
 \end{equation*}}
\noindent
and in paragraph \ref{parton_model_approach} 
is shown that, in the parton model context, functions $dF_{k}/dz$
have structure
{\small
\begin{equation*}
\frac{dF_{k}}{dz}(x,y,z)=
\sum_{a,h}\int_x^1\frac{d\xi}{\xi}\int_z^1\frac{d\zeta}{\zeta}
 \frac{d\hat{F}^{a}_{k}}{d\zeta}(\xi,y,\zeta)
 \mathfrak{f}^{a}\left(\frac{x}{\xi}\right)
 \mathfrak{D}_{h}\left(\frac{z}{\zeta}\right)
\end{equation*}}
\noindent
Finally, in paragraph \ref{QCD_approach}, the formalism has been
extended to hadrons containing heavy quarks and 
corrections originated from the mass of the initial state hadron
have been taken into account. 
}

\newpage
\section{Standard Approach}\label{standard_approach}

\begin{figure}[!h]
\centering\epsfig{file=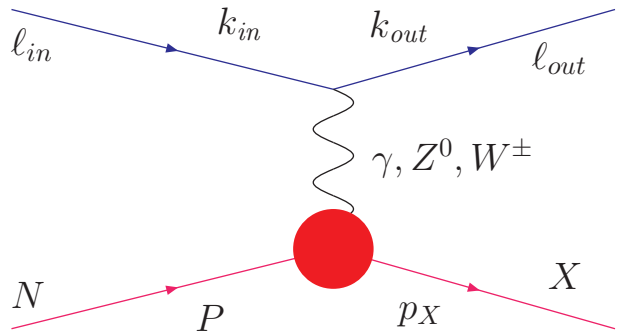} \caption{
\footnotesize{Deep Inelastic Scattering (DIS).}} \label{DIS_fig}
\end{figure}

The formalism introduced at the beginning of DIS investigations 
is now standard and well consolidated;
it originated when QCD did not exist yet and it has been 
formulated in the context of Quantum Relativistic Field 
Theories.
DIS is a process involving scattering between leptons and hadrons
(usually nucleons) and producing both hadronic ($X$) and leptonic 
final states:
\begin{equation*}
\ell_{\it in}(k_{\it in})+N(P)\rightarrow\ell_{\it out}(k_{\it out})+X(p_{X})
\end{equation*}
Leptons $\ell_{\it in}$ and $\ell_{\it out}$ can be
of the same kind (Neutral Currents exchange (NC),
where mediating particles are $\gamma$ and/or $Z^{0}$)
or different (Charged Currents exchange (CC),
mediated by $W^{\pm}$). 
A graphical sketch of a DIS process is 
shown in Fig.\ref{DIS_fig}.

\noindent
The general expression for a cross section is
\begin{equation}\label{cross_section}
d\sigma=\frac{1}{4F}{\overline{|T_{fi}|^2}}(2\pi)^4\delta^{(4)}
 (P+k_{\it in}-k_{\it out}-p_{X})d\Phi
 \end{equation}
 where $F = \sqrt{(Pk_{\it in})^2-m^{2}_{N}m^{2}_{\ell_{\it in}}}$
 is the flux factor,
 $\overline{|T_{fi}|^2}$ is the squared modulus of the amplitude summed
 over all final and initial states, averaged on the latter, $d\Phi$
 is the phase-space of the final state (for hadrons and leptons).
 Four-momentum conservation is assured by the $\delta$-Dirac.\\
\\
 From now on, we consider CC DIS for the particular process
\begin{equation*}
\nu_{\ell}(k_{\it in})+N(P)\rightarrow\ell^{-}(k_{\it out})+X(p_{X})
\end{equation*}

\noindent
 with massless neutrino (initial state, $m_{\nu}=0$) and massive charged lepton
 (final state, $m_{\ell}$); the following formalism holds also for NC case
 with few modifications.
 We are able to write the general structure of $T_{fi}$ in a simple way,
 representing the process as interaction of the leptonic current  
 $j_{L}^{\nu}$ (Fig.\ref{leptonic_current}) with the hadronic one
 $J_{H}^{\tau}$ ($r$, $s$ are the spins of the leptons, the ones of the hadrons
have been understood):
  \begin{equation*}
 T_{fi}=\underbrace{
 {\overbrace{\left(-i\frac{g_{W}}{\sqrt{2}}\right)}^{\small \text{weak coupling}} 
 \langle k_{\it out},s| j_{L}^{\nu} |k_{\it in},r
 \rangle}}_{\text{leptonic current}}
 \underbrace{
 B_{\nu\tau}(0)
 }_{\text{W propagator}}
 \underbrace{
 {\overbrace{\left(-i\frac{g_{W}}{\sqrt{2}}\right)}^{\text{weak coupling}}
 \langle p_{X}| J_{H}^{\tau} |P
 \rangle}}_{\text{hadronic current}}
 \end{equation*}
 where $B_{\nu\tau}(0)$ is the $W$ propagator in {\it unitary} gauge
 ($\eta=0$, appendix \ref{boson_propagator}), so $\sum |T_{fi}|^2$ is
 
\begin{equation}\label{squared_modulus}
 \begin{split}
 \sum_{\text{pol,X}}|T_{fi}|^{2}&=
 \frac{g^{4}_{W}}{4}
 \overbrace{\sum_{r,s}\langle k_{\it in},r| j_{L}^{\dagger\mu} |k_{\it out},s \rangle
 \langle k_{\it out},s| j_{L}^{\nu} |k_{\it in},r \rangle
 }^{L^{\mu\nu}}\\
 &\underbrace{\left[\frac{-g_{\mu\rho}+q_{\mu}q_{\rho}/M_{W}^{2}}{q^{2}-M_{W}^{2}}\right]
 \left[\frac{-g_{\nu\tau}+q_{\nu}q_{\tau}/M_{W}^{2}}{q^{2}-M_{W}^{2}}\right]}
 _{T_{\mu\rho\nu\tau}/(q^{2}-M_{W}^{2})^{2}} \underbrace{
 \sum_{\text{pol,X}}
 \langle P| J_{H}^{\dagger\rho}| p_{X} \rangle
 \langle p_{X}| J_{H}^{\tau} |P \rangle
 }_{W^{\rho\tau}}\\
 &=\frac{8 G_{F}^{2}M_{W}^{4}}{\left(q^{2}-M_{W}^{2}\right)^{2}}
 L^{\mu\nu}T_{\mu\rho\nu\tau}W^{\rho\tau}
 \end{split}
 \end{equation}
 using the relation $\frac{g_{W}^2}{8M_{W}^{2}}=\frac{G_{F}}{\sqrt{2}}$
 and introducing the leptonic ($L^{\mu\nu}$) and hadronic ($W^{\rho\tau}$) tensors.
 Next step consists in averaging on initial states:
 with regard to leptonic state, there is a neutrino
 (only one helicity state allowed), whereas for the hadronic one
 there is a nucleon (spin $1/2$),
 then resulting in an overall factor $1/2$ (tensors $L^{\mu\nu}$ and
 $W^{\rho\tau}$ are not averaged, as explicitly declared in appendices
 \ref{leptonic_tensor} and \ref{generic_hadronic_tensor}).
 We insert a further normalization factor $4\pi$ according to
 the usual definition of hadronic tensor ($\hat{W}^{\rho\tau}=W^{\rho\tau}/4\pi$).
 Therefore the cross section \ref{cross_section} can be written as
 \begin{equation} \label{generic_cross_section}
 d\sigma=\frac{G_{F}^{2}}{ME}
 \frac{M_{W}^{4}}{\left(Q^{2}+M_{W}^{2}\right)^{2}}L^{\mu\nu}T_{\mu\rho\nu\tau}
 \left[4\pi \hat{W}^{\rho\tau}(2\pi)^{4}\delta^{(4)}(P+q-p_{X})\right] d\Phi
 \end{equation}
 choosing the reference system where the nucleon is at rest
 (that is $P=(M,0,0,0)$), defining $Q^{2}=-q^{2}=-(k_{\it in}-k_{\it out})^{2}$
 and $E$ is the energy of the incoming neutrino.\\

\newpage
\noindent
 For CC case, the leptonic tensor is (\ref{leptonic_tensor})
 \begin{equation}
 L^{\mu\nu}=L^{\mu\nu}_{CC}=2\left[k_{out}^{\mu}k_{in}^{\nu}+k_{in}^{\mu}k_{out}^{\nu}-(k_{in}k_{out})g^{\mu\nu}\right]
 +2i\epsilon^{\mu\alpha\nu\beta}k_{\alpha}^{in}k_{\beta}^{out}
 \end{equation}
 Within this approach, the nature of the interaction between nucleon 
 and electroweak vector boson is supposed to be unknown, so that $\hat{W}^{\rho\tau}$
 is constructed as the most generic second rank {\it Lorentz-invariant} tensor.  
 In appendix \ref{generic_hadronic_tensor} we derive the expression 
 for this tensor and we report some interesting remarks. 
 \begin{equation*}
 \begin{split}
 \hat{W}^{\rho\tau}(2\pi)^{4}\delta^{(4)}(P+q-p_{X})&\equiv
 H^{\rho\tau}= -(2Pq)H_{1}g^{\rho\tau}
 +4H_{2}P^{\rho}P^{\tau}\\
 &-2iH_{3}\epsilon^{\rho\tau}_{\:\:\:\:\alpha\beta}P^{\alpha}q^{\beta}
 +2H_{4}q^{\rho}q^{\tau}
 +2H_{5}(P^{\rho}q^{\tau}+q^{\rho}P^{\tau})
 \end{split}
 \end{equation*}
 \begin{figure}[!h]
 \centering\epsfig{file=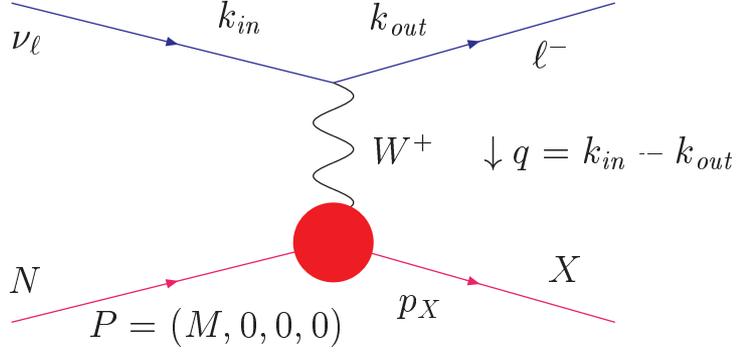}
 \caption{\footnotesize DIS kinematics.} \label{Kinematics_DIS}
 \end{figure}

 \noindent 
 At this point it is useful to describe DIS kinematics;
 through notation shown in Fig.\ref{Kinematics_DIS} for momenta of the particles,
 we can finally carry out some relations.
 Remember that we have chosen to work in the reference system where 
 the nucleon is at rest.
 \begin{align}
 Q^{2}&=-q^{2} & P^{2}&=M^{2}&\notag\\
 k_{\it in}q&=-\frac{Q^{2}+m_{\ell}^{2}}{2}& k_{\it out}q&=\frac{Q^{2}-m_{\ell}^{2}}{2}&\notag\\
 Pq&=M(E-E')\equiv \upsilon& k_{\it in}k_{\it out}&=\frac{Q^{2}+m_{\ell}^{2}}{2}&\label{Kine_DIS}\\
 Pk_{\it in}&=ME& Pk_{\it out}&=ME'&\notag\\
 x&=\frac{Q^{2}}{2\upsilon}&  y=\frac{qP}{k_{\it in}P}&=1-\frac{E'}{E}&\notag
 \end{align}
 where $q=k_{\it in}-k_{\it out}$ is the momentum carried by $W$ boson, $M$ and $m_{\ell}$
 are respectively the mass of the nucleon and the charged lepton,
 whereas $E$, $E'$ are respectively the energies of incoming ($\nu_{\ell}$) 
 and outgoing ($\ell^{-}$) leptons; all the external particles have been considered on-mass-shell.\\
 We can separate the phase-space of \ref{generic_cross_section}
 in hadronic ($d\Phi_X$) and leptonic part 
 \begin{equation} \label{ps_factorized}
 \begin{split}
 &d\Phi=d\Phi_{X}\left[\frac{d^4k_{\it out}}{(2\pi)^3}
 \delta\left(k_{\it out}^{2}-m_{\ell}^{2}\right)\right]
 =d\Phi_{X}\frac{yME}{2(2\pi)^3}d\phi dxdy \\
& d\Phi_{X}= d\hat{\Phi}_{X}\prod_{i}^{j}d\alpha_{i}
 \end{split}
 \end{equation}
 rewriting the leptonic phase-space in terms of $x$ and $y$ variables
 as explained in appendix \ref{contraction} (eq.\ref{muon_phase_space}).
 Moreover phase-space $d\Phi_X$ has been factorised in order to highlight
 the quantities $\alpha_i$: it is potentially possible to 
 remain differential on them, if the explicit dependence is needed.\\
 Making use of \ref{Kine_DIS}, in appendix \ref{boson_propagator} 
 we have calculated the contraction $L^{\mu\nu}_{CC}T_{\mu\rho\nu\tau}H^{\rho\tau}$ 
 for a generic gauge parameter in $W$ propagator, finally choosing the unitary one;
 substituting this result (\ref{first}) and \ref{ps_factorized} into
 \ref{generic_cross_section},
 then the cross section can be written in a differential way as
 \begin{equation}\label{full_contraction}
 \boxed{
 \begin{split}
 \frac{d^{2+j}\sigma}{dxdy\prod_{i}^{j}d\alpha_{i}}&=
 \frac{G_{F}^{2}ME}{\pi}\frac{M_{W}^{4}}{\left(Q^{2}+M_{W}^{2}\right)^{2}}
 \bigg\{\left[xy^{2}+\frac{m_{\ell}^{2}y}{2ME}\right](1+\chi_{1}(x,y))
 \frac{d^jF_{1}}{\prod_{i}^{j}d\alpha_{i}}\\
 &+\left[\left(1-\frac{m_{\ell}^{2}}{4E^{2}}\right)-\left(1+\frac{M}{2E}x\right)y
 \right](1+\chi_{2}(x,y))\frac{d^jF_{2}}{\prod_{i}^{j}d\alpha_{i}}\\
 &+\left[xy\left(1-\frac{y}{2}\right)-\frac{m_{\ell}^{2}y}{4ME}\right](1+\chi_{3}(x,y))
 \frac{d^jF_{3}}{\prod_{i}^{j}d\alpha_{i}}\\
& +\left[\frac{m_{\ell}^{2}(m_{\ell}^{2}+Q^2)}{4M^2E^2x}\right](1+\chi_{4}(x,y))
 \frac{d^jF_{4}}{\prod_{i}^{j}d\alpha_{i}}\\
 &-\left[\frac{m_{\ell}^{2}}{ME}\right](1+\chi_{5}(x,y))
 \frac{d^jF_{5}}{\prod_{i}^{j}d\alpha_{i}}
 \bigg.\bigg\}
 \end{split}}
 \end{equation}
having integrated over flat angular variable $\phi$ $\in [0,2\pi)$,
because no terms depend on such a quantity.
Factor $d\hat{\Phi}_{X}$ has been included in the definition 
of functions
\begin{equation}\label{F_definition}
  \frac{d^{j}F_{k}}{\prod_{i}^{j}d\alpha_{i}}(x,y,\alpha_i)
  \equiv A_k \int d\hat{\Phi}_{X} H_k
\end{equation}
with $A_{1,5}=4\upsilon$, $A_{2,3}=8\upsilon$ and $A_4=4x\upsilon$, 
because of the chosen parametrisation, in accordance with
appendix \ref{contraction}.
 Result \ref{full_contraction} agrees with literature
 for the inclusive case \cite{KRRE};
 with abuse of notation (for sake of convenience and for historical reasons),
 $Q^2$ variable has been let expressed in the second side of relation \ref{full_contraction} 
 and for analogous quantities, nevertheless it would be more correct to
employ the equality $Q^2=2MExy$, showing explicitly the full dependence
 on $x$ and $y$.
 Functions $\chi_k(x,y)$ are reported in appendix \ref{boson_propagator}.\\
\\
\\
\noindent
\underline{Inclusive Cross Section}\\
\\
In \ref{full_contraction} is shown a cross section semi-exclusive on hadronic
states, but it is possible to choose to obtain a fully inclusive result:
\begin{equation}\label{full_inclusive_cross}
\frac{d^{2}\sigma}{dxdy}=
 \frac{G_{F}^{2}ME}{\pi}
 \frac{M_{W}^{4}}{\left(Q^{2}+M_{W}^{2}\right)^{2}}
 \left[
 \sum_{k=1}^{5}c_{k}(x,y)F_{k}
 \right]
 \end{equation}
obviously identifying 
$$F_k(x,y)=\int\prod_{i}^{j}d\alpha_i
\left(\frac{d^jF_k}{\prod_{i}^{j}d\alpha_i}\right)
= A_k\int d\Phi_{X} H_k$$
and $c_k$ coefficients are the ones in \ref{full_contraction}.\\
This does not change considerably the structure of previous results.\\
\\
\noindent
\underline{Massless Case}\\
\\
For the massless charged lepton case (obtained setting 
$m_{\ell}=0$ in \ref{full_contraction}) the expression for coefficients 
$c_k$ becomes simpler
 \begin{equation}\label{full_contraction_massless}
\begin{split}
 \frac{d^{2+j}\sigma}{dxdy\prod_{i}^{j}d\alpha_i}
&=\frac{G_{F}^{2}ME}{\pi} \frac{M_{W}^{4}}{\left(Q^{2}+M_{W}^{2}\right)^{2}}\\
&\bigg\{
 xy^{2}\frac{d^jF_{1}}{\prod_{i}^{j}d\alpha_i}
 +\left(1-y+\frac{M}{2E}xy\right)\frac{d^jF_{2}}{\prod_{i}^{j}d\alpha_i}
 +xy\left(1-\frac{y}{2}\right)\frac{d^jF_{3}}{\prod_{i}^{j}d\alpha_i}
 \bigg\}
\end{split}
 \end{equation}
Notice that terms containing $H_{4}$ and $H_{5}$ (i.e. $F_{4}$, $F_{5}$)
disappear: this because coefficients $c_4$ and $c_5$ cancel out;
even a potential term $H_{6}$ (eq.\ref{hadronic_tensor}) is not present
as explained in appendix \ref{generic_hadronic_tensor}.
furthermore $H_{4}$ and $H_{5}$ exist only for the case of massive lepton,
whereas $H_{6}$ does not contribute in any case.\\

\newpage
\noindent
\underline{Range of variability for x and y}\\
\\
The range for $x$ and $y$ variables is
carried out from kinematics according to \ref{Kine_DIS}
\cite{KRRE_TM_1}:
\begin{equation*}
\begin{split}
\frac{m_{\ell}^2}{2M(E-m_{\ell})}&\leq x \leq 1
\qquad \qquad
y_1-y_2\leq y \leq y_1+y_2\\
&y_1=\frac{1-m_{\ell}^2\left( \frac{1}{2MEx}+\frac{1}{2E^2} \right)  }
{2\left( 1+\frac{Mx}{2E}\right)}\\
&y_2=\frac{\sqrt{\left(1-\frac{m_{\ell}^2}{2MEx}\right)^2-\frac{m_{\ell}^2}{E^2}}}
{2\left( 1+\frac{Mx}{2E}\right)}
\end{split}
\end{equation*}
\\
\underline{Charged Currents in general}\\
\\
Result \ref{full_contraction}
(and similarly \ref{full_inclusive_cross}, \ref{full_contraction_massless})
can be extended to the case of generic DIS process with charged current.
As highlighted at the end of appendix \ref{leptonic_tensor},
swapping particles for antiparticles,
the only variation for the whole study is a change of sign
for $c_{3}$ term.
A charged lepton in the initial states needs a factor $1/2$ 
because of the average on spin (instead of only one helicity state 
allowed for a neutrino).
Then 
\begin{equation}\label{CC_exclusive}
\begin{split}
\frac{d^{2+j}\sigma^{CC}}{dxdy\prod_{i}^{j}d\alpha_i}
=&\frac{G_{F}^{2}ME}{\pi}\frac{M_{W}^{4}}{\left(Q^{2}+M_{W}^{2}\right)^{2}}
\left[\left(\frac{E^2}{E^2-m_{\ell}^2}\right)
\left(\frac{1-(-1)^bP_{\ell}}{2}\right)\right]^a \\
&\left[
\sum_{k=1,2,4,5}c_{k}(x,y)\frac{d^jF_{k}}{\prod_{i}^{j}d\alpha_i}
+\left(-1\right)^b c_{3}(x,y)\frac{d^jF_{3}}{\prod_{i}^{j}d\alpha_i}
 \right]
 \end{split}
\end{equation}
where $c_k$ coefficients are obviously those contained in 
\ref{full_contraction} and $a$ and $b$ exponents
have to be evaluated following the table:
\begin{center}
\begin{tabular}{|c||c|c||}\hline
CC Process &$a$  &$b$  \\\hline
$\nu_{\ell}+N\rightarrow \ell^- + X$     &0  &0   \\\hline
$\ell^- +N\rightarrow \nu_{\ell^-} + X$     &1  &1   \\\hline
$\overline{\nu}_{\ell} +N\rightarrow \ell^+ + X$     &0  &1  \\\hline
$\ell^+ +N\rightarrow \overline{\nu}_{\ell} + X$     &1  &0   \\\hline
\end{tabular}
\end{center}
Term $E^2/(E^2-m_{\ell}^2)$ comes from flux factor
and phase-space (\ref{muon_phase_space}),
considering a massive lepton ($m_{\ell}$) in initial state.
We have also introduced the term $P_{\ell}$ to potentially 
take into account the polarisation degree of the (charged) 
lepton beams: $P_{\ell}=(N_L-N_R)/(N_L+N_R)$ where $N_L$ and $N_R$
are respectively the number of particles with  
left-handed and right-handed polarisation. \\
Unpolarized case corresponds to $P_{\ell}=0$.\\
Setting $a=b=0$, result in \ref{full_contraction} is recovered.

\vspace{1cm}
\noindent
We have initially parametrised the unidentified physics of the 
hadronic vertex with a set \{$H_{k}$\} of unknown functions and
consequently the result can not be predictive but it depends on them;
comparing the experimentally measured cross section 
to the parametrisation \ref{full_contraction} 
(or a simplified expression: in fact the one for massless leptons,
as shown in paragraph \ref{HP_correspondence})
functions $d^jF_{k}/\prod_i^jd\alpha_i$ can be carried out.\\
No hypothesis has been yet proposed on the nature of interactions 
behind the hadronic tensor, so some physical model 
has to be employed to describe it.
Eventually we remark that there are many other ways to parametrise 
the cross section: all are comparable and they differ in expressions
relating factors $H_{k}$ to $d^jF_{k}/\prod_i^jd\alpha_i$,
or in choosing ($x$, $y$, $\alpha_i$) observables
in the phase-space (appendix \ref{contraction}).

\section{Parton Model}\label{parton_model_approach}

\subsection{Partonic Cross Section}
\begin{figure}[!h]
\centering\epsfig{file=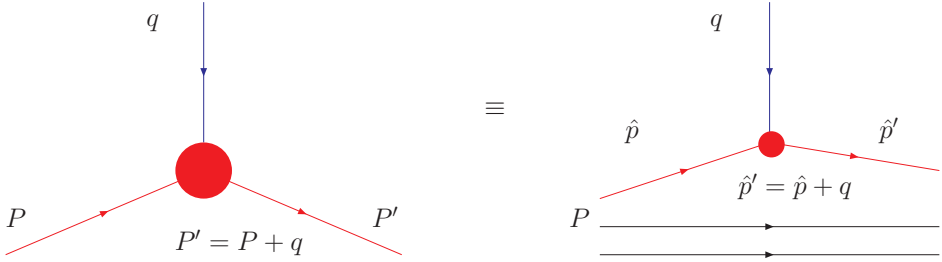, width=13 cm}
\caption{\footnotesize Parton Model for DIS.}\label{Parton_Model}
\end{figure}

 Cross section \ref{full_contraction} holds in general in the context
 of Quantum Relativistic Field Theories, with the support of 
 Weinberg-Salam-Glashow Electroweak theory.
 Applying the {\it parton model} and {\it QCD}, we are trying to predict
 and describe the functional form of terms $d^jF_k/\prod_i^j d\alpha_i$.\\
 The naive parton model is independent of QCD, having been hypothesised
 before than the latter had been formulated (\cite{P_MODEL_A},\cite{P_MODEL_B}).
 In fact it originated as quasi-classical model for DIS, based on the idea 
 that hadrons could be described as a collection of independent point-like particles 
 (so-called partons), having a small transverse momentum (and therefore 
 negligible) and interacting with the incident leptons by exchanging a 
 vector boson. 
 Identifying quarks and gluons with partons, the {\it quark-parton-model} 
 is made, so quantum numbers are assigned to partons in order to 
 reproduce the ones for known hadrons.
 The interactions vertex between $W$ boson and the nucleon can then be interpreted
 as in representation of Fig.\ref{Parton_Model}.

\noindent
We can consequently work at partonic level exactly as in previous section,
simply substituting for the momentum $P$ of the nucleon, 
the momentum $\hat{p}$ of the parton contained in it and 
interacting with the $W$ boson.\\
We introduce the {\it partonic tensor}, whose properties and structure are
identical to the hadronic one:
\begin{equation}
\begin{split}\label{partonic_tensor}
 h_{\mu\nu}^a= -(2\hat{p}q)h_{1}^a&g_{\mu\nu}
 +4h_{2}^a\hat{p}_{\mu}\hat{p}_{\nu}\\
 &-2ih_{3}^a\epsilon_{\mu\nu}^{\:\:\:\:\:\rho\tau}\hat{p}_{\rho}q_{\tau}
 +2h_{4}^aq_{\mu}q_{\nu}
 +2h_{5}^a(\hat{p}_{\mu}q_{\nu}+q_{\mu}\hat{p}_{\nu})
\end{split}
\end{equation}
where the $a$ index indicates which kind of parton is taken into account.\\
We define the partonic quantities
 \begin{equation}\label{parton_variables}
 \hat{x}=\frac{Q^2}{2\hat{p}q}\quad\quad\quad
 \hat{y}=\frac{\hat{p}q}{\hat{p}k}\quad\quad\quad
 \hat{\upsilon}=\hat{p}q
 \end{equation}
analogously to \ref{Kine_DIS} in order to contract \ref{partonic_tensor}
with the leptonic tensor \ref{leptonic_tensor};
also the phase-space of final state contains a parton, therefore
it has to be expressed by means of partonic quantities
($d\Phi_X\rightarrow d\varphi_X$):
$$d\varphi_X=d\hat{\varphi}_X\prod_{i}^{j}d\hat{\alpha}_{i}$$
Temporarily neglecting modifications to flux factor $F$ 
(introduced in \ref{cross_section}), for the sake of simplicity, 
and going back over all steps of previous section, we obtain

 \begin{equation}
 \begin{split}\label{partonic_full_cross_section}
 \frac{d^{2+j}\hat{\sigma}}{d\hat{x}d\hat{y}\prod_i^j d\hat{\alpha}_i}=&
 \frac{G_{F}^{2}ME}{\pi}
 \frac{M_{W}^{4}}{\left(Q^{2}+M_{W}^{2}\right)^{2}}
 \sum_a\bigg\{\bigg.
\left[\hat{x}\hat{y}^{2}\!+\!\frac{m_{\ell}^{2}\hat{y}}{2ME}\right](1\!+\!\chi_{1}(\hat{x},\hat{y}))
\frac{d^j\hat{F}^{a}_{1}}{\prod_i^j d\hat{\alpha}_i}\\
&+\left[\left(1-\frac{m_{\ell}^{2}}{4E^{2}}\right)-\hat{y}\right](1+\chi_{2}(\hat{x},\hat{y}))
\frac{d^j\hat{F}^{a}_{2}}{\prod_i^j d\hat{\alpha}_i}\\
&+\left[\hat{x}\hat{y}\left(1-\frac{\hat{y}}{2}\right)-\frac{m_{\ell}^{2}\hat{y}}{4ME}\right]
(1+\chi_{3}(\hat{x},\hat{y}))\frac{d^j\hat{F}^{a}_{3}}{\prod_i^j d\hat{\alpha}_i}\\
&+\left[\frac{m_{\ell}^{2}(m_{\ell}^{2}+Q^2)}{4M^2E^2\hat{x}}\right](1+\chi_{4}(\hat{x},\hat{y}))
\frac{d^j\hat{F}^{a}_{4}}{\prod_i^j d\hat{\alpha}_i}\\
&-\left[\frac{m_{\ell}^{2}}{ME}\right](1+\chi_{5}(\hat{x},\hat{y}))
\frac{d^j\hat{F}^{a}_{5}}{\prod_i^j d\hat{\alpha}_i}
 \bigg.\bigg\}
 \end{split}
 \end{equation}
 where, analogously to \ref{F_definition}, it is has been defined
\begin{equation}\label{dF_parton}
  \frac{d^{j}\hat{F}_{k}^a}{\prod_{i}^{j}d\hat{\alpha}_{i}}(\hat{x},\hat{y},\hat{\alpha}_i)
  \equiv \hat{A}_k\int d\hat{\varphi}_{X} h_k^a 
\end{equation}
being $\hat{A}_{1,5}=4\hat{\upsilon}$, $\hat{A}_{2,3}=8\hat{\upsilon}$,
$\hat{A}_4=4x\hat{\upsilon}$.

\subsection{Semi-Exclusive Processes}
We leave now the general formalism to consider a specific case.
We want to study a CC DIS with heavy quark production in the final state.
According to perturbative QCD at LO (Fig.\ref{born})
the phase-space contains only one particle (the heavy quark)
whereas at NLO (Fig.\ref{Real_Quark} and \ref{Real_Gluon}) 
there are two (an additional light quark or gluon).
As shown in appendix \ref{partonic_phase_space}, for the NLO case
there is only one degree of freedom, whilst obviously none at LO.
Then in the following we will consider cross sections with at the most
$j=1$ in \ref{full_contraction} and \ref{partonic_full_cross_section},
i.e. neglecting the trivial dependence (through $\delta$-Dirac)
on non-free variables.
We define $z$ ($\equiv\alpha_1$ in \ref{full_contraction})
and $\zeta$ ($\equiv\hat{\alpha}_1$ in \ref{partonic_full_cross_section})
respectively the variables of interest for the hadronic and partonic level. 
In particular we remark that, under these assumptions,
starting from \ref{dF_parton}, it is possible to obtain
(with a little abuse of notation)
\begin{equation}\label{partonic_dF}
  \frac{d\hat{F}_{k}}{d\zeta}(\hat{x},\hat{y},\zeta)=
  A_k \int d\hat{\varphi}_{X} h_k(\hat{x},\hat{y},\hat{\varphi}_{X},\zeta)
  \equiv A_k h_k(\hat{x},\hat{y},\zeta) \int d\hat{\varphi}_{X}
\end{equation}
implying that integration is performed on flat variables
(that is $h_k$ is independent of them) and on all those ones
being not degrees of freedom; 
obviously in $h_k(\hat{x},\hat{y},\hat{\varphi}_{X},\zeta)$ 
the dependent variables ($\hat{\varphi}_{X}$) are replaced
by relations carried out from $\int d\hat{\varphi}_{X}$.
We call this approach {\it semi-exclusive},
in order to highlight that the considered quantities are as
differential (exclusive) as possible, compatibly 
with the degrees of freedom.\\

\subsection{Parton-Hadron Correspondence}\label{HP_correspondence}
\begin{figure}[!h]
\centering\epsfig{file=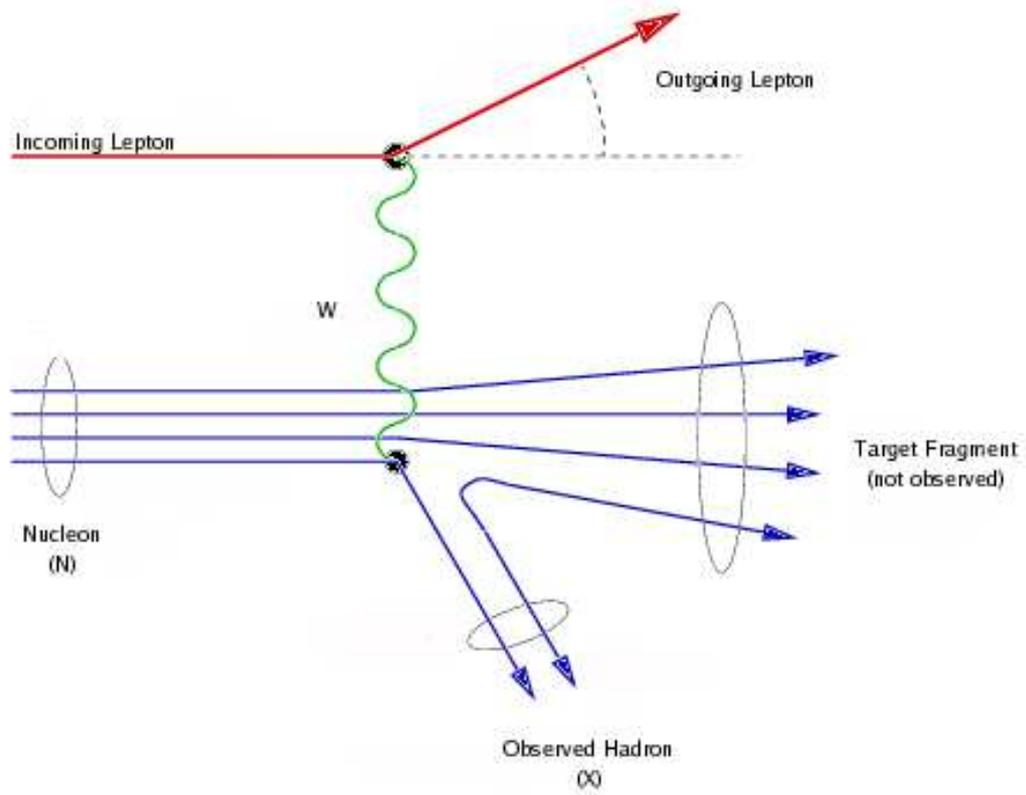,width=14truecm}
\caption{\footnotesize{Typical configuration of a final state for 
a DIS process.}}
\label{final_state}
\end{figure}

\noindent
In nature only hadrons are experimentally accessible physical states,
whereas partons have not been observed.
Assuming partons on-mass-shell and collinear to the momentum of nucleon 
containing them, the hadronic cross section has the factorised form 
\begin{equation}
 d\sigma(P)\sim\sum_a\int_{0}^{1}d\xi \mathfrak{f}^{(a)}(\xi)
 d\hat{\sigma}_{a}(\hat{p}_a= \xi P)
\end{equation}
Parameter $\xi\in(0,1)$ is the fraction of nucleon momentum carried by the 
parton of kind $a$, whilst $\mathfrak{f}^{a}$ describes 
how $\xi$ is distributed and in practice it could be interpreted 
as probability density of interaction (through boson exchange) with a parton
of kind $a$ and momentum $\hat{p}_a=\xi P$.
Quantity $\mathfrak{f}^{a}(\xi)$ is called Partonic Distribution Function
(PDF) and it is non-perturbatively calculable.\\
In the final state we have imposed the presence of partons, then, 
similarly to PDF, also Fragmentation Functions (FF) have been 
introduced to phenomenologically describe the transitions 
from quarks to hadrons. 
Using the correspondence parton-hadron in a more rigorous way,
flux factor $F$ becomes $F\xi$ and it holds
$$\hat{x}=\frac{x}{\xi} \:\:\:\:\:\: \hat{y}=y \:\:\:\:\:\:
\hat{\upsilon}=\xi\upsilon \:\:\:\:\:\:
\chi_k(\hat{x},\hat{y})=\chi_k(x,y) $$
so we are able to write the hadronic cross section as
convolution of the partonic one \ref{partonic_full_cross_section}
with PDF and FF:
\begin{equation}\label{F_j_relations}
 \begin{split}
&\frac{d^{3}\sigma}{dxdydz}=
\frac{G_{F}^{2}ME}{\pi}
 \frac{M_{W}^{4}}{\left(Q^{2}+M_{W}^{2}\right)^{2}}
 \Bigg\{\\
&\sum_{k=1,3,4,5}c_{k}(x,y)
 \left[\sum_a\sum_{h}\int_x^1\frac{d\xi}{\xi}\int_z^1\frac{d\zeta}{\zeta}
 \frac{d\hat{F}^{a}_{k}}{d\zeta}(\xi,y,\zeta)
 \mathfrak{f}^{a}\left(\frac{x}{\xi}\right)
 \mathfrak{D}_{h}\left(\frac{z}{\zeta}\right)
 \right]\\
& + \tilde{c}_{2}(x,y)
 \left[\sum_a\sum_{h}\int_x^1\frac{d\xi}{\xi}\int_z^1\frac{d\zeta}{\zeta}
 \xi\frac{d\hat{F}^{a}_{2}}{d\zeta}(\xi,y,\zeta)
 \mathfrak{f}^{a}\left(\frac{x}{\xi}\right)
 \mathfrak{D}_{h}\left(\frac{z}{\zeta}\right)
 \right]
 \Bigg\}
\end{split}
\end{equation}
where $c_k$ and $\tilde{c}_2$ are respectively the coefficients 
of terms $d^jF_k$ and $d^jF_2$ in \ref{partonic_full_cross_section},
whilst $\mathfrak{D}_{h}$ is a FF describing the manner of 
forming a hadron of kind $h$, starting from a final state parton.
In this investigation we assume that the hadron containing 
the heavy quark is originated directly from such a quark and 
not from other partons, being not predominant channels.
Comparing cross section \ref{F_j_relations}
with the one purely hadronic previously carried out
(eq.\ref{full_contraction})
\begin{equation}
\frac{d^{3}\sigma}{dxdydz}=\frac{G_{F}^{2}ME}{\pi}
 \frac{M_{W}^{4}}{\left(Q^{2}+M_{W}^{2}\right)^{2}}
 \left\{
 \sum_{k=1}^{5}c_{k}(x,y)\frac{dF_{k}}{dz}(x,y,z)
 \right\}
\end{equation}
it is clear an explicit and immediate correspondence
between hadronic and partonic level
\begin{equation}\label{adron_parton_correspondence}
 \begin{split}
&\frac{dF_{k}}{dz}(x,y,z)=
\sum_{a,h}\int_x^1\frac{d\xi}{\xi}\int_z^1\frac{d\zeta}{\zeta}
 \frac{d\hat{F}^{a}_{k}}{d\zeta}(\xi,y,\zeta)
 \mathfrak{f}^{a}\left(\frac{x}{\xi}\right)
 \mathfrak{D}_{h}\left(\frac{z}{\zeta}\right)\:\:\:\:\: k=1,3,4,5\\
& \frac{dF_{2}}{dz}(x,y,z)=\frac{\tilde{c}_{2}(x,y)}{c_{2}(x,y)}
\sum_{a,b,h}\int_x^1\frac{d\xi}{\xi}\int_z^1\frac{d\zeta}{\zeta}
 \xi\frac{d\hat{F}^{a}_{2}}{d\zeta}(\xi,y,\zeta)
 \mathfrak{f}^{a}\left(\frac{x}{\xi}\right)
 \mathfrak{D}_{h}\left(\frac{z}{\zeta}\right)
\end{split}
\end{equation}
Notice that several $\hat{F}^{a}_k$ correspond to each $F_k$
because there are many kind of partons realising
the sub-process shown in Fig.\ref{Parton_Model},
and to each one there is associated an opportune distribution 
$\mathfrak{f}^{a}(\xi)$.
Furthermore it is very interesting that 
no terms $\frac{M}{2E}\hat{x}\hat{y}$ appear into 
$\tilde{c}_2$ coefficient for $\hat{F}_2$ \eqref{partonic_full_cross_section}:
this because for an on-mass-shell nucleon
it holds $P^2=M^2$, whereas for a parton $\hat{p}^2=0$.
However in QCD this fact is predicted by the factorization theorem 
asserting that only the {\it dominant} contribution can be reproduced
by the perturbative theory: 
are therefore excluded terms proportional to $M^2/Q^2$ \cite{PDF_Soper}
(term $\frac{M}{2E}xy$ is in fact equivalent to $\frac{M^2}{Q^2}x^2y^2$).
From the operative point of view there are no problems, because in practice
such a term is negligible, being typically $M\sim 1\: GeV$,
$30\lesssim E [GeV]\lesssim600$ (CCFR, NuTeV experiments \cite{DISSERTORI},
\cite{Data_1}) and also because of a suppression, being $x<1$.
Then we can tacitly assume $\tilde{c}_2(x,y)\sim c_2(x,y)$;
for the case of inclusive cross sections 
the difference is less than 2{\small{\%}}
(\cite{Tung_Aivazis_Olness}, page 3091). 
In view of this remark, one should be coherent with the chosen accuracy: 
in fact, with regard to experimentally accessible parameters, 
coefficients $c_4$ and $c_5$ are much smaller
than that last term we have neglected, like other contributions
inside $c_1$, $c_2$ and $c_3$.
So expression in \ref{full_contraction_massless} has to be 
used as parametrisation for cross sections to carry out quantitative results;
this also because functions $F_4$ and $F_5$ calculated in QCD (\ref{Coefficient_Functions_Semi-Esclusive})
are quantitatively comparable to $F_1$, $F_2$, $F_3$.
We remark that this fact does not come (only) from an issue of 
experimental precision, but it originates from the limit of the 
parton model well highlighted by the factorization theorem of
Perturbative QCD.
Therefore, even with an exact treatment \eqref{full_contraction} for DIS
considering massive charged lepton (the one coming directly by leptonic current
in DIS), we intrinsically can not obtain an accuracy higher than 
the massless case, at the state of the theory.\\
\\
For the case of heavy quark production in the final state, 
the convolution with $\mathfrak{D}_{h}$ does not prevent
from recovering directly the inclusive result; 
in fact thanks to the usual definition of FF normalization
(holding for heavy quarks)\begin{equation*}
\sum_h\int_0^1 dt \mathfrak{D}_{h}(t)= 1
\end{equation*}
it is possible to verify that
\begin{equation*}
\begin{split}
F_{k}(x,y)&=\int_0^1 dz \left[\frac{d{F}_{k}}{dz}(x,y,z)\right]\\
&=\int_0^1 dz \left[
\sum_{a,h}\int_x^1\frac{d\xi}{\xi}\int_z^1\frac{d\zeta}{\zeta}
 \frac{d\hat{F}^{a}_{k}}{d\zeta}(\xi,y,\zeta)
 \mathfrak{f}^{a}\left(\frac{x}{\xi}\right)
 \mathfrak{D}_{h}\left(\frac{z}{\zeta}\right)
 \right]
\\
&=\sum_{a}\int_x^1\frac{d\xi}{\xi}
\mathfrak{f}^{a}\left(\frac{x}{\xi}\right)
\left[\int d\zeta\frac{\hat{F}^{a}_{k}}{d\zeta}(\xi,y,\zeta)\right]\\
&=\sum_{a}\int_x^1\frac{d\xi}{\xi}
\mathfrak{f}^{a}\left(\frac{x}{\xi}\right)
\hat{F}^{a}_{k}(\xi,y)
\end{split}
\end{equation*}
We remark that the whole hadronic final state
consists in the remainder of nucleon not directly
participating to the process and in the one generated by
the interaction with $W$ boson.
The former is made of spectator-partons, obviously hadronizing
through interactions with particles carrying colour charge
in the process; nevertheless this contribution generally is 
not taken into account, because usually it remains in the beam pipe
for experiments not-a-rest or it does not reach the detectors. 
In any case it is not experimentally selected:
only the latter produces the 
\textquotedblleft observed\textquotedblright \, 
final state(Fig.\ref{final_state}).
\newpage

\newpage
\subsection{Slow Rescaling and Target Mass Correction}\label{QCD_approach}
The main result we have carried out till now consists in 
the correspondence between hadronic and partonic structure functions 
as reported in \ref{adron_parton_correspondence}.\\
In order to enable a smoother comparison with literature
and to use a more compact formalism, we adopt the convention
\begin{equation*}
\begin{split}
\frac{dF_k}{dz}(x,y,z)=&\mathscr{F}_k(x,y,z) \qquad k=1,4,5\\
\frac{dF_2}{dz}(x,y,z)=&2x\mathscr{F}_2(x,y,z)\\
\frac{dF_3}{dz}(x,y,z)=&2\mathscr{F}_3(x,y,z)
\end{split}
\end{equation*}
\begin{equation}\label{compact_F_j}
\mathscr{F}_k(x,y,z)=\sum_{a,h}
\int_{\chi}^1\frac{d\xi}{\xi}
\int_{z}^1\frac{d\zeta}{\zeta}
\hat{\mathscr{F}}_k^{a}(\xi,y,\zeta)
\mathfrak{f}^{a}\left(\frac{\chi}{\xi}\right)
\mathfrak{D}_h\left(\frac{z}{\zeta}\right)
\end{equation}
where we have profitably redefined  $\chi=x$,
$\hat{\mathscr{F}}_k^{a}=\frac{d\hat{F}_k^{a}}{d\zeta}$ for
$k=1,4,5$,
$\hat{\mathscr{F}}_2^{a}=\frac{\xi}{2\chi}\frac{d\hat{F}_2^{a}}{d\zeta}$
and
$\hat{\mathscr{F}}_3^{a}=\frac{1}{2}\frac{d\hat{F}_3^{a}}{d\zeta}$.
Relations \ref{compact_F_j} are not the arrival point of our
investigation and they need a more exhaustive examination.\\
\\
\underline{Heavy Quarks}\\
\\
The goal of our analysis is to study hadron production in CC DIS
processes; in particular we consider hadrons containing a heavy 
quark, then, to obtain a realistic and correct result, we assume
that the parton produced in the final state is the heavy quark
with mass $m$ becoming a hadron later on.
As shown in detail in the following (see remarks after \ref{double_diff_mass}),
it is natural to introduce an adimensional parameter
$\lambda$, defined as
\begin{equation}\label{lambda_definition}
\lambda=\frac{Q^2}{Q^2+m^2}
\end{equation}
where as usually $Q^2$ is the quantity in \ref{Kine_DIS}.
Relations \ref{compact_F_j}, expressed by means of $x$, 
properly hold only for the case of massless quark production;
investigating more in depth, we have found that the right link
between partonic and hadronic level is carried out substituting 
$x\rightarrow x/\lambda$, that is by means of a 
{\it slow-rescaling} \textquotedblleft prescription\textquotedblright\: 
to take into account also the mass of the produced quark. 
The massless quark case can be recovered for $\lambda=1$.\\
\newpage
\noindent
\underline{Target Mass Corrections}\\
\\
Relation $\hat{p}_a=\xi P$ is usually introduced in an  
\textquotedblleft infinite\textquotedblright\: reference system 
where masses of particles are negligible compared to momenta;
it is a connection between hadronic and partonic view,
correct if the mass of target-nucleon is disregarded. 
In fact, using light-cone formalism (appendix \ref{light_cone}), 
it is possible to demonstrate that $\hat{p}^{+}=\xi P^{+}$, but 
the partonic momentum $\hat{p}^{-}$ is not simply a rescaling 
of the \textquotedblleft small\textquotedblright\: momentum 
$P^{-}$ of the nucleon, i.e. $\hat{p}^{-}\neq\xi P^{-}$
\cite{KRRE},\cite{PDF_Soper}.\\
Consequently {\it Target Mass Corrections} (TMC) are needed;
they have non-perturbative nature.
We introduce the Natchmann $\eta$ \cite{NATCHMANN} variable defined by
\begin{equation}\label{natch_var}
\frac{1}{\eta}=\frac{1}{2x}\left(1+\rho\right)
\qquad
\rho=\sqrt{1+\left(\frac{2Mx}{Q}\right)^2}
\end{equation}
(it holds for massless partons inside nucleons)
and we describe the hadronic functions $dF_k/dz$ imposing 
the substitution $\chi\rightarrow\eta$ in relations \ref{compact_F_j}
(and not into the coefficient of $\mathscr{F}_2$, written on purpose
as $2x$).\\
In the collinear limit ($\hat{p}_{\perp}=0$) new relations have found \cite{Tung_Aivazis_Olness}
\begin{equation}\label{F_TMC}
\begin{split}
\frac{dF_1}{dz}(x,y,z)=&\mathscr{F}_1(x,y,z)\\
\frac{dF_2}{dz}(x,y,z)=&2\frac{x}{\lambda}\frac{\mathscr{F}_2(x,y,z)}{\rho^2}\\
\frac{dF_3}{dz}(x,y,z)=&2\frac{\mathscr{F}_3(x,y,z)}{\rho}\\
\frac{dF_4}{dz}(x,y,z)=&\frac{1}{\lambda}\frac{(1-\rho)^2}{2\rho^2}\mathscr{F}_2(x,y,z)
+\mathscr{F}_4(x,y,z)+\frac{(1-\rho)}{\rho}\mathscr{F}_5(x,y,z)\\
\frac{dF_5}{dz}(x,y,z)=&\frac{\mathscr{F}_5(x,y,z)}{\rho}
+\frac{(1-\rho)}{\lambda\rho^2}\mathscr{F}_2(x,y,z)
\end{split}
\end{equation}
where a mixing involving terms $k=4,5$ appears.
In this case $\mathscr{F}_k(x,y,z)$ is defined as
\begin{equation*}
\mathscr{F}_k(x,y,z)=\sum_{a,h}
\int_{\eta/\lambda}^1\frac{d\xi}{\xi}
\int_{z}^1\frac{d\zeta}{\zeta}
\hat{\mathscr{F}}_k^{a}(\xi,y,\zeta)
\mathfrak{f}^{a}\left(\frac{\eta/\lambda}{\xi}\right)
\mathfrak{D}_h\left(\frac{z}{\zeta}\right)
\end{equation*}
Furthermore $\lambda$ parameter \eqref{lambda_definition} 
has been introduced in order to include also the
treatment of massive quarks produced in final state.
Relations \ref{F_TMC} hold up to NLO for initial state partons collinear to
nucleon ($\hat{p}_{\perp}=0$) or up to LO, if $\hat{p}_{\perp}\neq0$; 
for the latter case it is possible an extension up to NLO
completing with further contributions \cite{KRRE_TM_2},
\cite{KRRE_TM_3} beyond the purpose of this investigation.\\
It can be trivially verified that, if $M\rightarrow 0$, 
situation \ref{compact_F_j} is recovered from \ref{F_TMC} 
with null TMC
\begin{equation}
\begin{split}
&\qquad\hat{p}_a\rightarrow\xi P\qquad
\eta\rightarrow x\qquad
\rho\rightarrow 1\\
\frac{dF_{1,4,5}}{dz}&\rightarrow\mathscr{F}_{1,4,5}\qquad
\frac{dF_2}{dz}\rightarrow2\frac{x}{\lambda}\mathscr{F}_2\qquad
\frac{dF_3}{dz}\rightarrow2\mathscr{F}_3
\end{split}
\end{equation}

\begin{figure}[!h]
\centering
{\includegraphics[angle=270,width=14cm]{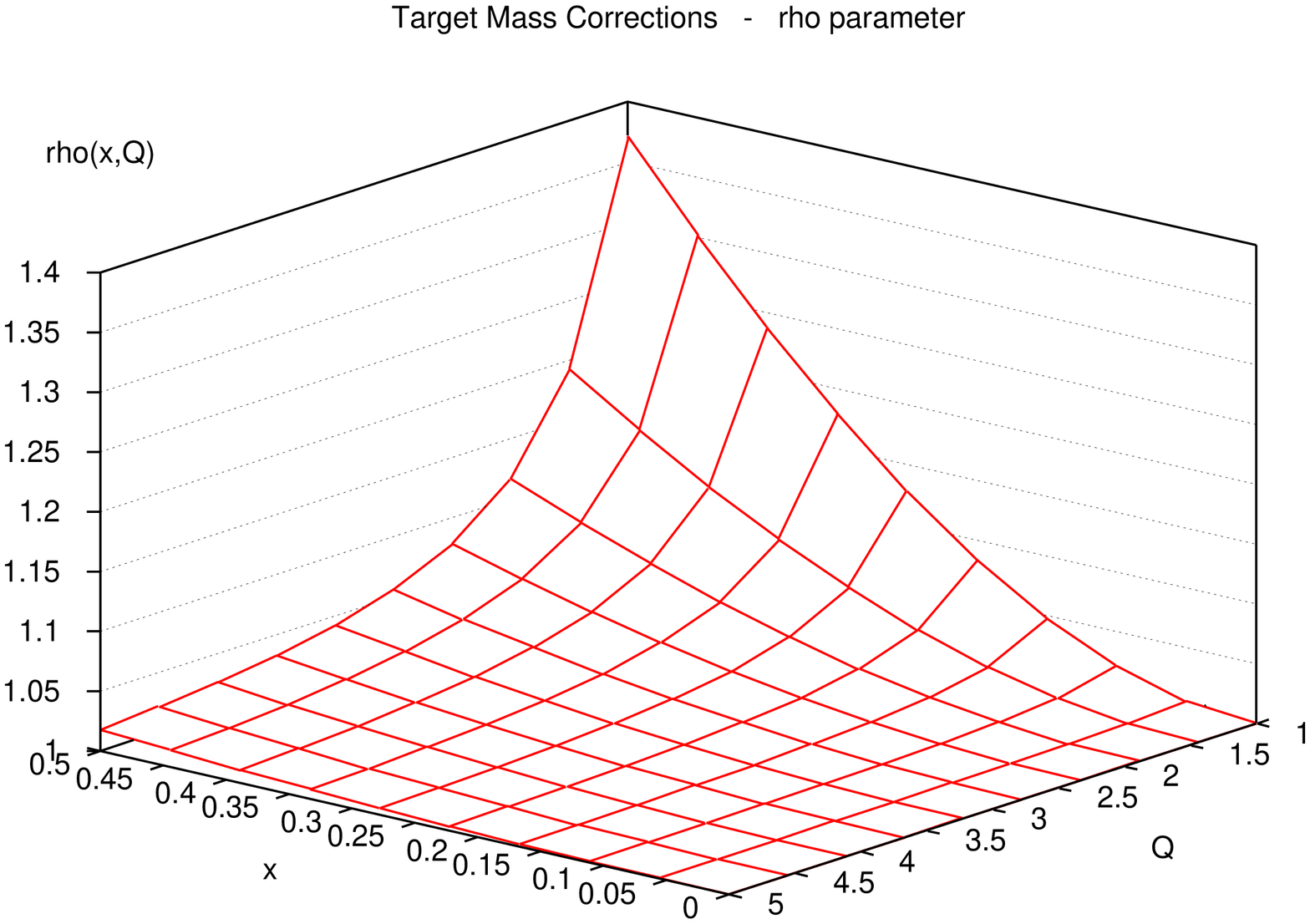}}
{\includegraphics[angle=270,width=13.5cm]{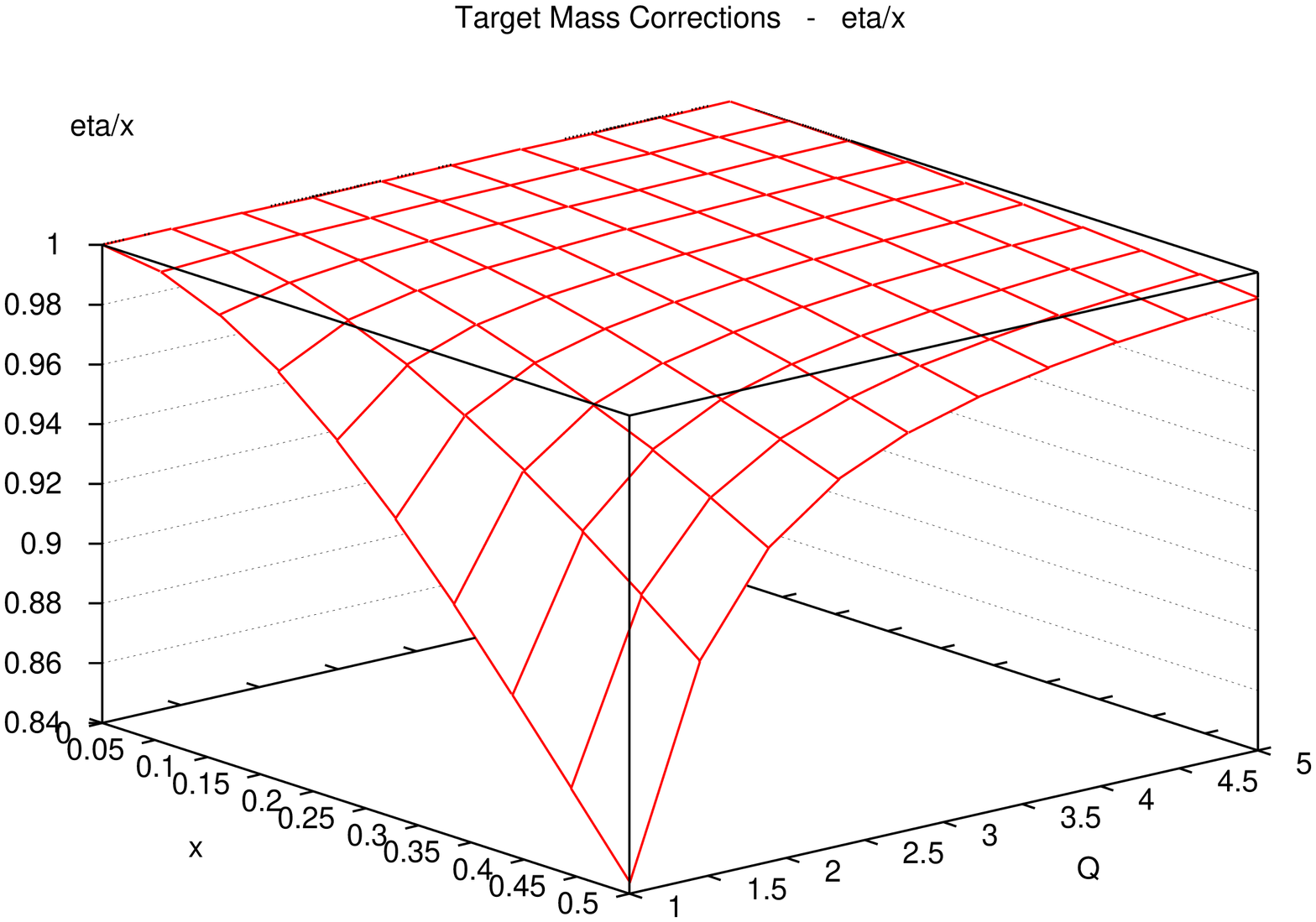}}
 \caption{\footnotesize Target Mass Corrections - Dependence on $x$ and $Q$.}
 \label{TMC_x_q}
\end{figure}

\vspace{2cm}
\noindent
In short, in accordance with the hypothesis on the 
mass of the target-nucleon ($M_{Nucl}$) and the produced quark ($m$), 
$\chi$ has to be chosen following the table
\begin{center}
\begin{tabular}{|c||c|c||}\hline
Quark Mass  &\multicolumn{2}{c||}{Nucleon Mass}\\
            &$M_{Nucl}=0$             &$M_{Nucl}\neq0$  \\\hline
$m=0$     &$\chi=x$                 &$\chi=\eta$   \\\hline
$m\neq0$  &$\chi=x/\lambda$         &$\chi=\eta/\lambda$ \\\hline
\end{tabular}
\end{center}
or, without loss of generality, setting $\chi=\eta/\lambda$:
opportune quantities are then included by default (according to the assumptions done) .\\
As reported in \cite{Tung_Aivazis_Olness}, 
being the Charm mass not so different from the one of a nucleon, 
if corrections for massive Charm quark are taken into account, 
then also TMC should be included.
Considering only the former would be not consistent;
in fact, in some configurations of kinematics,
the discrepancy can be even more than 25{\small{\%}}.
In Fig.\ref{TMC_x_q} it is illustrated the behaviour of $\rho$ and $\eta/x$ 
in the regions $0 \leq x\leq 0.5$ and $1 \leq Q[GeV]\leq 5$.

\newpage

\chapter{Coefficient Functions}
\label{chapter_Coeff_Funct}

{\it
In the previous chapter we have shown that the
cross section is proportional to structure 
functions $\mathscr{F}_{k}$ \eqref{compact_F_j},
having the form 
\begin{equation*}
\mathscr{F}_k=\sum_{a,h}
\int_{\chi}^1\frac{d\xi}{\xi}
\int_{z}^1\frac{d\zeta}{\zeta}
\hat{\mathscr{F}}_k^{a}(\xi,y,\zeta)
\mathfrak{f}^{a}\left(\frac{\chi}{\xi}\right)
\mathfrak{D}_h\left(\frac{z}{\zeta}\right)
\end{equation*}
Now, in the context of perturbative QCD,
we calculate the expression of the
\textquotedblleft Coefficient Functions\textquotedblright\:
up to Next-to-Leading Order, using the decomposition 
$$\hat{\mathscr{F}}_k^{a}=\hat{\mathscr{F}}_k^{a,LO}
+\hat{\mathscr{F}}_k^{a,NLO}$$}\\
\\
\\
\noindent
For the treatment of the generic partonic tensor,
unknown functions $h_k$ have been introduced
in order to parametrise the cross section;
assuming that QCD is the right theory to describe 
the structure of hadrons (and then in particular of the nucleons), we are carrying out 
the analytic form for this kind of functions.
Still referring to CC DIS process $\:\nu_{\ell}+N\rightarrow \ell^-+X$
with heavy quarks production, we obtain the tensor 
$h^{a,QCD}_{\mu\nu}$ from amplitudes of Feynman diagrams
with a parton of kind $a$ in the initial state.
We introduce a set of projectors $P_{k}^{\mu\nu}$
(appendix \ref{appendix_Projectors}) so that
$$P_{k}^{\mu\nu}h_{\mu\nu}^a=p_k h_{k}^a$$
where $h_{\mu\nu}^a$ is the partonic tensor 
in \ref{partonic_tensor} and $p_k$ are  
opportune coefficients \eqref{Projectors}. 
Therefore 
\begin{equation}\label{QCD_parton_h_k}
h_k^a= \frac{P_{k}^{\mu\nu}h_{\mu\nu}^a}{p_k}
\equiv
\frac{1}{4\pi}\frac{P_{k}^{\mu\nu}h_{\mu\nu}^{a,QCD}}{p_k}
\end{equation}
identifying the QCD tensor with the partonic one
and inserting the proper normalization factor $4\pi$,
in accordance with the convention introduced in \ref{generic_cross_section}.
Consequently it is possible to carry out the analytical form of
every term $h_k^a$ and then to obtain the expression of
$\hat{\mathscr{F}}_k^{a}$ through relation \ref{partonic_dF}
\begin{equation}\label{QCD_dF}
\hat{\mathscr{F}}_{k}^{a}(\xi,y,\zeta)
  =\hat{\mathscr{A}}_k h_k^a\int d\hat{\varphi}_{X}
  = \frac{\hat{\mathscr{A}}_k}{4\pi}
     \left(\frac{P_{k}^{\mu\nu}h_{\mu\nu}^{a,QCD}}{p_k}\right)
     \int d\hat{\varphi}_{X}
\end{equation}
with $\hat{\mathscr{A}}_{1,2,3,5}=4\hat{\upsilon}$,
$\hat{\mathscr{A}}_{4}=4\xi\hat{\upsilon}$.$^{1}$
\footnotetext[1]{
As a consequence of \ref{compact_F_j},  
$\hat{\mathscr{A}}_{2,3}=\hat{A}_{2,3}/2=\hat{A}_{1,5}$}
Deriving from perturbative QCD, tensor
$h_{\mu\nu}^{a,QCD}$ is calculable at every perturbative order;
for each parton $a$, from now on definitely identified
with quarks ($q$) or gluons ($g$), we can then expand
\begin{equation}
h_{\mu\nu}^{a,QCD}= h_{\mu\nu}^{a,LO}+h_{\mu\nu}^{a,NLO}+\textellipsis
\end{equation}
At Next-to-Leading Order we can coherently write
\begin{equation*}
\hat{\mathscr{F}}_{k}^{a}(\xi,y,\zeta)
  =\hat{\mathscr{F}}_{k}^{a, LO}(\xi,y,\zeta)+
   \hat{\mathscr{F}}_{k}^{a, NLO}(\xi,y,\zeta)
\end{equation*}
with obvious notation.
All calculations have been performed by hand and 
subsequently checked with the aid of FORM \cite{FORM} 
or MATHEMATICA \cite{MATHEMATICA}.

\newpage
\section{Tools for Calculations}
\subsection{Dimensional Regularization}\label{Dimension}

We have chosen the {\it dimensional regularization} approach to deal with 
ultraviolet and infrared divergences, met respectively in 
virtual and real contributions at NLO:
we have also chosen to work in a Minkowski space-time with 
$D=4+2\varepsilon$ dimensions.
In order to obtain convergence of the expressions, $\varepsilon<0$ is formally assumed for the ultraviolet case,
whereas $\varepsilon>0$ for the infrared one;
in particular there are poles of kind $1/\varepsilon^n$ that, when
$\varepsilon\rightarrow 0$, indicate the presence of a divergence in 
four dimensions.
Unlike other regularization techniques, the dimensional one allows
to maintain gauge and Lorentz invariance.
Some prescriptions are needed in order to work in a $D$-dimensional space-time:
\begin{itemize}
\item rescaling the coupling constants by means of a parameter
with dimension of a mass, to maintain the right 
dimensionality of the Lagrangian:
$g\rightarrow g\mu^{\frac{4-D}{2}}$.
\item using relations holding for an Algebra in 
$D$ dimensions (appendix \ref{Clifford}) and extending 
to $D$ dimensions also the phase-space (appendix \ref{partonic_phase_space}).
\item paying attention particularly to $\gamma_5$ matrix 
inside the electroweak vertex
$\gamma^{\mu}\left(\frac{1-\gamma_5}{2}\right)$.
\end{itemize}
This last point is very delicate:
in fact $\gamma_5=i\epsilon_{\mu\nu\rho\tau}
\gamma^{\mu}\gamma^{\nu}\gamma^{\rho}\gamma^{\tau}/4!$,
however the Levi-Civita tensor is not well defined beyond
four dimensions. In this case it is necessary to 
adopt a particular scheme of regularization to avoid
lacks of foundation
\cite{DISSERTORI},\cite{G5_Weinzierl},\cite{G5_Dim_Reg},\cite{G5_PV}.
We have chosen the HVBM scheme \cite{KRST},\cite{G5_HV},\cite{G5_BM},
where the prescription is
\begin{equation}\label{Gamma5_prescription}
\{\gamma^{\mu},\gamma_5\}=0\:\:\:\:\:\mu=0,1,2,3\qquad\qquad\qquad
[\gamma^{\mu},\gamma_5]=0\:\:\:\:\text{otherwise}
\end{equation}
Such a trick splits the $D$-dimensional space-time in two 
sub-spaces of $4$ and $D-4$ dimensions respectively:
the relation on the left of \ref{Gamma5_prescription} is applied for the former,
whereas the one on the right for the latter.
In accordance with remarks in \cite{KRST}, for the case of 
of unpolarised DIS some simplifications appear, so $\gamma_5$ 
contributions are only those arising from the 4-dimensional sub-space.

\newpage
\subsection{Expansions in the Sense of Distributions}\label{Distrib_pgf}

It must be paid particular attention to the manner to isolate 
singularities for the encountered quantities;
in fact, to highlight the pole structure through $\varepsilon$ parameter, 
it is necessary to perform expansions in the sense of mathematical distributions
(appendix \ref{DISTRIBUTIONS}).
In order to maintain the analysis as general as possible and subsequently 
to recover the massless case, we have to deal in an accurate way with 
the contributions not containing poles for massive case, but
containing for the massless one. For example, the term
$$\frac{(1-\xi)^{1+2\varepsilon}}{(1-\lambda\xi)^{2+\varepsilon}}$$
is not singular for the massive case ($\lambda\neq1$, $\lambda<1$),
so a Taylor expansion apparently seems to be opportune 
$$\frac{(1-\xi)^{1+2\varepsilon}}{(1-\lambda\xi)^{2+\varepsilon}}
=\frac{(1-\xi)}{(1-\lambda\xi)^2}\bigg[1+
\varepsilon \bigg(2\log(1-\xi)-\log(1-\lambda\xi)\bigg)
+O(\epsilon^{2})\bigg]
$$
When $\varepsilon\rightarrow 0$ we would obtain $\frac{(1-\xi)}{(1-\lambda\xi)^2}$,
however if $\lambda=1$, that is recovering the massless limit, it becomes 
$(1-\xi)^{-1}$, clearly diverging when integrated up to $\xi=1$.
Strategy of expansion therefore consists in considering the quantity
$C=A+B$, where $A$ and $B$ are the exponents of the generic expression
$(1-\xi)^{A}(1-\lambda \xi)^{B}$;
if $C{\small(\varepsilon=0)}<0$ we need an expansion in the sense of mathematical distributions,
else such a term would be definitely finite also for the massless case.
With regard to the previous example, $C(\varepsilon=0)=-1$ so,
expanding in distributional sense, (about the definition of $K_A$, see paragraph \ref{NLO})
\begin{equation*}
\begin{split}
(1-\xi)^{1+2\epsilon}(1-\lambda \xi)^{-2-\epsilon}=
&\delta(1-\xi)\left[\frac{1}{\epsilon}-\frac{(1-\lambda)^{\epsilon}}{\epsilon}
-\frac{1+\lambda}{\lambda}K_{A}-\frac{(1-\lambda)^{\epsilon}}{\lambda}\right]\\
&+\left[\frac{1-\xi}{(1-\lambda \xi)^2}\right]_{+}+O(\epsilon)
\end{split}
\end{equation*}
from which immediately both massless
\begin{equation*}
(1-\xi)^{-1+\varepsilon}=\frac{1}{\epsilon}\delta(1-\xi)+\frac{1}{(1-\xi)_{+}}
+O(\epsilon)
\end{equation*}
and massive case
\begin{equation*}
\begin{split}
\frac{(1-\xi)}{(1-\lambda \xi)^2}
=
-\delta(1-\xi)\left[\log(1-\lambda)+\frac{1+\lambda}{\lambda}K_{A}+\frac{1}{\lambda}\right]
+\left[\frac{1-\xi}{(1-\lambda \xi)^2}\right]_{+}
\end{split}
\end{equation*}
can be recovered.

\newpage
\subsection{Fragmentation Variable}\label{FRAGMENTATION}

In paragraph \ref{HP_correspondence} we have shown 
that the hadronic cross section is given by a convolution
of the partonic one with Parton Distribution Functions
and Fragmentation Functions.
For both PDF and FF the link between partonic and hadronic
points of view occurs imposing a prescription connecting 
a hadronic observable quantity to an analogous partonic variable (not observable).
For FF the choice of carrying out such a correspondence is quite
free.
As an example, a way consists in considering the energy 
of the final state hadron and imposing that is a 
fraction of the one carried by the originating parton. 
The same arguments hold for momentum or any components.
Working with high momenta, all choices are almost equivalent,
whereas when masses and energies become comparable,
the schemes remarkably differ.
In any case it has to be defined a {\it fragmentation variable},
that is the quantity of interest allowing a link between
partonic and hadronic level.
For our investigation we introduce the observable \cite{KRST}
\begin{equation}\label{macro_z}
z=\frac{P_H\cdot P_N}{q\cdot P_N}
\end{equation}
where $P_H$ is the four-momentum of the produced hadron,
$P_N$ is the one of the initial state nucleon and $q$
the one of $W$ boson \eqref{Kine_DIS}.
In the reference system where the nucleon is at rest,
previous expression reduces to \cite{KMO}
\begin{equation*}
z=\frac{E_H}{E_W}
\end{equation*}
where $E_H$ and $E_W$ are respectively the energies
of the produced hadron and $W$ boson.
We can introduce an analogous quantity at parton level
\begin{equation}\label{micro_z}
\zeta=\frac{\hat{p}_b\cdot \hat{p}_a}{q\cdot \hat{p}_a}
\end{equation}
where $\hat{p}_b$ is the momentum of the final state parton
generating the hadron and $\hat{p}_a$ the one of the initial state;
imposing the prescription defined through $z$, a fraction
$t\in(0,1)$ of $\zeta$, the typical convolution as in 
\ref{compact_F_j} is recovered
\begin{equation*}
\int^1_{\zeta_{min}}\!\!\!\!\!\! d\zeta
\hat{\mathscr{F}}_k^{a}(\xi,y,\zeta)
\int_0^1\!\! dt \mathfrak{D}_{h}(t)
\delta(z-t\zeta)
=\int_{max\left\{z,\zeta_{min}\right\} }^1
\!\!\!\!\!\!\!\!\!\!\!\!\!\!\!\!\!\!\!\!\!\!\!\!
\hat{\mathscr{F}}_k^{a}(\xi,y,\zeta)
\mathfrak{D}_{h}\left(\frac{z}{\zeta}\right)\frac{d\zeta}{\zeta}
\end{equation*}
Notice that $z=t\zeta$ correctly holds if 
$\hat{p}=\xi P_N$ (that is neglecting 
Target Mass Corrections \eqref{QCD_approach}) 
and that momentum $P_H$ of the final state hadron 
is a fraction $t$ of the originating parton ($P_H=t \hat{p}_b$); 
this last hypothesis entails that masses are negligible
with respect to the considered momenta, being
\begin{equation*}
\hat{p}_b=\left(\sqrt{m_b^2+|\underline{p}_b|^2},\underline{p}_b\right)
\qquad
P_H=\left(\sqrt{m_H^2+|\underline{P}_H|^2},\underline{P}_H\right)
\end{equation*}


\noindent
After the choice of the hadronic fragmentation variable,
the Coefficient Functions have to be rewritten by means
of the corresponding partonic quantity.
Dealing with the partonic phase-space (appendix \ref{partonic_phase_space}) 
we have introduced the variable
$\hat{w}=(1+\cos\theta)/2$ where $\theta$ is the scattering angle 
between the produced heavy quark (mass $m$) and the straight line
formed by the momentum ($\underline{q}$) of $W$ and of the
initial state parton ($\underline{p}$), in the reference system 
of the centre of mass ($\underline{p}+\underline{q}=0)$.
In the following paragraphs we calculate 
the functional form for Coefficient Functions in 
terms of $\hat{w}$ variable, obtaining then 
$\mathscr{F}^a_k(\xi,y,\hat{w})$;
in order to convolve with FF, it is necessary to perform 
a change of variables to get $\zeta$ as defined in \ref{micro_z}:
\begin{equation}
\begin{split}
&\hat{w}=\left[\zeta -\frac{(1-\lambda)\xi}{(1-\lambda\xi)}\right]
\frac{(1-\lambda \xi)}{(1-\xi)}
\qquad\qquad
d\hat{w}=\frac{(1-\lambda \xi)}{(1-\xi)}d\zeta\\
&\hat{w}_{min}=0\rightarrow \zeta_{min}=\frac{(1-\lambda)\xi}{(1-\lambda\xi)}\\
&\hat{w}_{max}=1\rightarrow \zeta_{max}=1
\end{split}
\end{equation}
It is very important to pay attention to such an operation, 
because we are not dealing with ordinary functions, but with distributions;
in fact the following rules hold
\begin{equation}
\begin{split}
\delta(1-\hat{w})d\hat{w}&\rightarrow \delta(1-\zeta)d\zeta\\
\frac{1}{(1-\hat{w})_+}d\hat{w}&\rightarrow \frac{1}{(1-\zeta)_{\oplus}}d\zeta\\
\left[\frac{\log(1-\hat{w})}{1-\hat{w}}\right]_+d\hat{w}&\rightarrow
\left[\frac{\log(1-\zeta)/(1-\zeta_{min})}{1-\zeta}\right]_{\oplus}d\zeta
\end{split}
\end{equation}
introducing distribution $\oplus$ defined by
\begin{equation}
\int_{\zeta_{min}}^1d\zeta f(\zeta)\left[g(\zeta)\right]_{\oplus}\equiv
\int_{\zeta_{min}}^1d\zeta\left[f(\zeta)-f(1)\right]g(\zeta)
\end{equation}
Furthermore, being $\zeta_{min}\rightarrow 1$ when $\xi\rightarrow 1$,
\begin{equation}
\delta(1-\xi)f(\zeta,\xi)= \delta(1-\xi)\delta(1-\zeta)
\left[\int_{\zeta_{min}}^1f(t,\xi)dt\right]_{\xi=1}
\end{equation}


\newpage
\section{Leading Order}\label{leading_order}
\begin{figure}[!h]
 \centering\epsfig{file=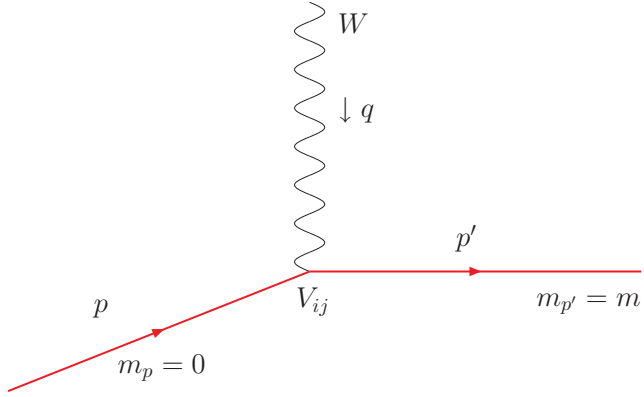, width=8.7 cm}
 \caption{\footnotesize CC DIS: parton contribution at Leading Order.}
 \label{born}
 \end{figure}

At the first perturbative order only the Feynman diagram
in Fig.\ref{born} contributes to the scattering amplitude.

 \noindent
Partonic vertex is then represented as point-like interaction 
of a quark with the electroweak boson, so the formulation
is the one explained in appendix \ref{leptonic_tensor},
because, like leptons, also quarks are elementary fermions. 
QCD tensor at Born level for this process is therefore
\begin{equation}\label{QCD_born}
h_{\mu\nu}^{q,LO}=\left\{2\left[p_{\mu}p'_{\nu}+p'_{\mu}p_{\nu}-(pp')g_{\mu\nu}\right]
-2i\epsilon_{\mu\nu}^{\:\:\:\:\alpha\beta}p_{\alpha}p'_{\beta}\right\}|V_{ij}|^2
\end{equation}
and from four-momentum conservation ($p'=p+q$), it follows
$h_{4}^{q,LO}=0$ and $h_{k=1,2,3,5}^{q,LO}=|V_{ij}|^2/4\pi$, 
in accordance with contractions in \ref{QCD_parton_h_k}.
Such a result is clear also at first sight comparing \ref{QCD_born}
to \ref{partonic_tensor}.
The phase-space for the final state of a massive quark is given by
\begin{equation*}
\begin{split}
d\varphi_X^{LO}
&=\frac{d^4p'}{(2\pi)^3}
\delta_+\left(p'^{2}-m^{2}\right)(2\pi)^{4}
\delta^{(4)}\left(p'-q-p\right)
\equiv d\hat{w} \int d\hat{\varphi}^{LO}_X\\
&=  2\pi \frac{\lambda}{Q^2}\delta\left(1-\xi\right)
\delta(1-\hat{w})  d\hat{w}
\end{split}
\end{equation*}
as referred in \ref{ps_lo_fact}.\\
Finally we obtain the expression of the Coefficient Functions,
according with \ref{QCD_dF}
\begin{equation}\label{LO_result}
\boxed{
\begin{split}
\hat{\mathscr{F}}_k^{a,LO}(\xi,y,\hat{w})
&=\frac{\mathscr{A}_k}{4\pi} h_k^{a,LO}\int d\hat{\varphi}_X^{LO}\\
&=(1-\delta_{k4})\delta_{aq}|V_{ij}|^2\delta\left(1-\xi\right)
\delta(1-\hat{w})
\end{split}}
\end{equation}
At LO it does not exist any contribution coming from 
gluons, then $h^{g,LO}_k=0$ $\forall k$: 
factor $\delta_{aq}$ takes it into account.

\section{Next-to-Leading Order}\label{NLO}

\begin{figure}[!h]
\centering
{\includegraphics[width=6cm]{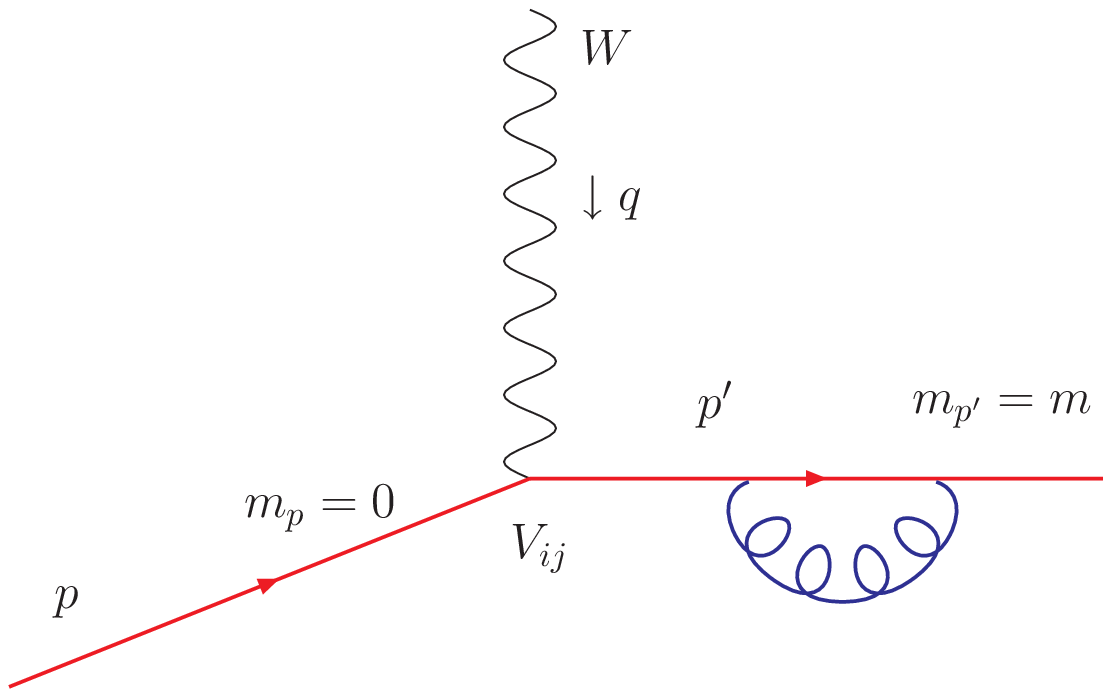}}
 \hspace{4mm}
{\includegraphics[width=6cm]{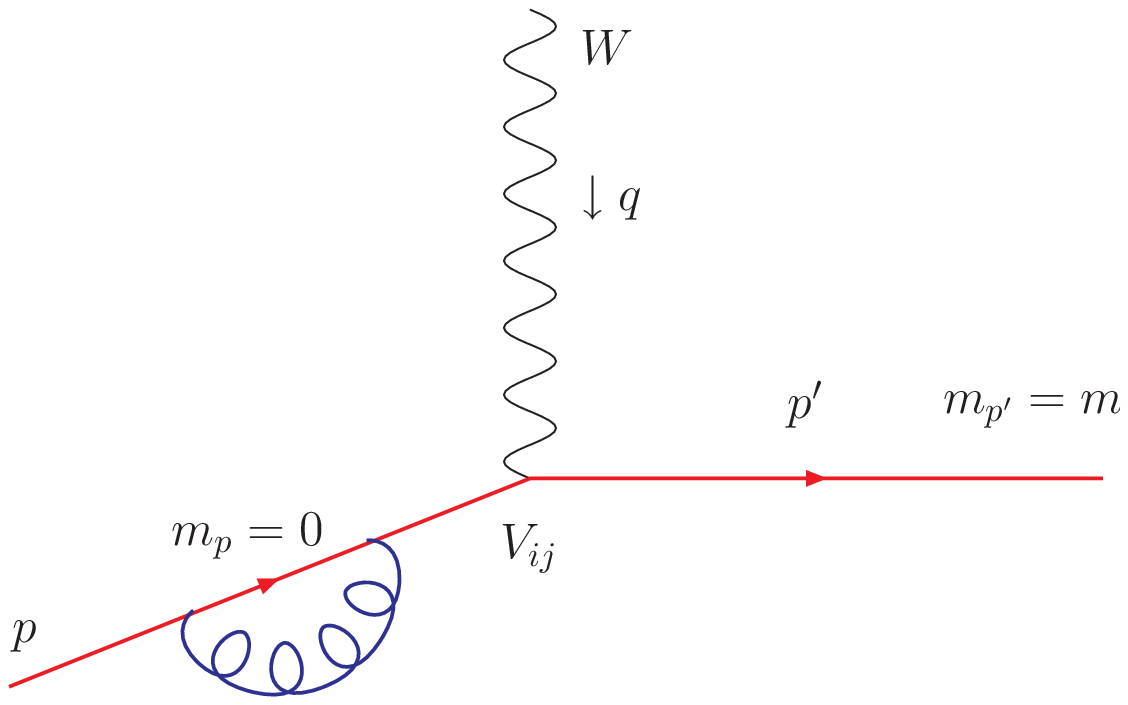}}
 \caption{\footnotesize CC DIS: Self-energy (Virtual NLO contribution).}
 \label{Virtual_Self}
\end{figure}

\begin{figure}[!h]
\centering
{\includegraphics[width=5cm]{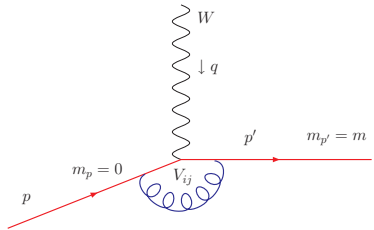}}
 \caption{\footnotesize CC DIS: Vertex correction (Virtual NLO contribution).}
 \label{Virtual_Vertex}
\end{figure}

\begin{figure}[!h]
\centering
{\includegraphics[width=6.5cm]{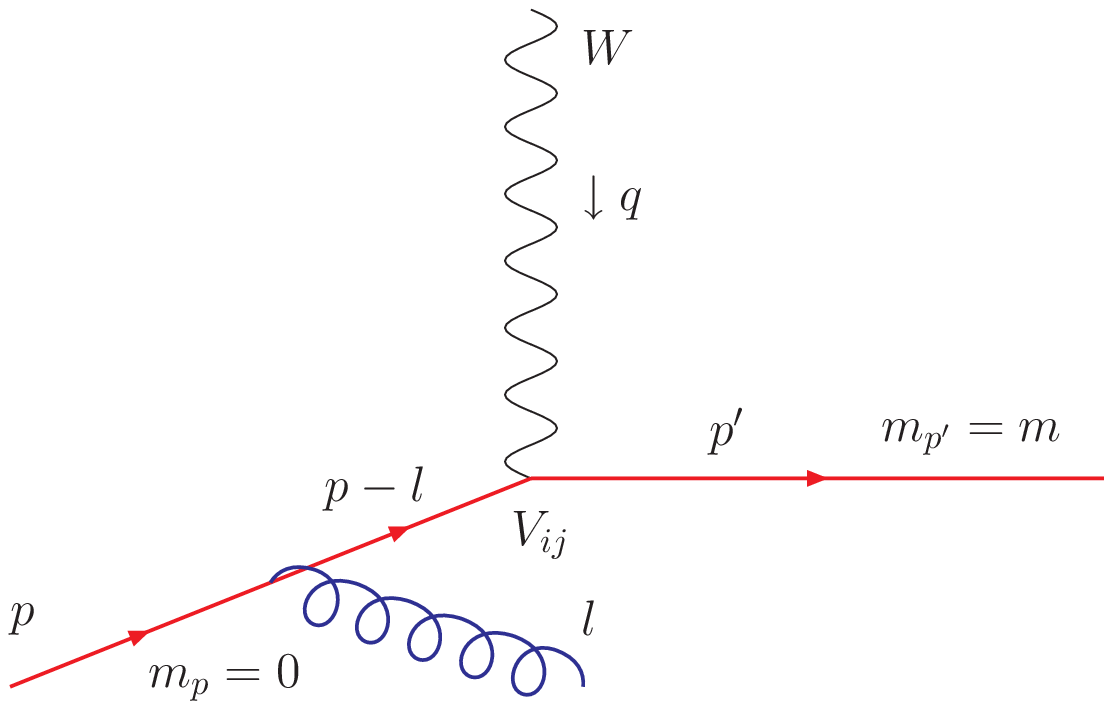}}
 \hspace{4mm}
{\includegraphics[width=6.5cm]{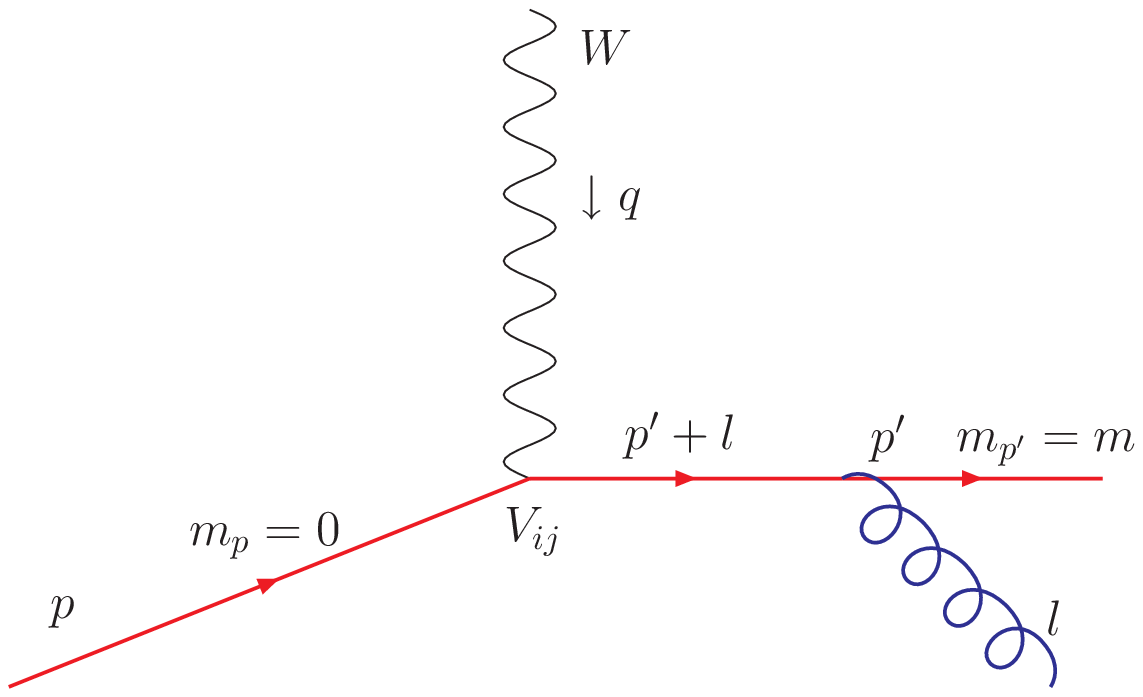}}
 \caption{\footnotesize CC DIS: quark-channel (NLO Real contribution).}
 \label{Real_Quark}
\end{figure}

\begin{figure}[!h]
\centering
{\includegraphics[width=6.5cm]{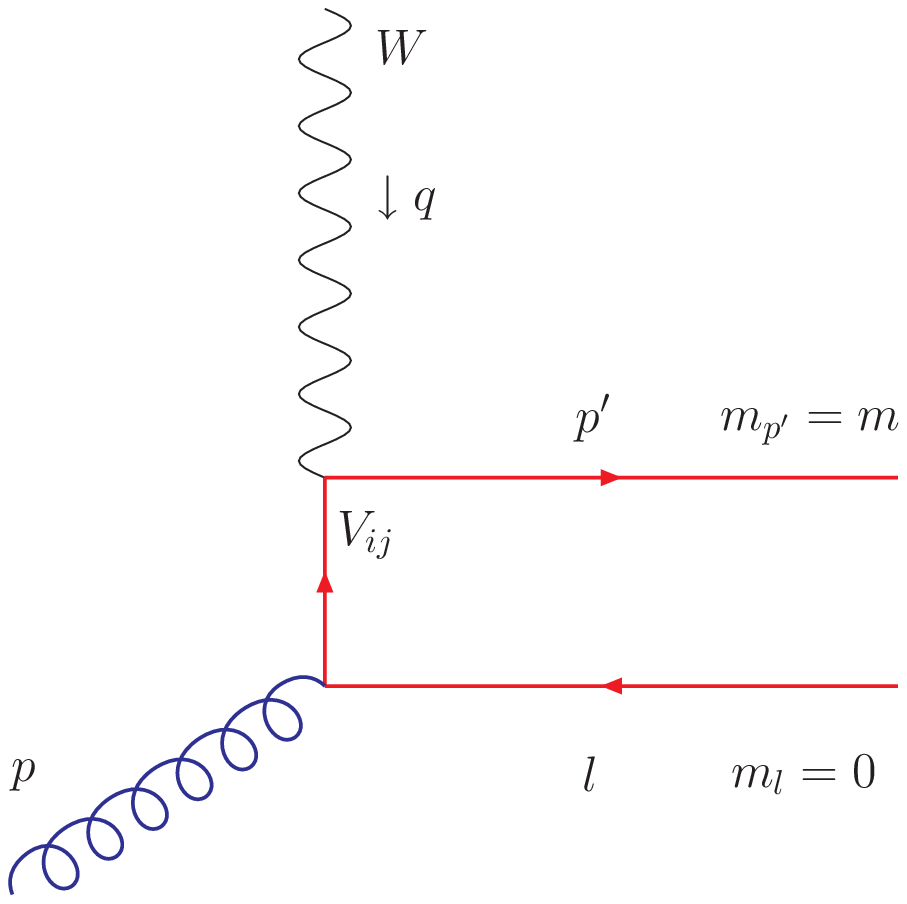}}
 \hspace{4mm}
{\includegraphics[width=6.5cm]{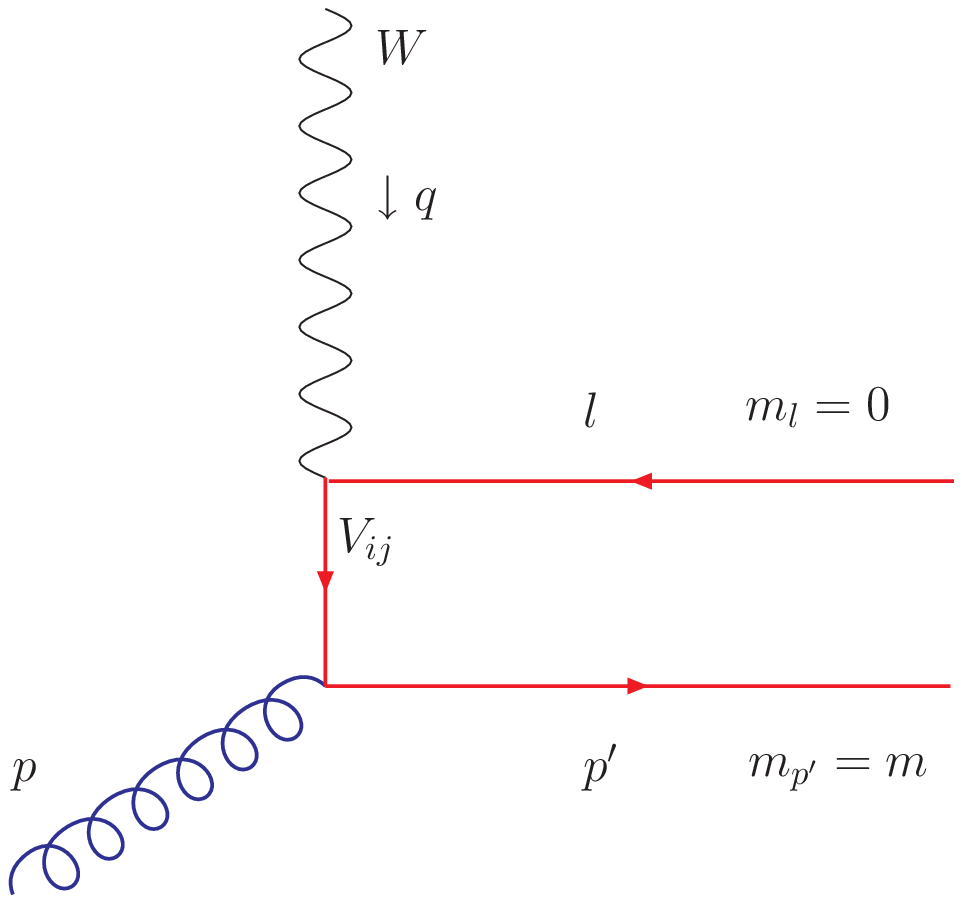}}
 \caption{\footnotesize CC DIS: gluon-channel (NLO Real contribution).}
 \label{Real_Gluon}
\end{figure}

At Next-to-Leading Order Feynman diagrams are of two kinds:
in addition to the quark channel already present at LO, 
a new channel appears where an initial state gluon generates 
a quark/antiquark pair by splitting (Fig.\ref{Real_Gluon});
one of these partons interacts then with $W$ boson.
We have therefore to distinguish between quark-channel and gluon-channel.
The contribution directly coming from quarks is illustrated in Fig.\ref{Real_Quark}
and consists in a radiative correction of LO, that is in gluon emission from a quark.\\
To obtain the QCD partonic tensor at NLO, we have to sum, for each production channel, the amplitudes of two Feynman diagrams and calculate the squared modulus.
As expected in Quantum Field Theory, these contributions diverge if integrated
on the whole phase-space, because of soft and collinear singularities;
introducing the dimensional regularization formalism (paragraph
\ref{Dimension}) divergences are carried out as $\varepsilon$ poles  
and can be analytically treated. 
Obviously at NLO also virtual diagrams in Fig.\ref{Virtual_Self} 
and Fig.\ref{Virtual_Vertex} have to be included;
the latter cancel soft divergences of quark channel
leaving only the collinear ones, according to
KLN theorem (\cite{K}, \cite{LN}).\\
In accordance with appendix \ref{partonic_phase_space},
the phase-space for real corrections at NLO is
\begin{equation}\label{phase_space_hat}
\begin{split}
d\varphi_X^{NLO}&=\frac{1}{8\pi}\left(\frac{Q^{2}+m^{2}}{4\pi}\right)^{\epsilon}\frac{1}{\Gamma(1+\epsilon)}
\hat{w}^{\epsilon}(1-\hat{w})^{\epsilon}\xi^{-\epsilon}(1-\xi)^{1+2\epsilon}
(1-\lambda\xi)^{-1-\epsilon}d\hat{w}\\
&=\left[\int d\hat{\varphi}_X^{NLO}\right]d\hat{w}
\end{split}
\end{equation}
expressed through $\xi$ and $\hat{w}$ variables defined 
in appendix \ref{partonic_phase_space}
$$\xi=\frac{\hat{x}}{\lambda}\:\:\:\:\:\:\lambda=\frac{Q^2}{Q^2+m^2}
\:\:\:\:\:\:\hat{w}=\frac{1+\cos\theta}{2}
$$
where $\theta$ is the scattering angle between the heavy quark (mass $m$)
and the straight line of momentum ($\underline{q}$) of $W$ boson and 
initial state parton ($\underline{p}$), in the centre of mass of the system of reference ($\underline{p}+\underline{q}=0)$;
$\hat{x}$ and $\lambda$ have been respectively introduced in 
\ref{parton_variables} and \ref{lambda_definition}.\\
Using the same variables for the phase-space, some invariants can be rewritten as
\begin{align}
2pp'&=\frac{Q^2}{\lambda}\left[
\frac{\hat{w}}{\xi}+\frac{(1-\lambda)(1-\hat{w})}{1-\lambda\xi}  \right] \notag\\
2pl &=\frac{Q^2}{\lambda}\frac{(1-\xi)(1-\hat{w})}{\xi(1-\lambda \xi)} \label{scalar_products} \\
2p'l&=\frac{Q^2}{\lambda}\frac{(1-\xi)}{\xi}\notag
\end{align}
in agreement to kinematics convention shown in  figures 
\ref{Real_Quark} and \ref{Real_Gluon}.
For sake of completeness, we introduce other definitions and
expressions used in the following
\begin{align*}
\hat{s}=&\left(p+q\right)^2=\frac{Q^2}{\lambda}\left(\frac{1-\lambda\xi}{\xi}\right)&
m^2=&Q^2\left(\frac{1-\lambda}{\lambda}\right)&\\
K_A=&\frac{(1-\lambda)}{\lambda}\log(1-\lambda)&
\text{Li}_2(\lambda)=&-\int_0^1dt\frac{\log(1-\lambda t)}{t}&
\end{align*}
Having carried out calculations using $\hat{w}$ variable,
it is possible to move to the fragmentation variable $\zeta$ 
in accordance with paragraph \ref{FRAGMENTATION}.

\newpage
\subsection{Virtual Contribution}
\begin{figure}[!h]
\centering
{\includegraphics[width=11cm]{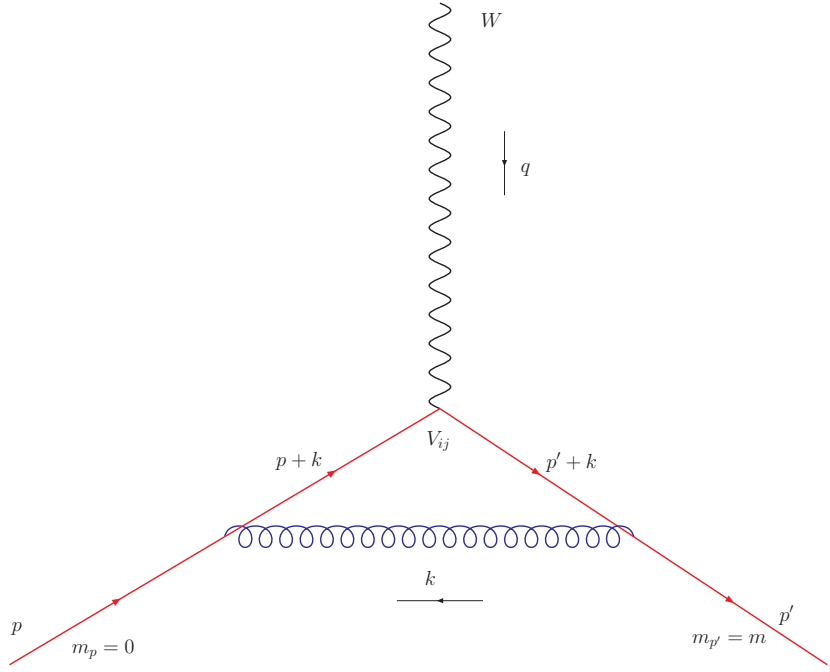}}
 \caption{\footnotesize Kinematics of Vertex-correction.}
 \label{Vertex}
\end{figure}

In this section we carry out the contribution of virtual diagrams
appearing at NLO. We can distinguish between 
self-energy (Fig.\ref{Virtual_Self}) or vertex (Fig.\ref{Virtual_Vertex})
contributions.
For diagram in Fig.\ref{Vertex}

\noindent
we can write the amplitude as
\begin{equation*}
\mathcal{M}^V_{\nu}=
-i\frac{g_w}{\sqrt{2}}V_{ij}
\overline{u}_{out}^{(s)}(p')\Lambda_{\nu}\frac{(1-\gamma_5)}{2}\overline{u}_{in}^{(r)}(p)
\end{equation*}
identifying
\begin{equation*}
\Lambda_{\nu}\equiv
\left(\frac{g_s}{\mu^{\varepsilon}}\right)^2C_F\int\frac{d^Dk}{(2\pi)^D}\frac{1}{k^2}
\gamma^{\rho}\frac{(\slh{p}'+\slh{k}+m)}{(p+k)^2-m^2}\gamma_{\nu}
\frac{(\slh{p}+\slh{k})}{(p+k)^2}\gamma_{\rho}
\end{equation*}
and using the dimensional regularization as usual
($D=4+2\varepsilon$, $g_s\rightarrow g_s\mu^{-\varepsilon}$, $d^4k\rightarrow d^Dk$);
we have chosen the Feynman gauge for the propagator of the virtual gluon
$-i g_{\rho\sigma}/k^2$, because the result is unchanged if all external lines
are on-mass-shell \cite{GOTT}, even using a more general gauge of the form
\ref{propagator}.\\
The structure of numerator inside the integral comprises scalar, vector
and tensorial (of second rank) terms;
this situation is considerably more complicated than the case for
massless particles, not simply because of the appearance of a 
mass for the outgoing particle, but also because of
the mass-asymmetry between incoming and outgoing fermionic lines.
Calculating the integral, the result is 
\begin{equation*}
\Lambda_{\nu}=
\frac{\alpha_s}{4\pi}C_F\frac{1}{\Gamma(1+\varepsilon)}
\left(\frac{Q^2+m^2}{4\pi\mu^2} \right)^{\varepsilon}
\left[C_0\gamma_{\nu}+\frac{C_p}{m}p_{\nu}+\frac{C_q}{m}q_{\nu} \right]
\end{equation*}
being
\begin{equation}
\begin{split}
C_0&=-\frac{2}{\varepsilon^2}+\frac{3}{\varepsilon}-8-K_A
+\frac{(1-\lambda)^{\varepsilon}}{\varepsilon^2}\left[
1-2\varepsilon+\left(4+2\text{Li}_2(\lambda)-\frac{\pi^2}{6}\right)\varepsilon^2
\right]\\
C_p&=2K_A\\
C_q&=-2\left(\frac{1-\lambda+K_A}{\lambda}\right)
\end{split}
\end{equation}
This is the result for the vertex correction in Fig.\ref{Vertex}.
Now self-energy contribution in Fig.\ref{Virtual_Self}
has to be calculated;
a way to include it into the previous treatment consists 
in renormalization of fermionic wave function.
Renormalization constant is defined on-mass-shell:
\begin{equation*}
z_i^{-1}\equiv 1-\frac{d\Sigma}{d\slh{p}_i}\bigg|_{\slh{p}_i=m_i}
\end{equation*}
being $\Sigma$ the Green function of self-energy, whereas
$p_i$, $m_i$ are respectively momenta and masses of the
considered particles.\\
For the outgoing fermionic (massive) line
\begin{equation*}
z_{p'}=1+\frac{\alpha_s}{4\pi}C_F\left( \frac{m^2}{4\pi\mu^2}\right)^{\varepsilon}
\frac{1}{\Gamma(1+\varepsilon)}\left[\frac{3}{\varepsilon}-4 \right]
\end{equation*}

\noindent 
whereas for the (massless) one incoming, trivially $z_p=1$.\\
Consequently the renormalized vertex is carried out by
\begin{equation*}
\sqrt{z_p z_{p'}}\left(\gamma_{\nu}+\Lambda_{\nu}\right)\frac{(1-\gamma_5)}{2}
\equiv
\left(\gamma_{\nu}+\Lambda^R_{\nu}\right)\frac{(1-\gamma_5)}{2}
\end{equation*}
\begin{equation*}
\Lambda_{\nu}^R=
\frac{\alpha_s}{4\pi}C_F\frac{1}{\Gamma(1+\varepsilon)}
\left(\frac{Q^2+m^2}{4\pi\mu^2} \right)^{\varepsilon}
\left[C_R\gamma_{\nu}+\frac{C_p}{m}p_{\nu}+\frac{C_q}{m}q_{\nu} \right]
\end{equation*}
paying attention to the fact that also the Born level diagram has to be taken into account
in renormalization
\begin{equation*}
C_R=
-\frac{2}{\varepsilon^2}+\frac{3}{\varepsilon}-8-K_A
+\frac{(1-\lambda)^{\varepsilon}}{\varepsilon^2}\left[
1-\frac{\varepsilon}{2}+\left(2+2\text{Li}_2(\lambda)-\frac{\pi^2}{6}\right)\varepsilon^2
\right]
\end{equation*}
\begin{figure}[!h]
\centering
{\includegraphics[width=6.5cm]{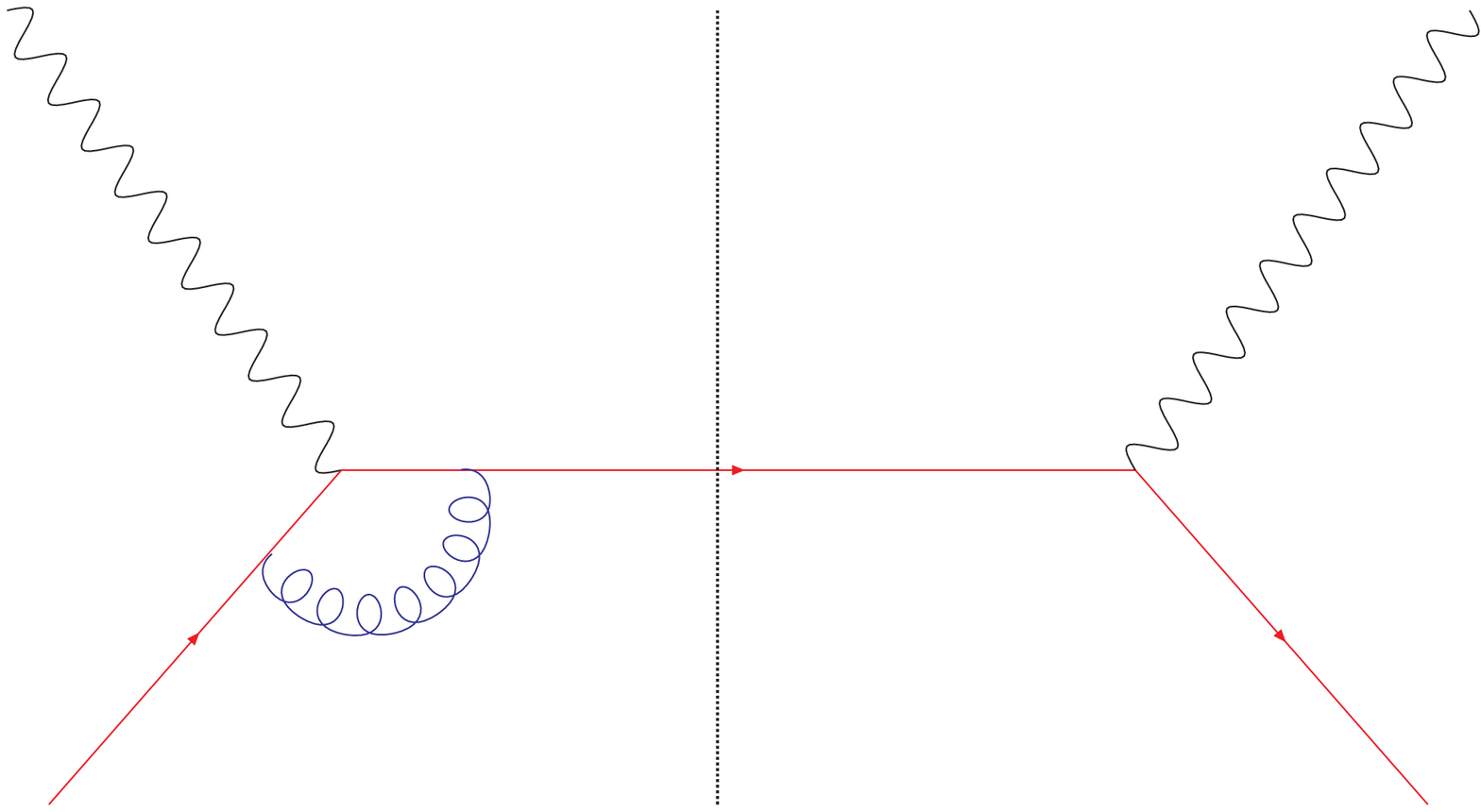}}
 \hspace{4mm}
{\includegraphics[width=6.5cm]{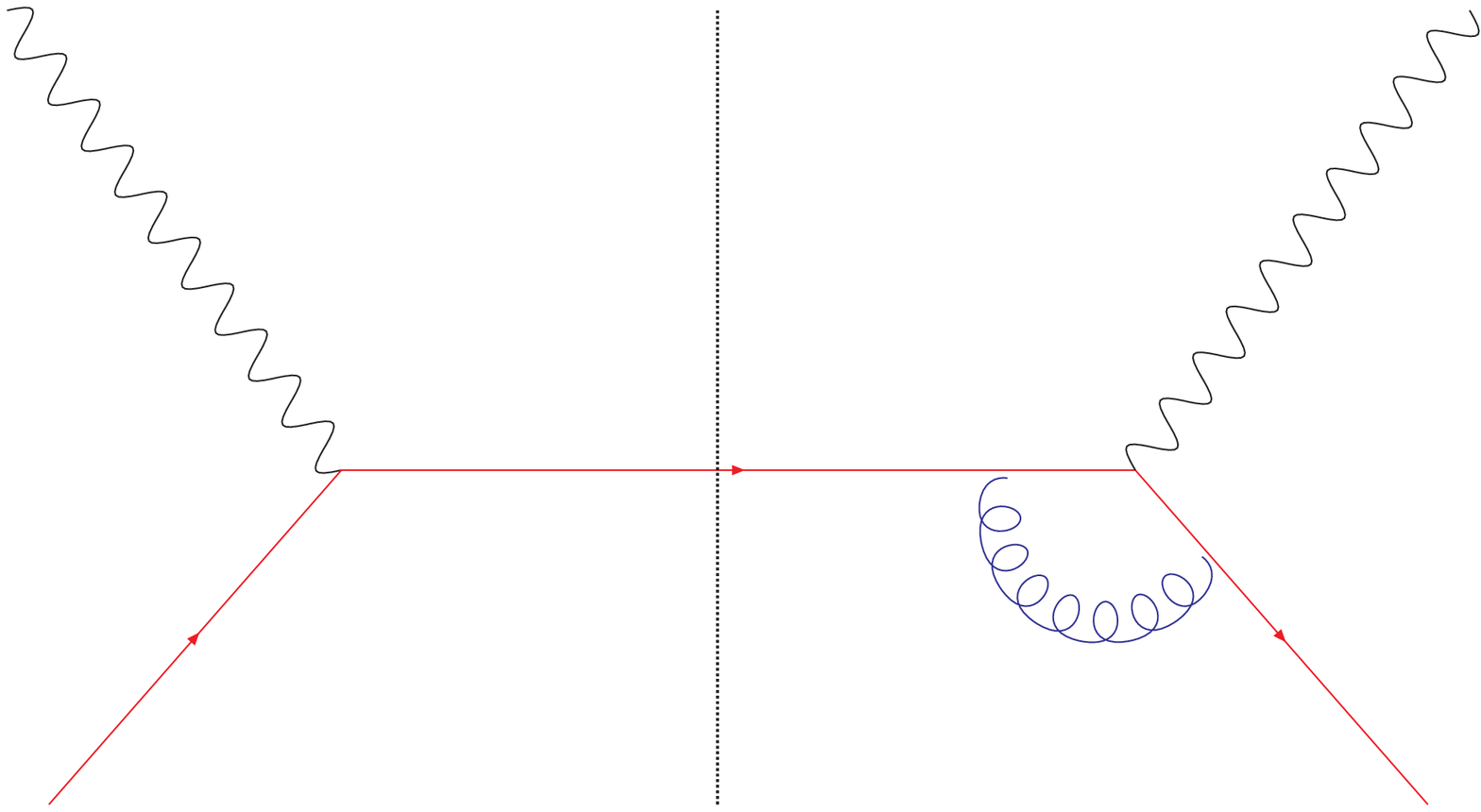}}
 \caption{\footnotesize Cut diagrams for the virtual contribution, 
 with renormalized vertex.}
 \label{Cuts}
\end{figure}

\noindent
In order to obtain the whole virtual contribution at NLO, 
we can describe the QCD tensors corresponding to cut diagrams 
in Fig.\ref{Cuts};
being one the complex conjugate of the other, their sum
is twice the real contribution of each one.
Then QCD tensor for the virtual contribution is
\begin{equation}\label{virtual_tensor}
h_{\mu\nu}^{V}=2|V_{ij}|^2\text{Re}\left\{\text{Tr}\left[
\slh{p}
\gamma_{\mu}\frac{(1-\gamma_5)}{2}
(\slh{p}'+m)
\Lambda^R_{\nu}\frac{(1-\gamma_5)}{2}
\right]\right\}
\end{equation}
neglecting the electroweak coupling constant 
as in other QCD tensors, because such a factor
has been already taken into account in eq.\ref{squared_modulus}. \\
All contractions with projectors in \ref{Projectors} 
can be calculated and the virtual contribution of Coefficient Functions
is carried out according to \ref{QCD_dF}; 
the phase-space is the same as for the Born case \ref{phase_space_LO}.
\begin{equation}\label{Coeff_funct_V}
\boxed{
\hat{\mathscr{F}}^{a,V}_{k}(\xi,y,\hat{w})=\delta_{aq}C_F\frac{\alpha_s}{2\pi}\frac{1}{\Gamma(1+\varepsilon)}
\left( \frac{Q^2+m^2}{4\pi\mu^2} \right)^{\varepsilon}
|V_{ij}|^2
\mathscr{C}_k
\delta(1-\xi)\delta(1-\hat{w})}
\end{equation}
with
\begin{center}
\begin{tabular}{|c||c|c|c|c|c||}\hline
$k$ &$1$  &$2$  &$3$  &$4$ &$5$    \\\hline
$\mathscr{C}_k$   &$C_R$  &$C_R+C_p/2$ &$C_R$ &$0$ &$C_R+C_q/2$ \\\hline
\end{tabular}
\end{center}
Our result essentially agrees with the one referred in \cite{GOTT}$^2$.

\footnotetext[2]{
In this article there are two mistakes about
virtual contributions for $k=2$ and $k=5$. 
In eq.A.12 and A.15 of \cite{GOTT} the wrong coefficients $1/4$ 
must be substituted for $1/2$;
this correction also holds for the first column of table I inside the same paper.}

\newpage
\subsection{Quark Channel}\label{Q_channel}
\noindent
The scattering amplitudes sum for the quark-channel case is
\begin{equation*}
\begin{split}
\mathcal{M}^Q&=-ig_{s}T^{a}_{kk'}V_{ij}
\bar{u}_{out}^{(s)}\left(p'\right)\\
&\bigg[
\gamma_{\mu}\frac{\left(1-\gamma_{5}\right)}{2}\frac{i\left(\slh{p}-\slh{l}\right)}{\left(p-l\right)^{2}}\gamma^{\rho}
+\gamma^{\rho}\frac{i\left(\slh{p}'+m+\slh{l}\right)}{\left(p'+l\right)^{2}-m^{2}}\gamma_{\mu}\frac{\left(1-\gamma_{5}\right)}{2}
\bigg]u_{in}^{(r)}\left(p\right)\epsilon_{\rho}^{(t)}\left(l\right)
\end{split}
\end{equation*}
where $s$ and $r$ are spin indices of the quark and $t$ of the gluon,
$T^{a}_{kk'}$ is the colour matrix associated with gluon emission
from a quark and $g_s$ the strong coupling constant,
whereas $V_{ij}$ is the CKM matrix element for the transition 
from a quark of flavour $j$ to $i$. 
Notice that the electroweak coupling constant has not been enclosed 
because already taken into account in expression \ref{squared_modulus}.
Taking the squared modulus of $\mathcal{M}^Q$, summing over polarisations 
($r$,$s$,$t$) and colours of quarks ($k$, $k'$) and of emitted gluon ($a$),
then 
\begin{equation}
\begin{split}\label{quark_squared}
h^Q_{\mu\nu}&\equiv\sum_{\substack{t,s,r\\a,k,k'}}\left|\mathcal{M}^Q\right|^{2}=
|V_{ij}|^{2}\alpha_{s}4\pi C_{F}\\
&\text{Tr}\bigg\{\bigg.
\left(\slh{p}'+m\right)\left[\gamma_{\mu}\frac{\left(1-\gamma_{5}\right)}{2}\frac{\left(\slh{p}-\slh{l}\right)}{\left(-2pl\right)}\gamma^{\rho}
+\gamma^{\rho}\frac{\left(\slh{p}'+m+\slh{l}\right)}{\left(2p'l\right)}\gamma_{\mu}\frac{\left(1-\gamma_{5}\right)}{2}\right]\\
&\left(\sum_{t}\epsilon_{\rho}^{(t)}\left(l\right)\overline{\epsilon}_{\tau}^{(t)}\left(l\right)
\right)\\
&\slh{p}\left.
\left[\gamma^{\tau}\frac{\left(\slh{p}-\slh{l}\right)}{\left(-2pl\right)}\gamma_{\nu}\frac{\left(1-\gamma_{5}\right)}{2}
+\gamma_{\nu}\frac{\left(1-\gamma_{5}\right)}{2}\frac{\left(\slh{p}'+m+\slh{l}\right)}{\left(2p'l\right)}\gamma^{\tau}
\right] \right\}
\end{split}
\end{equation}
Choosing an axial gauge for the gluon field, i.e. of the form
\begin{equation*}
\sum_{t}\epsilon_{\rho}^{(t)}\left(l\right)\overline{\epsilon}_{\tau}^{(t)}\left(l\right)=
-g_{\rho\tau}+\frac{n_{\rho}l_{\tau}+l_{\rho}n_{\tau}}{nl}
\end{equation*}
where $n$ is a generic vector with $n^2=0$, $np=0,$ and $l$ is the momentum of
the gluon, it is easy to verify that the terms in \ref{quark_squared} 
give null contribution when contracted with elementary tensors 
contained in projectors \ref{Projectors}.\\
We can therefore take into account only the contribution coming from $-g_{\rho\tau}$:
\begin{equation}
\begin{split}\label{quark_tensor}
h_{\mu\nu}^{Q}&=-|V_{ij}|^{2}\alpha_{s}4\pi C_{F}\text{Tr}\bigg\{\bigg.\\
&\left(\slh{p}'+m\right)\left[\gamma_{\mu}\frac{\left(1-\gamma_{5}\right)}{2}\frac{\left(\slh{p}-\slh{l}\right)}{\left(-2pl\right)}\gamma^{\rho}
+\gamma^{\rho}\frac{\left(\slh{p}'+m+\slh{l}\right)}{\left(2p'l\right)}\gamma_{\mu}\frac{\left(1-\gamma_{5}\right)}{2}\right]\\
&\slh{p}\left.
\left[\gamma_{\rho}\frac{\left(\slh{p}-\slh{l}\right)}{\left(-2pl\right)}\gamma_{\nu}\frac{\left(1-\gamma_{5}\right)}{2}
+\gamma_{\nu}\frac{\left(1-\gamma_{5}\right)}{2}\frac{\left(\slh{p}'+m+\slh{l}\right)}{\left(2p'l\right)}\gamma_{\rho}
\right] \right\}
\end{split}
\end{equation}
At this point all contractions between $h_{\mu\nu}^{Q}$ and elementary tensors 
$g^{\mu\nu}$, $p^{\mu}p^{\nu}$, $(p^{\mu}q^{\nu}+p^{\nu}q^{\mu})$,
$q^{\mu}q^{\nu}$, $\epsilon^{\mu\nu}_{\:\:\:\:\alpha\beta}p^{\alpha}q^{\beta}$
can be evaluated and each term $P_k^{\mu\nu}h_{\mu\nu}^{Q}$ can be reconstructed as linear
combination of them.\\
\\
\underline{Example of Contraction}\\
\\
We illustrate the contraction $-g^{\mu\nu}h_{\mu\nu}^Q$ as an example,
because this is the most complicated contribution and is very rich of particularities.
\begin{equation*}
\begin{split}
-g^{\mu\nu}&h_{\mu\nu}^Q=
|V_{ij}|^{2}\alpha_{s}\mu^{-2\varepsilon}4\pi C_{F}
4\left(1+\epsilon\right)\\
&\bigg[
\left(1+\epsilon\right)
\left(\frac{2pl}{2p'l}+\frac{2p'l}{2pl}\right)
-2\epsilon
+2\frac{\left(2pp'+2pl-2p'l\right)}{2pl\:2p'l}\left(2pp'-m^{2}\frac{2pl}{2p'l}
\right)\bigg]\\
&\:\:\:\:\:\:=|V_{ij}|^{2}\alpha_{s}\mu^{-2\varepsilon}4\pi C_{F}
4\left(1+\epsilon\right) \\
&\left[
\left(1+\epsilon\right)
\left(\frac{1-\hat{w}}{1-\lambda \xi}+\frac{1-\lambda \xi}{1-\hat{w}}\right)
-2\epsilon
+2\frac{\hat{w}\xi}{(1-\hat{w})(1-\xi)}\frac{(1-\lambda \xi)}{(1-\xi)}
\right]
\end{split}
\end{equation*}
Trivially the previous expression can be connected to that in 
\cite{DISSERTORI},\cite{STRM} for the massless case ($m\rightarrow0$ 
or equivalently $\lambda\rightarrow 1$).
Adding the phase-space $\int \hat{\varphi}_X^{NLO}$ \eqref{phase_space_hat},
terms of the form $(1-\xi)^a(1-\lambda\xi)^b$ and $(1-\hat{w})^c$ are generated:
they have to be managed as mathematical distributions
according to previous remarks (paragraph \ref{Distrib_pgf}) in order to isolate singularities.
Expanding up to $\varepsilon$ order (excluded)
\newpage
\begin{equation*}
\begin{split}
-&g^{\mu\nu}h_{\mu\nu}^Q \int\hat{\varphi}_X^{NLO}=
\frac{\alpha_s}{2\pi}C_F|V_{ij}|^{2}\frac{4\pi(1+\varepsilon)}{\Gamma(1+\varepsilon)}
\left(\frac{Q^2+m^2}{4\pi\mu^2}\right)^{\varepsilon}\bigg\{
\\
&\left(1-\hat{w}\right)\left[\frac{1-\xi}{(1-\lambda \xi)^2}\right]_{+}
+\frac{(1-\xi)}{(1-\hat{w})_{+}}
+2\frac{\hat{w}}{(1-\hat{w})_{+}}\frac{\xi}{(1-\xi)_{+}}\\
&+\delta(1-\xi)\delta(1-\hat{w})\left[
\frac{2}{\epsilon^{2}}-\frac{3}{\epsilon}
-\frac{(1-\lambda)^{\epsilon}}{\epsilon^{2}}\left(1-\frac{\epsilon}{2}-2\epsilon^{2}+\frac{\pi^{2}}{6}\epsilon^{2}
+2\text{Li}_{2}(\lambda)\epsilon^{2}
\right)
\right]\\
&+\delta(1-\hat{w})\bigg\{
(1+\xi^{2})\left[\frac{2\log(1-\xi)-\log(1-\lambda \xi)}{1-\xi}\right]_{+}
-(1+\xi^{2})\frac{\log\xi}{1-\xi}\\
&\qquad\qquad+(1-\xi)+\frac{1}{\epsilon}\left[\frac{1+\xi^{2}}{1-\xi}\right]_{+}
\bigg\}\\
&+\left[1-(1-\lambda)^{\epsilon} \right]\delta(1-\xi)\bigg\{
(1+\hat{w}^{2})\left[\frac{\log(1-\hat{w})}{1-\hat{w}}\right]_{+}
+(1+\hat{w}^{2})\frac{\log \hat{w}}{(1-\hat{w})}\\
&\qquad\qquad+(1-\hat{w})+\frac{1}{\epsilon}\left[\frac{1+\hat{w}^{2}}{1-\hat{w}}\right]_{+}\bigg\}\\
&+(1-\lambda)^{\epsilon}\delta(1-\xi)\left\{
\left[\hat{w}\frac{\log(1-\hat{w})}{1-\hat{w}}\right]_{+}
+\left[\frac{\hat{w}\log \hat{w}}{1-\hat{w}}\right]_{+}
+\frac{1}{\epsilon}\left[\frac{\hat{w}}{1-\hat{w}}\right]_{+}
\right\}\\
&-(1-\hat{w})\delta(1-\xi)\left[\frac{1+\lambda}{\lambda}K_{A}
+\frac{(1-\lambda)^{\epsilon}}{\lambda}\right]
\:\:\bigg\}
\end{split}
\end{equation*}
Soft singularities cancellation is worked out
adding virtual contributions;
contraction of tensor $-g^{\mu\nu}$ with the virtual 
one \eqref{virtual_tensor} gives
\begin{equation*}
\begin{split}
-g^{\mu\nu}h_{\mu\nu}^{V}\int\hat{\varphi}_X^{LO}=&
C_F\frac{\alpha_s}{2\pi}\frac{4\pi(1+\varepsilon)}{\Gamma(1+\varepsilon)}
\left( \frac{Q^2+m^2}{4\pi\mu^2} \right)^{\varepsilon}
|V_{ij}|^2\delta(1-\xi)\delta(1-\hat{w})\\
&\bigg\{
-\frac{2}{\epsilon^{2}}+\frac{3}{\epsilon}-8-K_A
+\frac{1-\lambda+K_A}{\lambda}\\
&-\frac{(1-\lambda)^{\epsilon}}{\epsilon^{2}}\left(-1+\frac{\epsilon}{2}-2\epsilon^{2}+\frac{\pi^{2}}{6}\epsilon^{2}
-2\text{Li}_{2}(\lambda)\epsilon^{2}
\right)
\bigg\}
\end{split}
\end{equation*}
then

\begin{equation*}
\begin{split}
-&g^{\mu\nu}\left(h_{\mu\nu}^Q\int\hat{\varphi}_X^{NLO}
+h_{\mu\nu}^{V}\int\hat{\varphi}_X^{LO}\right)=
C_F\frac{\alpha_s}{2\pi}\frac{4\pi(1+\varepsilon)}{\Gamma(1+\varepsilon)}
\left( \frac{Q^2+m^2}{4\pi\mu^2} \right)^{\varepsilon}
|V_{ij}|^2\bigg\{\\
&\left(1-\hat{w}\right)\left[\frac{1-\xi}{(1-\lambda \xi)^2}\right]_{+}
+\frac{(1-\xi)}{(1-\hat{w})_{+}}
+2\frac{\hat{w}}{(1-\hat{w})_{+}}\frac{\xi}{(1-\xi)_{+}}\\
&-\delta(1-\xi)\delta(1-\hat{w})
\left\{8+K_A-\frac{1-\lambda+K_A}{\lambda}
+(1-\lambda)^{\epsilon}\left(-4+\frac{\pi^2}{3}\right)
\right\}\\
&+\delta(1-\hat{w})\bigg\{
(1+\xi^{2})\left[\frac{2\log(1-\xi)-\log(1-\lambda \xi)}{1-\xi}\right]_{+}
-(1+\xi^{2})\frac{\log\xi}{1-\xi}\\
&\qquad\qquad+(1-\xi)+\frac{1}{\epsilon}\left[\frac{1+\xi^{2}}{1-\xi}\right]_{+}
\bigg\}\\
&+\left[1-(1-\lambda)^{\epsilon} \right]\delta(1-\xi)\bigg\{
(1+\hat{w}^{2})\left[\frac{\log(1-\hat{w})}{1-\hat{w}}\right]_{+}
+(1+\hat{w}^{2})\frac{\log \hat{w}}{(1-\hat{w})}\\
&\qquad\qquad+(1-\hat{w})+\frac{1}{\epsilon}\left[\frac{1+\hat{w}^{2}}{1-\hat{w}}\right]_{+}\bigg\}\\
&+(1-\lambda)^{\epsilon}\delta(1-\xi)\left\{
\left[\hat{w}\frac{\log(1-\hat{w})}{1-\hat{w}}\right]_{+}
+\left[\frac{\hat{w}\log \hat{w}}{1-\hat{w}}\right]_{+}
+\frac{1}{\epsilon}\left[\frac{\hat{w}}{1-\hat{w}}\right]_{+}
\right\}\\
&-(1-\hat{w})\delta(1-\xi)\left[\frac{1+\lambda}{\lambda}K_{A}+\frac{(1-\lambda)^{\epsilon}}{\lambda}\right]
\bigg\}
\end{split}
\end{equation*}
\underline{\bf Massless Result}\\
\\
\noindent
For $m=0$ ($\Rightarrow$ $\lambda=1$, $\xi=\hat{x}$)
the previous expression becomes
\begin{equation}\label{double_diff_massless}
\begin{split}
&-g^{\mu\nu}\left(h_{\mu\nu}^Q\int\hat{\varphi}_X^{NLO}
+h_{\mu\nu}^{V}\int\hat{\varphi}_X^{LO}\right){\bigg|}_{m=0}\!\!\!\!\!\!\!\!=
C_F\frac{\alpha_s}{2\pi}\frac{4\pi(1+\varepsilon)}{\Gamma(1+\varepsilon)}
\left( \frac{Q^2}{4\pi\mu^2} \right)^{\varepsilon}
|V_{ij}|^2\Bigg\{\\
&\frac{1-\hat{w}}{(1-\hat{x})^2_+}
+\frac{1-\hat{x}}{(1-\hat{w})_{+}}
+2\frac{\hat{w}}{(1-\hat{w})_{+}}\frac{\hat{x}}{(1-\hat{x})_{+}}
-8\delta(1-\hat{x})\delta(1-\hat{w})\\
&+\delta(1-\hat{w})\bigg\{
(1+\hat{x}^{2})\left[\frac{\log(1-\hat{x})}{1-\hat{x}}\right]_{+}
-(1+\hat{x}^{2})\frac{\log\hat{x}}{1-\hat{x}}\\
&\qquad\qquad+(1-\hat{x})+\frac{1}{\epsilon}\left[\frac{1+\hat{x}^{2}}{1-\hat{x}}\right]_{+}
\bigg\}\\
&+\delta(1-\hat{x})\bigg\{
(1+\hat{w}^{2})\left[\frac{\log(1-\hat{w})}{1-\hat{w}}\right]_{+}
+(1+\hat{w}^{2})\frac{\log \hat{w}}{(1-\hat{w})}\\
&\qquad\qquad+(1-\hat{w})+\frac{1}{\epsilon}
\left[\frac{1+\hat{w}^{2}}{1-\hat{w}}\right]_{+}\bigg\}\Bigg\}
\end{split}
\end{equation}
We note that some poles remain; their coefficient is proportional
to the Altarelli-Parisi splitting functions 
$$P^{(0)}_{qq}(t)=C_F\left[\frac{1+t^2}{1-t}\right]_+$$
It is easy to verify that these singularities are collinear 
and therefore can be assimilated into PDF and FF
(respectively for poles of $P^{(0)}_{qq}(\hat{x})$ and $P_{qq}^{(0)}(\hat{w})$)
thanks to {\it collinear factorization} theorem \cite{Collinear_Fact_Th}.
We can rewrite$^3$
\begin{equation*}
\frac{1}{\Gamma(1+\varepsilon)}
\left( \frac{Q^2}{4\pi\mu^2} \right)^{\varepsilon}\frac{P^{(0)}_{qq}(t)}{\varepsilon}
=P^{(0)}_{qq}(t)\left[\frac{1}{\varepsilon}+\gamma_e-\log{4\pi}+
\log\left( \frac{Q^2}{\mu^2} \right)\right]+O(\varepsilon)
\end{equation*}
and we can choose the $\overline{MS}$ scheme prescribing for removing the term 
\begin{equation*}\label{MS_bar}
P^{(0)}_{qq}(t)\left[\frac{1}{\varepsilon}+\gamma_e-\log{4\pi}\right]
\end{equation*}
leaving consequently the dependence on two arbitrary factorization scales $\mu$  
($\mu_F^{PDF}$ and $\mu_F^{FF}$) inside the term
$-g^{\mu\nu}\left(h_{\mu\nu}^Q\int\hat{\varphi}_X^{NLO}+
h_{\mu\nu}^{V}\int\hat{\varphi}_X^{LO}\right)$,
finite from now on.
For sake of consistency, the convolutions obtained in
\ref{compact_F_j} must involve PDF and FF defined 
using such a scheme.
\footnotetext[3]{
Apparently in \ref{double_diff_massless} there would be a multiplicative factor
$(1+\varepsilon)$ not yet taken into account: when expanded, it seems to be able to
generate other finite contributions mixing with the singular terms.
Actually this fact does not happen because coefficients in \ref{Projectors_h_j} 
exactly cancel this factor.}\\
\underline{\bf Massive Result}: $m\neq0$ ($\lambda\neq 1$)
\begin{equation*}
\begin{split}
&-g^{\mu\nu}\left(h_{\mu\nu}^Q\int\hat{\varphi}_X^{NLO}
+h_{\mu\nu}^{V}\int\hat{\varphi}_X^{LO}\right)=
C_F\frac{\alpha_s}{2\pi}\frac{4\pi(1+\varepsilon)}{\Gamma(1+\varepsilon)}
\left( \frac{Q^2+m^2}{4\pi\mu^2} \right)^{\varepsilon}
|V_{ij}|^2\Bigg\{\\
&\left(1-\hat{w}\right)\left[\frac{1-\xi}{(1-\lambda \xi)^2}\right]_{+}
+\frac{(1-\xi)}{(1-\hat{w})_{+}}
+2\frac{\hat{w}}{(1-\hat{w})_{+}}\frac{\xi}{(1-\xi)_{+}}\\
&-\delta(1-\xi)\delta(1-\hat{w})
\left\{4+\frac{\pi^2}{3}+K_A-\frac{1-\lambda+K_A}{\lambda}
\right\}\\
&+\delta(1-\hat{w})\bigg\{
(1+\xi^{2})\left[\frac{2\log(1-\xi)-\log(1-\lambda \xi)}{1-\xi}\right]_{+}
-(1+\xi^{2})\frac{\log\xi}{1-\xi}\\
&\qquad\qquad+(1-\xi)+\frac{1}{\epsilon}\left[\frac{1+\xi^{2}}{1-\xi}\right]_{+}\bigg\}\\
&+\delta(1-\xi)\bigg\{
-\log(1-\lambda)\left(
\left[\frac{1+\hat{w}^{2}}{1-\hat{w}}\right]_{+}
+\left[\frac{\hat{w}}{1-\hat{w}}\right]_{+}\right)
+\left[\hat{w}\frac{\log(1-\hat{w})}{1-\hat{w}}\right]_{+}\\
&+\left[\frac{\hat{w}\log \hat{w}}{1-\hat{w}}\right]_{+}
+\frac{1}{\epsilon}\left[\frac{\hat{w}}{1-\hat{w}}\right]_{+}
-(1-\hat{w})\left[\frac{1+\lambda}{\lambda}K_{A}+\frac{1}{\lambda}\right]
\bigg\}\Bigg\}
\end{split}
\end{equation*}
As for the massless case, it remains a collinear singularity 
for the term $\left[\frac{1+\xi^{2}}{1-\xi}\right]_{+}$: 
it is included into PDF as shown before.
In this case, however, it does not exist an analogous term
to be assimilated into FF and then an 
arbitrary scale $\mu_F^{FF}$ does not appear; 
at phenomenological level this implies that it is possible to use 
FF not depending on an energetic scale.\\
Nevertheless it remains apparently uncancelled a pole of the form 
\begin{equation}\label{anomalus_pole}
\frac{1}{\epsilon}\delta(1-\xi)
\left(\frac{\hat{w}}{1-\hat{w}}\right)_{+}
\end{equation}
The explanation of this fact comes from the structure of 
the phase-space at Born level: as referred in appendix \ref{partonic_phase_space},
\textquotedblleft angular\textquotedblright $ \hat{w}$ variable 
is not well defined when the final state quark is produced at rest
($p'=0\Rightarrow \xi=1$).
It is therefore necessary to give meaning to this term 
either passing to a better defined $\zeta$ variable
\eqref{zeta_definition} or by means of prescription \ref{t_hat_distribution},
so
\begin{equation*}
\delta(1-\xi)g(\xi,\hat{w})=\delta(1-\xi)\delta(1-\hat{w})
\left[\int_0^1d\alpha g(\xi,\alpha)\right]_{\xi=1}
\end{equation*}
and since
\begin{equation*}
\int_0^1d\alpha g_+(\alpha)=0
\end{equation*}
the anomalous pole disappears: this is therefore a phase-space effect.\\
After $\overline{MS}$ subtraction, it is then possible to rewrite$^4$
\begin{equation}\label{double_diff_mass}
\begin{split}
&-g^{\mu\nu}\left(h_{\mu\nu}^Q\int\hat{\varphi}_X^{NLO}
+h_{\mu\nu}^{V}\int\hat{\varphi}_X^{LO}\right)\bigg|_{\overline{MS}}=
C_F\frac{\alpha_s}{2\pi}4\pi(1+\varepsilon)
|V_{ij}|^2\bigg\{\\
&\left(1-\hat{w}\right)\left[\frac{1-\xi}{(1-\lambda \xi)^2}\right]_{+}
+\frac{(1-\xi)}{(1-\hat{w})_{+}}
+2\frac{\hat{w}}{(1-\hat{w})_{+}}\frac{\xi}{(1-\xi)_{+}}\\
&-\delta(1-\xi)\delta(1-\hat{w})
\left\{4+\frac{\pi^2}{3}+\frac{1}{2\lambda}
+\frac{(1+3\lambda)}{\lambda}K_A-\frac{1-\lambda+K_A}{\lambda}
\right\}\\
&+\delta(1-\hat{w})\bigg\{
(1+\xi^{2})\left[\frac{2\log(1-\xi)-\log(1-\lambda \xi)}{1-\xi}\right]_{+}
-(1+\xi^{2})\frac{\log\xi}{1-\xi}\\
&\qquad\qquad+(1-\xi)+\log\left(\frac{Q^2+m^2}{\mu^2}\right)\left[\frac{1+\xi^{2}}{1-\xi}\right]_{+}\bigg\}
\Bigg\}
\end{split}
\end{equation}
\footnotetext[4]{Also for this case, it holds what 
explained in note 3.}

\noindent
We want to remark that it is not possible to recover the massless result
\eqref{double_diff_massless} simply taking the limit of the massive one
\eqref{double_diff_mass};
this because of the non-commutativity of limits
$$
\lim_{\lambda\rightarrow 1^-}\lim_{\epsilon\rightarrow 0^+}(1-\lambda)^{\epsilon}=1
\qquad\qquad
\lim_{\epsilon\rightarrow 0^+}\lim_{\lambda\rightarrow 1^-}(1-\lambda)^{\epsilon}=0
$$
Notice that $\hat{x}$ is the variable of distributions that 
we identify with Altarelli-Parisi splitting functions for the massless case,
whereas $\xi$ for the massive case: this suggests that the relevant 
quantity at parton level for the massive case is a {\it slow rescaling} 
of $\hat{x}$, that is $\xi=\hat{x}/\lambda$ \cite{Tung_Aivazis_Olness},\cite{GOTT}.\\
\\
\newpage
\underline{\bf Quark Coefficient Functions}\\
\\
NLO quark-channel contribution to Coefficient Functions
can be expressed in the form
\begin{equation}\label{Q_coefficient_functions}
\boxed{
\hat{\mathscr{F}}_{k}^{q,NLO}(\xi,y,\zeta)
  =\frac{\alpha_s(\mu_R^2)}{2\pi}
\overbrace{  \bigg[
  (1-\delta_{k4}){\bf\mathscr{H}_0^q}(\xi,y,\zeta,\mu_F^2,\lambda)+
  {\bf\Delta_k^q}(\xi,\zeta,\lambda)
\bigg]}^{\mathscr{H}_k^q(\xi,y,\zeta,\mu_F^2,\lambda)}|V_{ij}|^2}
\end{equation}

\begin{equation*}
\begin{split}
{\bf \mathscr{H}_0^q}(&\xi,y,\zeta,\mu_F^2,\lambda)=\delta(1-\zeta)\bigg\{\bigg.
P_{qq}^{(0)}(\xi)\log\left(\frac{Q^2+m^2}{\mu_F^2}\right)\\
&+C_F\left[
1-\xi+(1+\xi^2)\left[\frac{2\log(1-\xi)-\log(1-\lambda\xi)}{(1-\xi)}\right]_+
\!\!\!\!-\frac{(1+\xi^2)}{(1-\xi)_+}\log\xi
\right]\!\bigg\}\\
&-C_F\delta(1-\xi)\delta(1-\zeta)\left[ 4+\frac{\pi^2}{3}+\frac{1}{2\lambda}
+\frac{(1+3\lambda)K_A}{2\lambda}
 \right]\\
&+C_F\bigg\{
\frac{1-\xi}{(1-\zeta)_\oplus}+(1-\zeta)\left(\frac{1-\lambda\xi}{1-\xi}\right)^2
\left[\frac{1-\xi}{(1-\lambda\xi)^2}\right]_+\\
&+2\frac{\xi}{(1-\xi)_+}\frac{1}{(1-\zeta)_\oplus}
\left[1-(1-\zeta)\frac{1-\lambda\xi}{1-\xi}\right]
+2\xi\left[1-(1-\zeta)\frac{1-\lambda\xi}{1-\xi}\right]
\!\bigg\}
\end{split}
\end{equation*}

\begin{equation*}
\begin{split}
{\bf \Delta_1^q}(\xi,\zeta,\lambda)=&0\\
{\bf \Delta_2^q}(\xi,\zeta,\lambda)=&
2C_F\bigg\{\bigg.\delta(1-\xi)\delta(1-\zeta)\frac{K_A}{2}\\
&\qquad\qquad\qquad-\left[\xi(1-3\lambda)\left( 1-(1-\zeta)
\frac{1-\lambda\xi}{1-\xi}\right)+(1-\lambda) \right]
\!\bigg\}\\
{\bf \Delta_3^q}(\xi,\zeta,\lambda)=&2C_F\bigg\{
(1-\xi)\left[1-(1-\zeta)\frac{1-\lambda\xi}{1-\xi}\right]-(1-\lambda\xi)\bigg\}\\
\\
{\bf \Delta_4^q}(\xi,\zeta,\lambda)=&2C_F\zeta\\
\\
{\bf \Delta_5^q}(\xi,\zeta,\lambda)=&2C_F\bigg\{
\xi\left[1-(1-\zeta)\frac{1-\lambda\xi}{1-\xi}\right]+\lambda\xi\zeta\bigg\}
\end{split}
\end{equation*}

\newpage
\subsection{Gluon Channel}
\noindent
The sum of scattering amplitudes for this channel is
\begin{equation*}
\begin{split}
\mathcal{M}^G=&-ig_{s}T^{a}_{kk'}V_{ij}
\bar{u}_{out}^{(s)}\left(p'\right)\\
&\bigg[
\gamma_{\mu}\frac{\left(1-\gamma_{5}\right)}{2}\frac{i\left(\slh{p}-\slh{l}\right)}{\left(p-l\right)^{2}}\gamma^{\rho}
+\gamma^{\rho}\frac{i\left(\slh{p}'+m-\slh{p}\right)}{\left(p'-p\right)^{2}-m^{2}}\gamma_{\mu}\frac{\left(1-\gamma_{5}\right)}{2}
\bigg]v_{out}^{(r)}\left(l\right)\epsilon_{\rho}^{(t)}\left(p\right)
\end{split}
\end{equation*}
therefore the QCD tensor can be carried out as the previous case,
paying however attention to colour factors ($T_F=1/2$);
remarks on contributions coming from gluon (axial) gauge still hold. 
\begin{equation}
\begin{split}\label{gluon_tensor}
h_{\mu\nu}^{G}=&-|V_{ij}|^{2}\alpha_{s}4\pi T_{F}\text{Tr}\bigg\{\bigg.\\
&\left(\slh{p}'+m\right)\left[\gamma_{\mu}\frac{\left(1-\gamma_{5}\right)}{2}\frac{\left(\slh{p}-\slh{l}\right)}{\left(2pl\right)}\gamma^{\rho}
+\gamma^{\rho}\frac{\left(\slh{p}'+m-\slh{p}\right)}{\left(2p'p\right)}\gamma_{\mu}\frac{\left(1-\gamma_{5}\right)}{2}\right]\\
&\:\:\:\slh{l}\left.
\left[\gamma_{\rho}\frac{\left(\slh{p}-\slh{l}\right)}{\left(2pl\right)}\gamma_{\nu}\frac{\left(1-\gamma_{5}\right)}{2}
+\gamma_{\nu}\frac{\left(1-\gamma_{5}\right)}{2}\frac{\left(\slh{p}'+m-\slh{p}\right)}{\left(2p'p\right)}\gamma_{\rho}
\right] \right\}
\end{split}
\end{equation}
This result can be directly obtained also from \ref{quark_tensor},
making use of symmetry of Feynman diagrams under crossing
among particles of initial and final state:
$$h_{\mu\nu}^{G}(p,l,p')=-h_{\mu\nu}^{Q}(-l,-p,p')$$
At this point all contractions between the gluonic tensor 
$h_{\mu\nu}^{G}$ and the elementary ones contained in 
projectors \ref{Projectors} can be calculated;
subsequently relations \ref{scalar_products} and
phase-space \ref{phase_space_hat} are employed
(in this case the massless particle corresponds to a light quark).
Expansions (in the sense of distributions) encountered in this channel 
are easier than in quark-channel, in fact there are not infrared singularities
excepting the collinear one coming from the initial state parton.
It is possible to check that the only pole has the form
\begin{equation*}
\frac{1}{\Gamma(1+\varepsilon)}
\left( \frac{Q^2+m^2}{4\pi\mu^2} \right)^{\varepsilon}\frac{P^{(0)}_{qg}(t)}{\varepsilon}
=P^{(0)}_{qg}(t)\left[\frac{1}{\varepsilon}+\gamma_e-\log{4\pi}+
\log\left( \frac{Q^2+m^2}{\mu^2} \right)\right]
\end{equation*}
setting
$$P^{(0)}_{qg}(t)=T_F\left[t^2+(1-t)^2\right]$$
Adopting the $\overline{MS}$ scheme, such a singularity 
is removed, including it into PDF as shown in the previous
sections.
Finally functions $h_k^G$ have been carried out according to 
relations \ref{QCD_dF} and to the structure of projectors
\ref{Projectors}.
\\
\newpage
\underline{\bf Gluon Coefficient Functions}\\
\\
In the following we report the expression of NLO contribution
to Coefficient Functions for the gluon-channel, 
using the same formalism of the quark case.
\begin{equation}\label{G_coefficient_functions}
\boxed{
\hat{\mathscr{F}}_{k}^{g,NLO}(\xi,y,\zeta)
  =\frac{\alpha_s(\mu_R^2)}{2\pi}
\overbrace{
 \bigg[
(1-\delta_{k4}){\bf\mathscr{H}_0^g}(\xi,y,\zeta,\mu_F^2,\lambda)
+{\bf\Delta_k^g}(\xi,\zeta,\lambda)
\bigg]}
^{\mathscr{H}_k^g(\xi,y,\zeta,\mu_F^2,\lambda)}
|V_{ij}|^2}
\end{equation}

\begin{equation*}
\begin{split}
{\bf \mathscr{H}_0^g}(\xi,y,\zeta,\mu_F^2,\lambda)=\delta(1-\zeta)\bigg\{
&P_{qg}^{(0)}(\xi)\left[\log\left(\frac{Q^2+m^2}{\mu_F^2}\right)
+\log\frac{(1-\xi)^2}{\xi(1-\lambda\xi)}\right]\\
&+\xi(1-\xi)\bigg\}
+\left[\frac{1}{(1-\zeta)_{\oplus}}+\frac{1}{\zeta} \right]P_{qg}^{(0)}(\xi)
\end{split}
\end{equation*}

\begin{equation*}
\begin{split}
{\bf \Delta_1^g}(\xi,\zeta,\lambda)&=
-\frac{\xi^2}{\zeta^2}(1-\lambda)(1-2\lambda)+\frac{2\xi}{\zeta}(1-\lambda)(1-2\lambda\xi)
+2\xi\lambda(1-\lambda\xi)\!-\!1\\
{\bf \Delta_2^g}(\xi,\zeta,\lambda)&=
\frac{\xi^2}{\zeta^2}(1-\lambda)(1-6\lambda+6\lambda^2)
+\frac{6\xi\lambda}{\zeta}(1-\lambda)(1-2\lambda\xi)
\\
&\qquad\qquad\qquad\qquad\qquad\qquad
+\lambda\left[6\lambda\xi(1-\lambda\xi)-1\right]\\
{\bf \Delta_3^g}(\xi,\zeta,\lambda)&=\frac{\xi^2}{\zeta^2}(1-\lambda)(1-2\lambda)
-\frac{2}{\zeta}\left[P_{qg}^{(0)}(\xi)+\xi(1-\xi-\lambda\xi)(1-\lambda)\right]\\
{\bf \Delta_4^g}(\xi,\zeta,\lambda)&=2\bigg[
(1-\xi)-(1-\lambda)\xi\frac{1-\zeta}{\zeta}
\bigg]\\
{\bf \Delta_5^g}(\xi,\zeta,\lambda)&=\Delta_1^g(\xi,\zeta,\lambda)+2\xi\bigg\{
-\frac{(1-\lambda)^2\xi}{\zeta^2}
+\frac{(1-\lambda)}{\zeta}(1-2\lambda\xi)\\
&\qquad\qquad\qquad\:\:\:\:\:
+\lambda(1-\lambda\xi)+\lambda\left[
(1-\xi)-(1-\lambda)\xi\frac{(1-\zeta)}{\zeta}
\right]
\bigg\}
\end{split}
\end{equation*}

\newpage
\section{NLO Coefficient Functions}
\subsection{Semi-Exclusive Coefficient Functions}
\label{Coefficient_Functions_Semi-Esclusive}

We own of all the contributions to enunciate the complete 
result for semi-exclusive
Coefficient Functions up to Next-to-Leading Order:
\begin{equation*}
\hat{\mathscr{F}}_{k}^{a}(\xi,y,\zeta)
  =\hat{\mathscr{F}}_{k}^{a, LO}(\xi,y,\zeta)+
   \hat{\mathscr{F}}_{k}^{a, NLO}(\xi,y,\zeta)
\end{equation*}
where $\hat{\mathscr{F}}_{k}^{a, LO}(\xi,y,\zeta)$ has been 
carried out in \ref{LO_result}
\begin{equation*}
\hat{\mathscr{F}}_k^{a,LO}(\xi,y,\zeta)
=(1-\delta_{k4})\delta_{aq}\delta\left(1-\xi\right)
\delta(1-\zeta)|V_{ij}|^2
\end{equation*}
so $\hat{\mathscr{F}}_{k}^{a, NLO}(\xi,y,\zeta)$
can be expressed as
\begin{equation*}
\hat{\mathscr{F}}_{k}^{a,NLO}(\xi,y,\zeta)
  =\frac{\alpha_s(\mu_R^2)}{2\pi}
\mathscr{H}_k^a(\xi,y,\zeta,\mu_F^2,\lambda)|V_{ij}|^2
\end{equation*}
in accordance with the notation introduced in 
\ref{Q_coefficient_functions} and  \ref{G_coefficient_functions}.

\noindent
With respect to terms where $a=q,g$ and $k=1,2,3$,
these expressions fully agree with literature \cite{GKRR},
\cite{KRST}.
Inclusive Coefficient Functions can be recovered 
integrating over $\zeta$ (we obtained it also from 
a dedicated study).
In this way it has been possible to verify 
the correctness of the semi-exclusive results 
for $a=q,g$ with $k=4,5$. \cite{KRST},\cite{GOTT}$^5$\\
\footnotetext[5]{
In that paper the modern approach of dimensional regularization,
prescribing that gluon helicity has be set to {\it$D-2$}, has not be
taken into account.}

\subsection{Inclusive Coefficient Functions}\label{C_F_inclusive}

Inclusive Coefficient Functions $\hat{\mathscr{I}}_k^a$
are carried out integrating directly the semi-exclusive ones:
\begin{equation*}
\begin{split}
\hat{\mathscr{I}}_k^a(\xi,y)=&
\int_{\frac{(1-\lambda)\xi}{(1-\lambda\xi)}}^{1}
d\zeta\hat{\mathscr{F}}_k^a(\xi,y,\zeta)=
\hat{\mathscr{I}}_k^{a,LO}(\xi,y)+\hat{\mathscr{I}}_k^{a,NLO}(\xi,y)\\
=&\hat{\mathscr{I}}_k^{a,LO}(\xi,y)+\frac{\alpha_s(\mu_R^2)}{2\pi}
\hat{I}_k^{a,NLO}(\xi,y)|V_{ij}|^2
\end{split}
\end{equation*}
According with previous paragraph then
\begin{equation*}
\hat{\mathscr{I}}_k^{a,LO}=(1-\delta_{k4})\delta_{aq}\delta(1-\xi)|V_{ij}|^2
\end{equation*}
whereas at Next-to-Leading Order the following expressions hold \\

\vspace{0.5cm}
\noindent
\boxed{\text{\bf Inclusive Quark Coefficient Functions}}\\

\begin{equation*}
\begin{split}
\hat{I}_4^{q,NLO}=&
C_F\left[1-\frac{\xi^2(1-\lambda)^2}{(1-\lambda\xi)^2}\right]\\
\hat{I}_k^{q,NLO}=&C_F\bigg\{
a_k\delta(1-\xi)+b_{1,k}\left[ \frac{1}{1-\xi}\right]_+
\!\!\!\!+b_{2,k}\left[ \frac{1}{1-\lambda\xi}\right]_+
\!\!\!\!+b_{3,k}\left[ \frac{1-\xi}{(1-\lambda\xi)^2}\right]_+
\!\!\!\!+c^q
\bigg\}
\\
c^q=&\left[\frac{1+\xi^2}{1-\xi} \right]_+\log\left(\frac{Q^2+m^2}{\mu_F^2}\right)
-\bigg(4+\frac{1}{2\lambda}+\frac{\pi^2}{3}+\frac{1+3K_A}{2\lambda}\bigg)\delta(1-\xi)\\
&-(1+\xi^2)\frac{\log(\xi)}{1-\xi}
+(1+\xi^2)\left[\frac{2\log(1-\xi)-\log(1-\lambda\xi)}{1-\xi}\right]_+
\end{split}
\end{equation*}

\begin{center}
\begin{tabular}{|c||c|c|c|c||}\hline
k  &$a_k$   &$b_{1,k}$         &$b_{2,k}$      &$b_{3,k}$\\\hline
1  &$0$     &$1-4\xi+\xi^2$     &$\xi(1-\xi)$   &$1/2$ \\
2  &$K_A$     &$2-2\xi^2-2/\xi$     &$2/\xi-1-\xi$   &$1/2$  \\
3  &$0$     &$-1-\xi^2$     &$1-\xi$   &$1/2$  \\
5  &$-(1-\lambda+K_A)/\lambda$   &$-1-\xi^2$ &$3-2\xi-\xi^2$   &$\xi-1/2$ \\\hline
\end{tabular}
\end{center}

\vspace{1.5cm}
\noindent
\boxed{\text{\bf Inclusive Gluon Coefficient Functions}}

\begin{equation*}
\begin{split}
\hat{I}_4^{g,NLO}=&2(1-\xi)
-2(1-\lambda)\xi\log\left( \frac{1-\lambda\xi}{(1-\lambda)\xi}\right)\\
\hat{I}_k^{g,NLO}=&\bigg\{
\xi(1-\xi)C_{1,k}
+C_{2,k}
+(1-\lambda)\xi\log\left(\frac{1-\lambda\xi}{(1-\lambda)\xi}\right)
\left(C_{3,k}+\lambda\xi C_{4,k}\right)
+C^g_k
\bigg\}
\\
C^g_k=&\frac{[\xi^2+(1-\xi^2)]}{2}\left[
\log\left(\frac{Q^2+m^2}{\mu_F^2}\right)
+(-1)^{\delta_{k3}}\log\left(\frac{(1-\xi)^2}{(1-\lambda\xi)\xi}\right)
\right]
\end{split}
\end{equation*}

\begin{center}
\begin{tabular}{|c||c|c|c|c||}\hline
k  &$C_{1,k}$   &$C_{2,k}$     &$C_{3,k}$      &$C_{4,k}$\\\hline
1  &$4-4(1-\lambda)$     &$\frac{(1-\lambda)\xi}{(1-\lambda\xi)}-1$     &$2$   &$-4$ \\
2  &$8-18(1-\lambda)+12(1-\lambda)^2$     &$\frac{(1-\lambda)}{(1-\lambda\xi)}-1$     &$6\lambda$   &$-12\lambda$ \\
3  &$2(1-\lambda)$     &$0$     &$-2(1-\xi)$   &$2$ \\
5  &$8-10(1-\lambda)$     &$\frac{(1-\lambda)\xi}{(1-\lambda\xi)}-1$     &$4$  &$-10$ \\\hline
\end{tabular}
\end{center}


\rm
\chapter{Numerical Results and Analysis}
\label{chapter_results}

{\it
In the following we resume the obtained results
during the extensive study carried out in the previous chapters;
subsequently we will leave the general CC DIS case 
to address exclusively the events of kind
$$\nu_{\mu}+N\rightarrow\mu^{-}+H_{c}+X$$
where $H_c$ is a hadron containing Charm quark.
Such a process is suitable to investigate 
Strange quark distribution inside nucleons;
in particular we will introduce the observable 
\begin{equation*}
\frac{x}{\lambda}s_{\text{eff}}\left( \frac{x}{\lambda},y,z   \right)=
\frac{1}{2}
\frac{\pi}{G_{F}^{2}ME}
\frac{\left(Q^{2}+M_{W}^{2}\right)^{2}}{M_{W}^{4}}
|V_{cs}|^{-2}
\frac{d^{3}\sigma}{dxdydz}
\end{equation*}
and we will study the dependence on
some parameters.\\
\\
It will be shown that $s_{eff}$ 
is related to Strange quark distribution 
in nucleons and that an analogous quantity for 
antiquark distribution can be introduced;
in fact, because of properties of Charged Currents,
it is possible independently to probe  
the quark and antiquark distributions.
}

\newpage
\section{General Result} \label{general_result}

Semi-Exclusive cross section for the generic CC DIS process
$$ \nu_{\ell}+N\rightarrow\ell^{-}+H+X $$
where the charged lepton is assumed to be massive and the hadronic state $H$ 
comes from heavy quark production, can be expressed as
\begin{equation}\label{general_CC_DIS_semi}
\frac{d^{3}\sigma}{dxdydz}=
 \frac{G_{F}^{2}ME}{\pi}
 \frac{M_{W}^{4}}{\left(Q^{2}+M_{W}^{2}\right)^{2}}
\left[
 \sum_{k=1}^5c_{k}(x,y)\frac{dF_{k}}{dz}(x,y,z)
\right]
\end{equation}
being $x$, $y$, $M$, $E$ ($Q^2=2MExy$) 
as defined in relations \ref{Kine_DIS},
$z$ an observable quantity for the hadron containing 
the massive quark \eqref{macro_z} 
and coefficients $c_k(x,y)$ the same of functions $F_k$ 
inside expressions in \ref{full_contraction}.\\
Extension for the analogous CC DIS processes
\begin{equation*}
\begin{split}
\ell^{-}+N&\rightarrow\nu_{\ell}+H+X \\
\bar{\nu}_{\ell}+N&\rightarrow\ell^{+}+H+X\\
\ell^{+}+N&\rightarrow\bar{\nu}_{\ell}+H+X
\end{split}
\end{equation*}
is trivially performed changing the sign of $c_3$ and/or 
averaging on initial spin states as illustrated in 
paragraph \ref{standard_approach}.
According to remarks of section \ref{parton_model_approach},
the charged lepton $\ell^{\pm}$ should in practice be assumed massless for electron and muon,
being $m_e\sim 5\cdot 10^{-4}\:GeV$, $m_{\mu}\sim 0.1\:GeV$, 
whereas for the Tau lepton ($m_{\tau}\sim 1.8\:GeV$) some 
contributions coming from the mass are not necessarily negligible. 
Then, for electrons and muons, coefficients $c_k$ simplify
and terms $dF_k/dz$ disappear for $k=4,5$.\\
Functions $dF_k/dz$ can be parametrised as
\begin{equation*}
\begin{split}
\frac{dF_1}{dz}(x,y,z)=&\mathscr{F}_1\\
\frac{dF_2}{dz}(x,y,z)=&2\frac{x}{\lambda}\frac{\mathscr{F}_2}{\rho^2}\\
\frac{dF_3}{dz}(x,y,z)=&2\frac{\mathscr{F}_3}{\rho}\\
\frac{dF_4}{dz}(x,y,z)=&\frac{1}{\lambda}\frac{(1-\rho)^2}{2\rho^2}\mathscr{F}_2
+\mathscr{F}_4+\frac{(1-\rho)}{\rho}\mathscr{F}_5\\
\frac{dF_5}{dz}(x,y,z)=&\frac{\mathscr{F}_5}{\rho}+\frac{(1-\rho)}{\lambda\rho^2}\mathscr{F}_2
\end{split}
\end{equation*}
\begin{equation}\label{EXCLUSIVE_cross}
\mathscr{F}_k(x,y,z)=
\sum_a\int_{\chi}^1\frac{d\xi}{\xi}\int_{\text{max}\{z,\zeta_{min}\}}^1
\!\!\!\!\!\!\!\!\!\!\!\!\!\!\!\!\!\!\!\!\!\!\!
\hat{\mathscr{F}}_k^{a}(\xi,y,\zeta)
\mathfrak{f}^{a}\left(\frac{\chi}{\xi},\mu_F^2\right)
\mathfrak{D}_{h}\left(\frac{z}{\zeta}\right)\frac{d\zeta}{\zeta}
\end{equation}
being $\zeta_{min}=(1-\lambda)\xi/(1-\lambda\xi)$ whereas
$\chi$ and $\rho$ have been defined in section \ref{QCD_approach} 
depending on hypothesis about TMC.\\
It is possible to calculate parton structure functions
using perturbative QCD;
at Leading + Next-to-Leading Order we obtain 
\begin{equation*}
\hat{\mathscr{F}}_k^{a}(\xi,y,\zeta)=\left[
\delta_{aq}(1-\delta_{k4})\delta(1-\xi)\delta(1-\zeta)
+\frac{\alpha_s(\mu_R^2)}{2\pi}\mathscr{H}_k^a(\xi,\zeta,\mu_F^2,\lambda)
\right]|V_{ij}|^2
\end{equation*}
where parton $a$ is a quark or a gluon, and 
expressions $\mathscr{H}_k^a$ are given in
\ref{Q_coefficient_functions} and \ref{G_coefficient_functions}.
The inclusive formalism holds neglecting the dependence on $z$ 
in the previous expressions and with
\begin{equation}\label{inclusive_cross}
\mathscr{F}_k^a(x,y)=
\sum_a\int_{\chi}^1\frac{d\xi}{\xi}
\hat{\mathscr{F}}_k^{a}(\xi,y)
\mathfrak{f}^{a}\left(\frac{\chi}{\xi},\mu_F^2\right)
\end{equation}
where $\hat{\mathscr{F}}_k^{a}(\xi,y)=
\int_{\zeta_{min}}^1\!\!\!d\zeta\hat{\mathscr{F}}_k^{a}(\xi,y,\zeta)$,
so we can write
\begin{equation*}
\hat{\mathscr{F}}_k^{a}(\xi,y)=\left[
\delta_{aq}(1-\delta_{k4})\delta(1-\xi)
+\frac{\alpha_s(\mu_R^2)}{2\pi}\hat{I}_k^{a,NLO}(\xi,y)
\right]|V_{ij}|^2
\end{equation*}
as for the Inclusive Coefficient Functions $\hat{I}_k^{a,NLO}$
in paragraph \ref{C_F_inclusive}.

\newpage
\section{Charm Production and Strange Distribution in Nucleons}
\label{strange_result}
Exploiting the general and complete analytical result we can now 
dedicate ourselves to the study of a particular process.
We consider the production of $H_c$ hadrons, containing Charm quark(s),
through CC DIS events with muonic neutrinos in initial state,
supposing that Charm has mass $m_c$:
$$ \nu_{\mu}+N\rightarrow\mu^{-}+H_c+X $$
Neglecting the muon mass ($m_{\mu}=0$; see \ref{full_contraction_massless}),
the semi-exclusive cross-section is
\begin{equation}\label{effective_cross_section}
\begin{split}
\frac{d^{3}\sigma(\nu)}{dxdydz}=
 \frac{G_{F}^{2}ME}{\pi}
 \frac{M_{W}^{4}}{\left(Q^{2}+M_{W}^{2}\right)^{2}}
\bigg\{
 xy^{2}\mathscr{F}_{1}
\! +\left(1\!-\!y\right)\frac{2}{\rho^2}\frac{x}{\lambda}\mathscr{F}_{2}
\! +xy\left(1\!-\!\frac{y}{2}\right)\frac{2}{\rho}\mathscr{F}_{3}
 \bigg\}
\end{split}
\end{equation}
being $\mathscr{F}_k$ defined 
as reported in \ref{EXCLUSIVE_cross}.
We define the observable$^{6}$
\begin{equation}\label{observable}
\frac{x}{\lambda}s_{\text{eff}}\left( \frac{x}{\lambda},y,z   \right)=
\frac{1}{2}
\frac{\pi}{G_{F}^{2}ME}
\frac{\left(Q^{2}+M_{W}^{2}\right)^{2}}{M_{W}^{4}}
|V_{cs}|^{-2}
\frac{d^{3}\sigma^{\nu}}{dxdydz}
\end{equation}
experimentally measured \cite{Data_1}, \cite{Data_2}.
\footnotetext[6]{
Notice that \ref{observable} is not a simple choice
of normalization for the cross section in \ref{effective_cross_section}, 
because $Q^2$ depends on both $x$ and $y$, so 
the multiplicative factor in \ref{observable} is not a constant;
nevertheless in the kinematics region where $Q\ll M_W (\sim 80\:GeV)$ 
this dependence is negligible.}
 \\
First of all we analyse such a quantity taking into account 
only QCD contributions up to LO.
We have chosen to study events having a final state whose
the leptonic current has negative charge ($\mu^-$),
so we can use the convention that $W^+$ is emitted from 
the leptonic vertex and is absorbed into the partonic vertex.
Because of electric charge conservation, $W^+$ boson can interact
only with quarks having charge $-1/3$ (or antiquarks of charge $-2/3$)
turning it into a quark of charge $+2/3$ (or correspondingly
into an antiquark of charge $+1/3$); 
the transition probability from a flavour $j$ to $i$ 
is governed by the CKM matrix elements $V_{ij}$.
\begin{figure}[!h]
\centering\epsfig{file=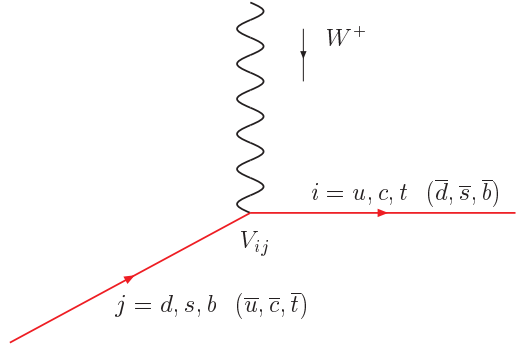,width=7cm}
\caption{\footnotesize{Allowed Transitions with a $W^+$ boson exchanged.}}
\label{kind_quark}
\end{figure}
Imposing that a Charm is produced in the final state,
then contributions coming from antiquarks with charge $-2/3$
are avoided;
in the initial state we can neglect Beauty quark,
since its presence inside nucleons is strongly suppressed  
because of the high mass ($m_b\sim 4.5\:GeV \gg M_{N}\sim 1 \:GeV$).
Then only {\it Down} and {\it Strange} quarks are probed by $W$ boson
(Fig.\ref{kind_quark}).\\

\begin{figure}[!h]
\centering
{
\includegraphics[angle=270,width=10.45cm]{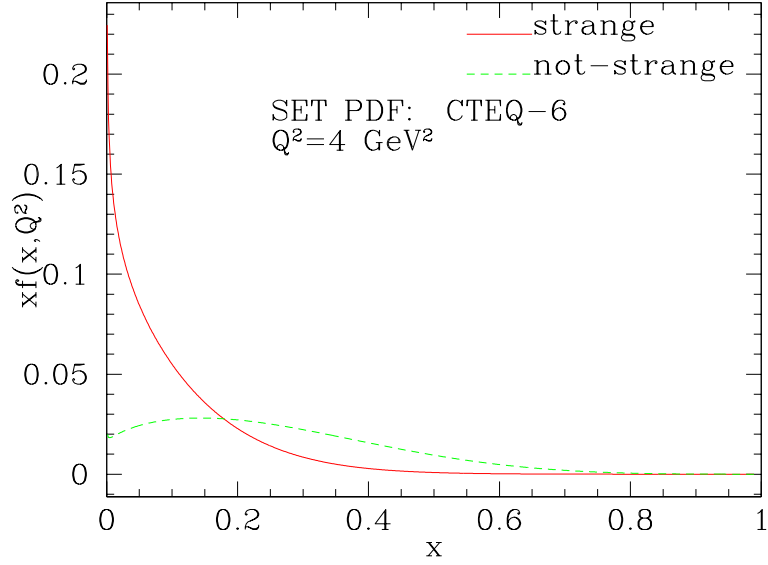}}
\caption{\footnotesize
For CTEQ 6 PDF set \cite{CTEQ_6}, with an energy scale $Q^2=4\:GeV^2$,
distributions $xs(x,Q^2)$ (continuous line) and 
$x\frac{|V_{cd}|^2}{|V_{cs}|^2}\left(\frac{d(x,Q^2)+u(x,Q^2)}{2}\right)$
(dashed line) are shown for the case of an isoscalar target.}
\label{s_eff_soppression_1}
\end{figure}

\begin{figure}[!h]
\centering
{\includegraphics[angle=270,width=10.45cm]{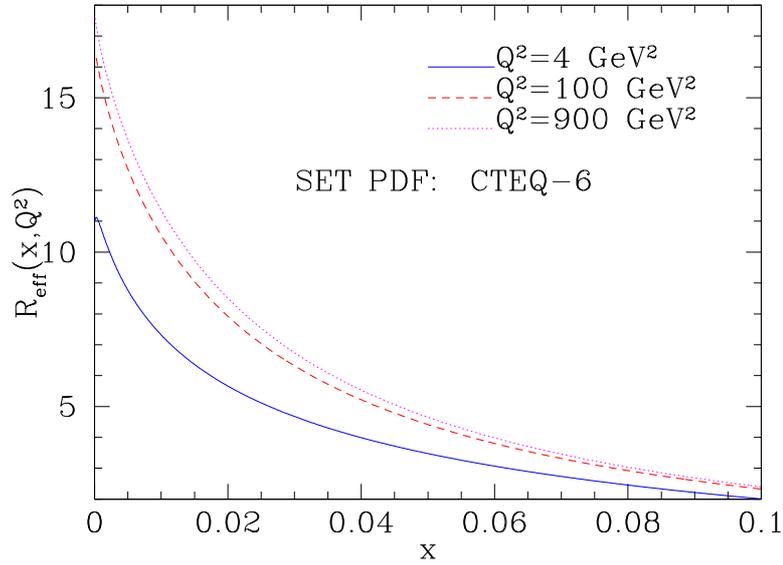}}
\caption{\footnotesize
Using CTEQ 6 PDF set \cite{CTEQ_6} the ratio 
$R_{eff}=s(x,Q^2)$/$\left[\frac{|V_{cd}|^2}{|V_{cs}|^2}\left(\frac{d(x,Q^2)+u(x,Q^2)}{2}\right)\right]$
is carried out varying the energetic scale $Q^2$ ($4\:GeV^2$, $100\:GeV^2$, $900\:GeV^2$)
in the region $x<0.1$.}
\label{s_eff_soppression_2}
\end{figure}

\newpage
\noindent
Taking into account the fact that, in general, a target is not simply made by only one kind of nucleon,
but rather it is a mixing of isotopes from many chemical elements (therefore having different mass and atomic numbers),
in practice it can be thought as a collection of protons and neutrons having average mass number $A$ and atomic number $Z$.
Supposing true and exact the strong isospin symmetry between 
proton and neutron (that is we assume that distributions 
$d_N$, $u_N$, $s_N$ for the neutron are respectively equal to 
$u_P$, $d_P$, $s_P$ for the proton),
we can finally rewrite observable \ref{observable} as
\begin{equation}\label{observable_born_TMC}
\begin{split}
\frac{x}{\lambda}s_{\text{eff}}^{LO}\left( \frac{x}{\lambda},y,z  \right)=
\frac{1}{\rho^2}\frac{x}{\lambda}
&\left[ \frac{|V_{cd}|^2}{|V_{cs}|^2}\left(\frac{Zd_P(\chi)+(A-Z)u_P(\chi)}{A}\right)
+ s_P\left(\chi\right)   \right]\\
&\cdot\left[1-y\left(1-\frac{\lambda\rho}{2} (3-\rho) \right) \right]
\mathscr{D}(z)
\end{split}
\end{equation}
because at LO $\mathscr{F}_{1,2,3}=\sum_{a}\mathfrak{f}^a(\chi)\mathscr{D}(z)$.
In CHARM experiment marble ($CaCO_3$) has been used: it corresponds to an
isoscalar target, that is $Z=(A-Z)$;
neglecting Target Mass Corrections ($\chi=x/\lambda$, $\rho=1$), then
\begin{equation}\label{observable_born}
\frac{x}{\lambda}s_{\text{eff}}^{LO}\left( \frac{x}{\lambda},y,z   \right)=
\frac{x}{\lambda}
\left[ \frac{|V_{cd}|^2}{|V_{cs}|^2}
\left(\frac{d_P+u_P}{2}\right)+ s_P\right]
\left[1-\frac{m^2_c\lambda}{2MEx}  \right]
\mathscr{D}(z)
\end{equation}
related to the quantity investigated in \cite{Data_1};
dependence of $d_P$, $u_P$ and $s_P$ on $x/\lambda$ 
and $\mu_F^2$ scale is understood.
Setting $|V_{cd}|\simeq 0.220$ and $|V_{cs}|\simeq 0.974$ \cite{PDG98},\cite{CONRAD}
it follows $|V_{cd}|^2/|V_{cs}|^2\sim 0.05$; 
$s_{eff}^{LO}$ can be therefore considered to all intents 
and purposes a Strange quark distribution inside nucleons.
Nevertheless we have to pay attention to special 
configurations (typically high $x$) where 
contribution of $u$ and $d$ could be not negligible
in spite of CKM suppression.
To this aim in Fig.\ref{s_eff_soppression_1} and \ref{s_eff_soppression_2} 
some comparisons between Strange and non-Strange 
component contained in \ref{observable_born} have been shown.

\newpage

\newpage
\subsection{Experimental Context}\label{pgf_exp}

Neutrino beams are usually generated by collisions of 
proton beams on a fixed target, typically Beryllium;
pions and kaons are produced in the event and
a large fraction of them decays into muons and muonic neutrinos,
through semi-leptonic channels.
Neutrino beam for CCFR experiment \cite{S_ASYMM_3},\cite{CCFR} 
is made of about $86.4${\small\%} $\nu_{\mu}$, $11.3${\small\%} $\overline{\nu}_{\mu}$,
$2.3${\small\%} $\nu_{e}$ and $\overline{\nu}_{e}$.
Energies of these beams cover a wide range: for CCFR from $30$ to $300$ GeV.
The target material, on which neutrinos are sent to investigate DIS, depends 
on the experiment: Iron for CCFR (NuTeV now) at Fermilab, and for CDHSW at CERN;
marble ($CaCO_3$) for CHARM, glass for CHARM II, Neon or Deuterium for BEBC, all at CERN.
Unlike marble and Deuterium, Iron is not an isoscalar target.
Nuclear effects should be included because the events typically consist in  
scattering on heavy targets: Fermi motion inside nucleons,
EMC effect and recombination of gluons among nucleons of close
nuclei (Nikolaev and Zakharov, 1975; Mueller and Qiu, 1986), \ldots \\
For DIS processes using neutrino beams, the typical range for $x$ and $Q^2$
are respectively $0.01<x<0.8$ and $0.1\:GeV^2 < Q^2 <100\:GeV^2$.\\
A review of measurements in experiments dedicated to DIS 
with neutrino beams is given in \cite{CONRAD}. \\
\\
Charm production through CC DIS is experimentally studied 
observing di-muonic events;
in fact Charm quark, straight after creation, hadronizes (typically)
into mesons ($D^0$, $D^+$, $D^{*0}$, \textellipsis) that can 
decay (also indirectly, 
$D^{*0}\rightarrow D^{0}\textellipsis\rightarrow \nu_{\mu}+\mu^{+}+\textellipsis$)
into neutrino and anti-muon:
\begin{equation}
\begin{split}
\nu_{\mu}+N \stackrel{W^{+}}{\longrightarrow} \mu^{-}+&H_c+X_1\\
&\hookrightarrow \nu_{\mu}+\mu^{+}+X_2
\end{split}
\end{equation}
The whole final state, made of a muon pair of opposite charge, 
is then a very strong signal of Charm production through CC DIS,
or correspondingly that Strange quark distribution inside nucleons
has been probed.

\newpage
\subsection{Test}
\begin{figure}
\centering
{\includegraphics[angle=270,width=12.0cm]{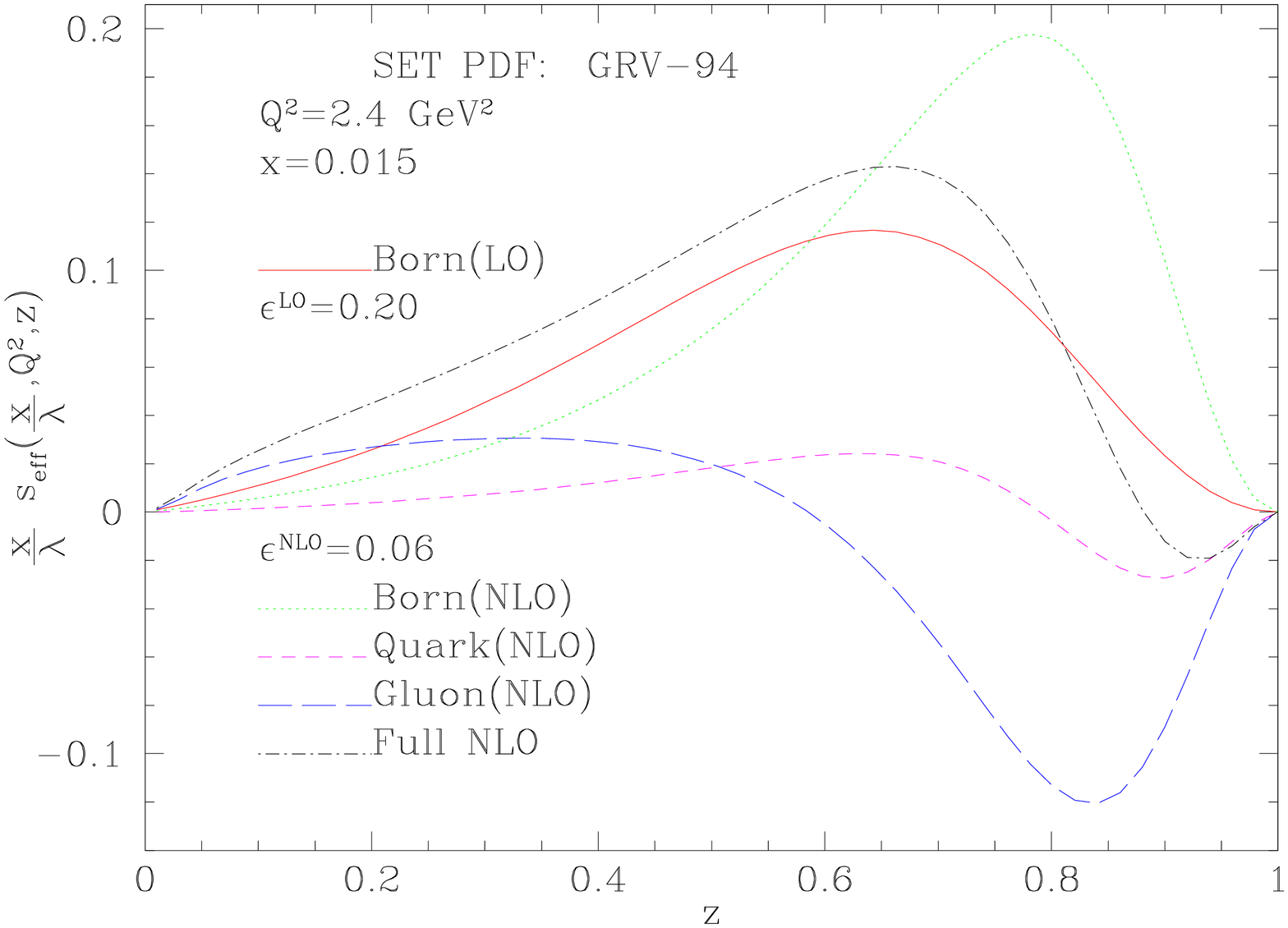}}
\caption{\footnotesize $\frac{x}{\lambda}s_{\text{eff}}\left(
\frac{x}{\lambda},y,z   \right)$ for $x=0.015$ and $Q^2=2.4\:GeV^2$.
$Born(LO)$ means that expression \ref{observable_born} 
has been used with the corresponding value 
$\epsilon_P^{LO}$ and PDF defined at LO. 
With notation $Born(NLO)$, it means that the analysis has been 
carried out at NLO using the definition in \ref{observable}, 
PDF at NLO and $\epsilon_P^{NLO}$.}\label{result_Kretzer_1}
\centering
{\includegraphics[angle=270,width=12.0cm]{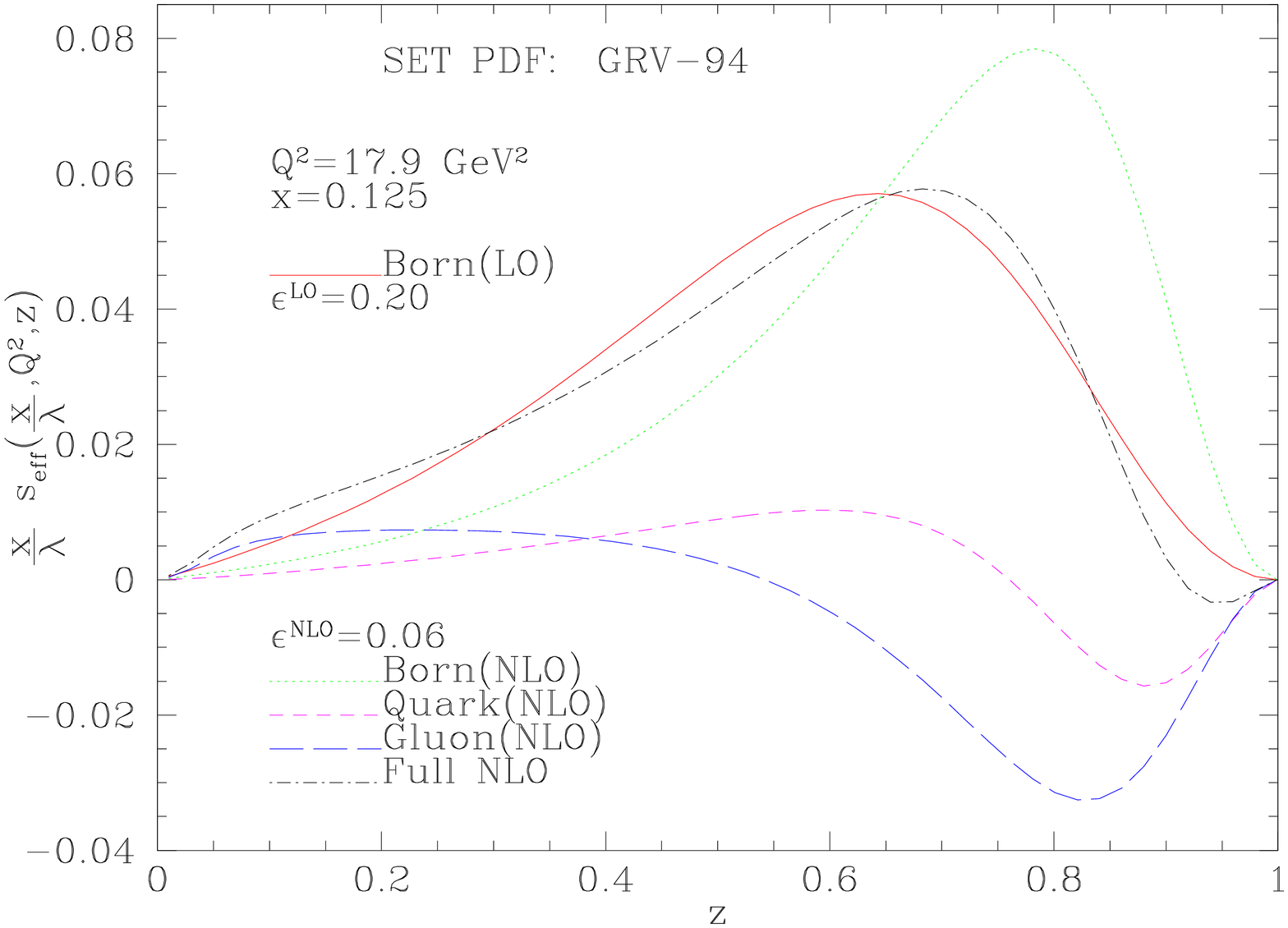}}
\caption{\footnotesize
$\frac{x}{\lambda}s_{\text{eff}}\left( \frac{x}{\lambda},y,z   \right)$
for $x=0.125$ and $Q^2=17.9\:GeV^2$.}\label{result_Kretzer_2}
\end{figure}

As initial test to check the correctness of our numerical implementation, 
we have tried to reproduce results in \cite{KR97};
we remind in the following all the employed assumptions:
\begin{itemize}
\item Charm mass set to $m_c=1.5\:GeV$
\item incident neutrino energy set to $E=192\:GeV$
\item renormalization scale ($\mu_R$) in the strong coupling constant 
      $\alpha_s$ chosen to be equal to the factorization one ($\mu_F$) for PDF
      and Coefficient Functions; both set to $\mu_F^2=\mu_R^2=Q^2+m_c^2$
\item GRV-94 PDF \cite{GRV}, with $\overline{MS}$ scheme
\item Peterson FF \cite{PETERSON}
      (see paragraph \ref{pgf_FF})
\item for LO result, PDF defined at LO have been coherently 
      employed, whereas the value $\epsilon_P^{LO}=0.20$ set for FF;\\ 
      for NLO result, PDF defined at NLO and $\epsilon_P^{NLO}=0.06$ 
      for FF
			
\item we have considered an isoscalar target; 
      contributions coming from {\it Up} and {\it Down}
      quarks have been neglected
\item CKM matrix elements have been set to
      $|V_{cd}|=0.220$ and $|V_{cs}|=0.974$;
      for coherence with the previous point, 
      $|V_{cd}|=0$ for the contribution coming from 
      the gluon (i.e. gluon splitting in $d$, $\bar{d}$ pair suppressed)
\item Target Mass Corrections disregarded
\end{itemize}
The obtained results have been illustrated in 
Fig.\ref{result_Kretzer_1}, \ref{result_Kretzer_2}:
they agree with literature \cite{KR97}.
Introducing also TMC \eqref{QCD_approach}, it can be 
verified that such a kind of corrections are absolutely 
negligible in the range of $x$ and $Q^2$ here explored.\\

\begin{figure}[!h]
\centering{
\includegraphics[angle=270,width=13cm]{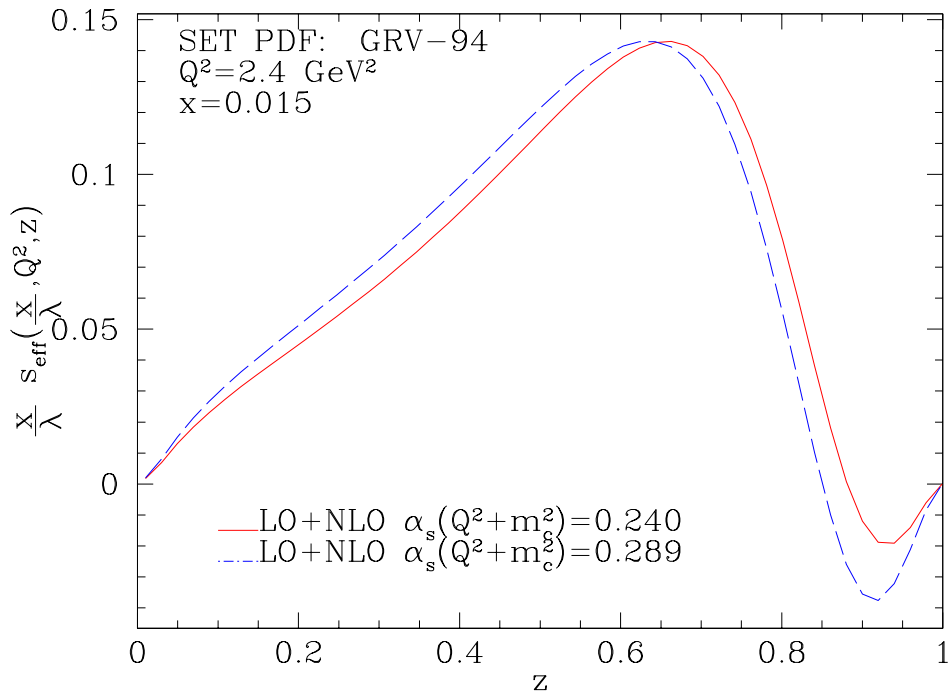}}
\caption{\footnotesize
Same plot as Fig.\ref{result_Kretzer_1}
for $\frac{x}{\lambda}s_{\text{eff}}\left( \frac{x}{\lambda},y,z \right)$
at LO+NLO, evaluated setting 
$\alpha_s(Q^2+m_c^2)\sim 0.240$ (continuous line)
and $\alpha_s(Q^2+m_c^2)\sim 0.289$ (dashed line).
}\label{grv_alpha}
\end{figure}

\newpage
\noindent
This simple test is satisfactory to point out 
the role of the appropriate value for $\alpha_s$ 
in order to compare the numerical results.
In \cite{GKRR},\cite{KR97} the routine used to calculate $\alpha_s$
at the scale of $Z^0$ mass gives a value$^6$ of $0.109$,
\footnotetext[7]{We would like to thank very much Dr.~Stephan~Kretzer 
to have provided us his numerical code to carry out a comparison.}
whereas our routine, using a modern value for $\Lambda_{QCD}$ 
(and being able to change thresholds, $\Lambda_{QCD}$ scales
and many other parameters) gives $\alpha_s(M_{Z^0}^2)\sim 0.118$.\\
At lower scales the discrepancy between results of the two routines
increases: for graph in Fig.\ref{result_Kretzer_1}
($Q^2=2.4\:GeV^2$, $m_c=1.5\:GeV$) we obtain
$\alpha_s(Q^2+m_c^2)\sim 0.240$, whereas using our routine
$\alpha_s(Q^2+m_c^2)\sim 0.289$.
In this case therefore we are under-evaluating NLO contributions of
about 20\%; in Fig.\ref{grv_alpha} the comparison 
between these last two results is shown.

\clearpage
\section{Analysis}\label{analysis}

Interpretation of $s_{eff}$ observable as quantity 
proportional to Strange quark distribution inside nucleons
does not hold beyond the Leading Order;
in fact Parton Distribution Functions are convoluted 
with Coefficient Functions (not trivial at NLO)
so besides contributions of light quarks $u$, $d$, $s$, 
even the one of gluon appears.\\
In order to study the distribution of Strange quark inside
nucleons by means of $s_{eff}$ observable, 
it must be subtracted from $s_{eff}^{exp}$, defined
as in \ref{observable} and experimentally measured,
the contribution coming from the gluon
(and even the one from $u$ and $d$ quarks, if not negligible)
predicted by the theory, using the Semi-Exclusive Coefficient Functions at NLO in \ref{G_coefficient_functions}
\eqref{Q_coefficient_functions}.\\
As pointed out at the beginning of this work, DIS mediated by 
Charged Currents allows to independently probe quark and
antiquark distributions;
all what we have written therefore holds also for the process
$$ \overline{\nu}_{\mu}+N\rightarrow\mu^{+}+H_{\overline{c}}+X $$
where $H_{\overline{c}}$ is a hadron containing Charm.\\
In this case it is possible to define the observable
$\overline{s}_{eff}$ analogously to \ref{observable}
\begin{equation*}
\frac{x}{\lambda}\overline{s}_{\text{eff}}
\left( \frac{x}{\lambda},y,z   \right)=
\frac{1}{2}
\frac{\pi}{G_{F}^{2}ME}
\frac{\left(Q^{2}+M_{W}^{2}\right)^{2}}{M_{W}^{4}}
|V_{cs}|^{-2}
\frac{d^{3}\sigma(\overline{\nu})}{dxdydz}
\end{equation*}
where the cross section is derived from the observations
on CC DIS in general (paragraph \ref{standard_approach})
\begin{equation*}
\begin{split}
\frac{d^{3}\sigma(\overline{\nu})}{dxdydz}=
 \frac{G_{F}^{2}ME}{\pi}
 \frac{M_{W}^{4}}{\left(Q^{2}+M_{W}^{2}\right)^{2}}
\bigg\{
 xy^{2}\mathscr{F}_{1}
\! +\left(1\!-\!y\right)\frac{2}{\rho^2}\frac{x}{\lambda}\mathscr{F}_{2}
\! -xy\left(1\!-\!\frac{y}{2}\right)\frac{2}{\rho}\mathscr{F}_{3}
 \bigg\}
\end{split}
\end{equation*}
By means of CC DIS is then possible to study 
{\it independently} distributions of 
Strange and anti-Strange quarks;
this is useful to point out an eventual 
asymmetry between the two distributions,
as hypothesised by recent experimental results
\cite{Data_3},\cite{CONRAD},\cite{S_ASYMM_3}.\\
From Fig.\ref{result_Kretzer_1} and \ref{result_Kretzer_2}
it has been found that NLO contribution coming from gluon-channel
({\it Gluon(NLO)}) is comparable to the one at LO 
coming directly from Strange quark ({\it Born(NLO)});
this because the gluon distribution inside nucleons is quantitatively high
and it balances the suppression originated from the powers of $\alpha_s$.
Finally we remark that $s_{eff}$ (and then also the cross section)
becomes negative when $z\gtrsim 0.85$:
quantitatively such a behaviour is obviously non-physical, 
but it indicates the need of resumming terms of higher order.\\

\newpage
\noindent
In the following we show the behaviour for observable
$(x/\lambda) s_{eff}(x/\lambda,Q^2,z)$
with respect to parameters and models; 
as setting, we assume
\begin{itemize}
\item $m_c=1.5\:GeV$ , $E=192\:GeV$, $|V_{cd}|=0.220$, $|V_{cs}|=0.974$
\item instead of GRV-94 PDF we use a more modern set:
      CTEQ-6 \cite{CTEQ_6} with $\overline{MS}$ scheme;
      we do not have a definition at LO, so the
      result carried out is only the one at NLO
\item we use Peterson FF \eqref{FF_peterson},
      with $\epsilon_P^{NLO}=0.06$
\item $\mu_F^2=\mu_R^2=Q^2+m_c^2$
\item isoscalar target and contributions from $u$, $d$
\item $\alpha_s$ generated by the same LHA-PDF package \cite{LHA}
 used for PDF;\\
for CTEQ-6:
$\alpha_s(M_{Z^0}^2)=0.118$, $\alpha_s(Q_0^2+m_c^2)=0.271$ \\
with $Q_0^2=2.4\:GeV^2$ and $m_c=1.5\:GeV$
\item Target Mass Corrections disregarded
\end{itemize}
Furthermore we consider only the case $x=0.015$ and $Q^2=2.4\:GeV^2$.\\
Numerical error-bars have not been shown in graphs because not visible,
being negligible ($<1${\small\%}).

\newpage
\subsection{PDF Dependence}

We want to verify the hypothesis that Strange 
quark distribution is dominant for this process 
with respect to one for other quarks, at least in the range of $x$ and $Q^2$ here taken into account: 
distributions of Up and Down ($|V_{cd}|=0.220$) have been added to results of Fig.\ref{result_Kretzer_1}.
\begin{figure}[!h]
\centering{
\includegraphics[angle=270,width=6.7cm]{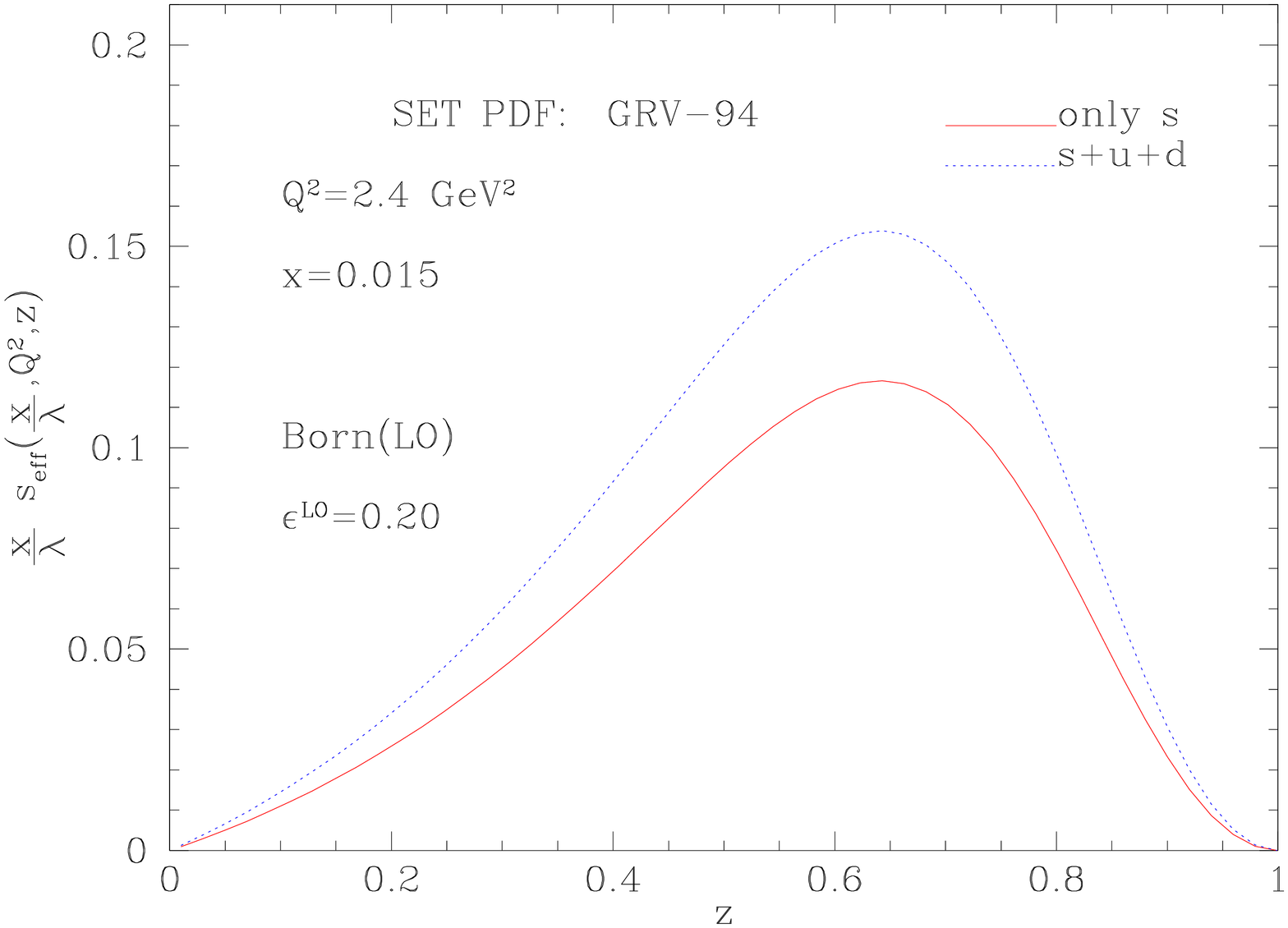}
\includegraphics[angle=270,width=6.7cm]{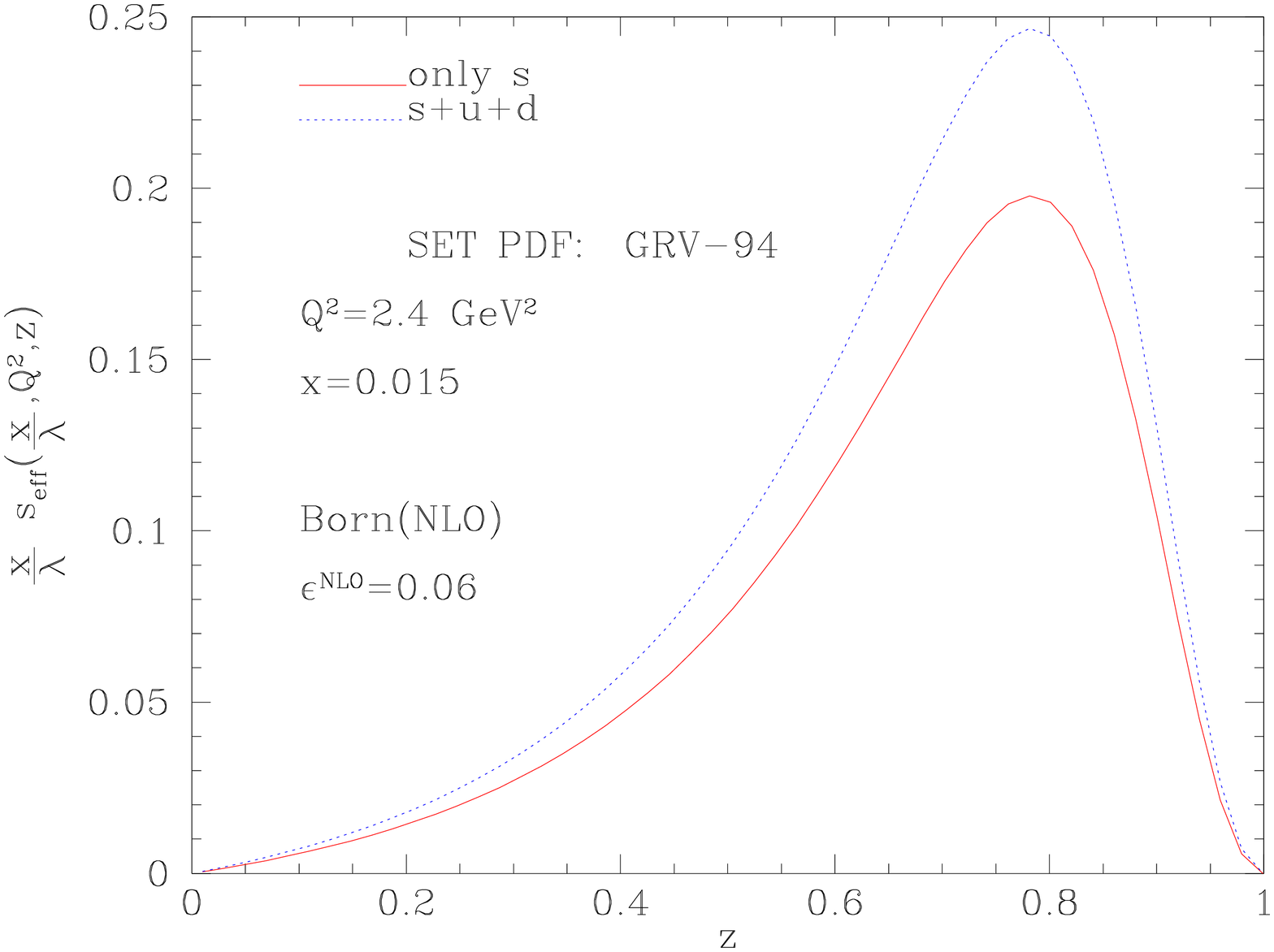}
\includegraphics[angle=270,width=6.7cm]{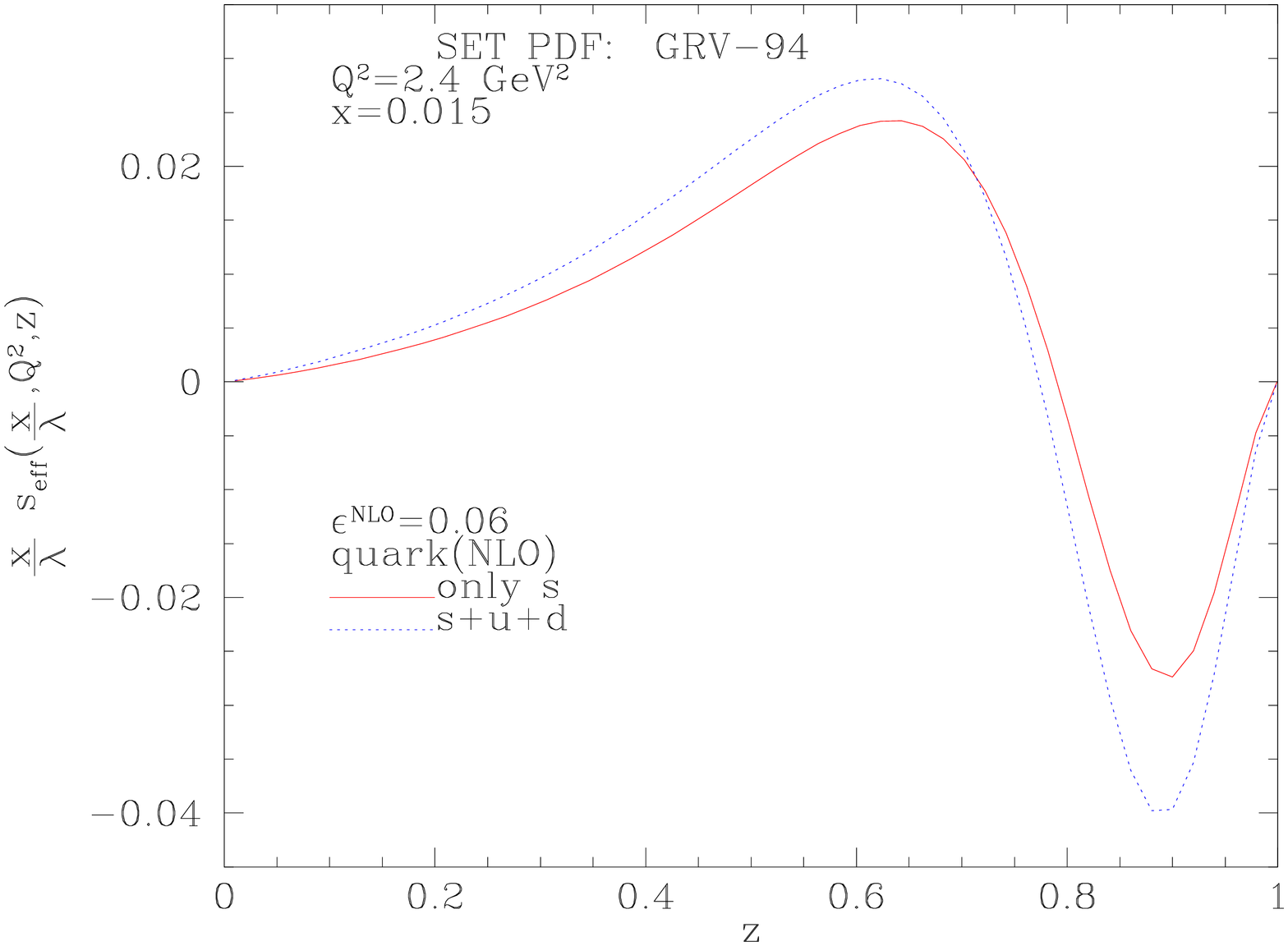}
\includegraphics[angle=270,width=6.7cm]{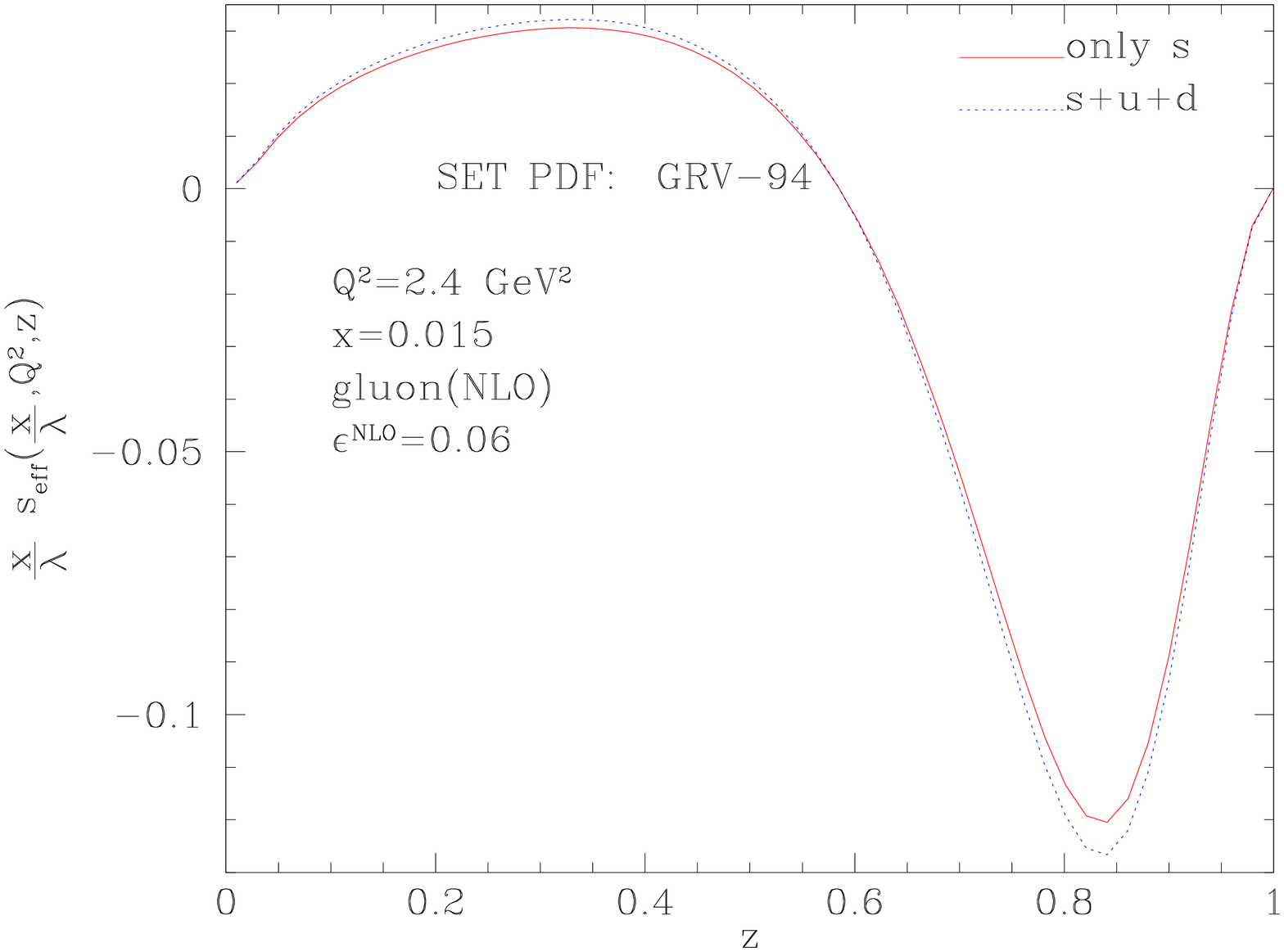}
}
\caption{\footnotesize Contributions to $\frac{x}{\lambda}s_{eff}\left(\frac{x}{\lambda},Q^2,z\right)$
for GRV-94 PDF set: comparison between the case without $Up$ and $Down$ quarks
($|V_{cd}|=0$, continuous line) and the case including them 
($|V_{cd}|=0.220$, dashed line).
}\label{GRV_updown}
\end{figure}

\begin{figure}[!h]
\centering{\includegraphics[angle=270,width=11.5cm]{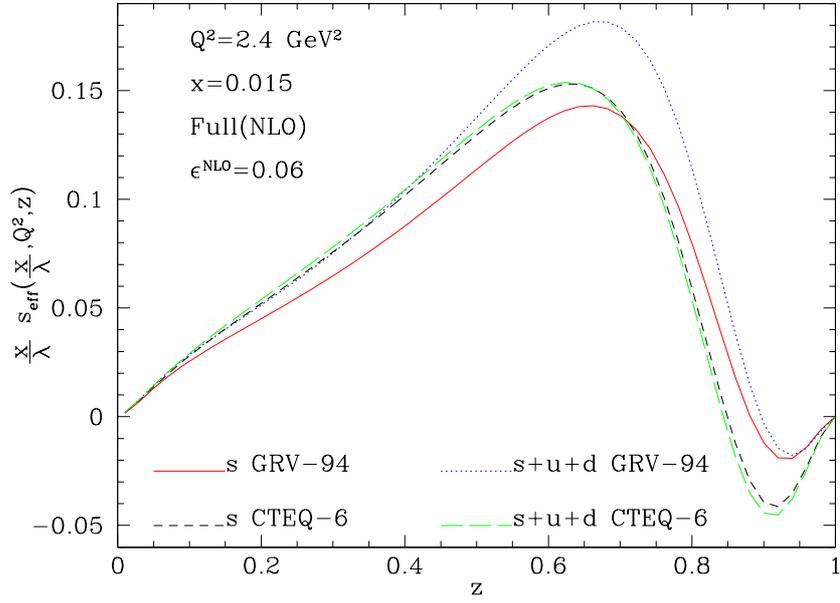}}
\caption{\footnotesize Graphs of $\frac{x}{\lambda}s_{eff}\left(\frac{x}{\lambda},Q^2,z\right)$
with and without contributions coming from distributions for $Up$ and $Down$ quarks, using GRV-94 and CTEQ-6 PDF sets.}\label{updown_full}
\end{figure}

\noindent
As illustrated in Fig.\ref{GRV_updown}, contributions coming from Up and Down
for GRV-94 PDF are not at all negligible: an accurate analysis 
must take them into account.
In Fig.\ref{s_eff_soppression_1} and \ref{s_eff_soppression_2} 
we have shown that for CTEQ-6 PDF (for the chosen values  
$x$ and $Q^2$) Up and Down are suppressed; 
this agrees with results in Fig.\ref{updown_full}.

\begin{figure}[!h]
\centering{\includegraphics[angle=270,width=11.5cm]{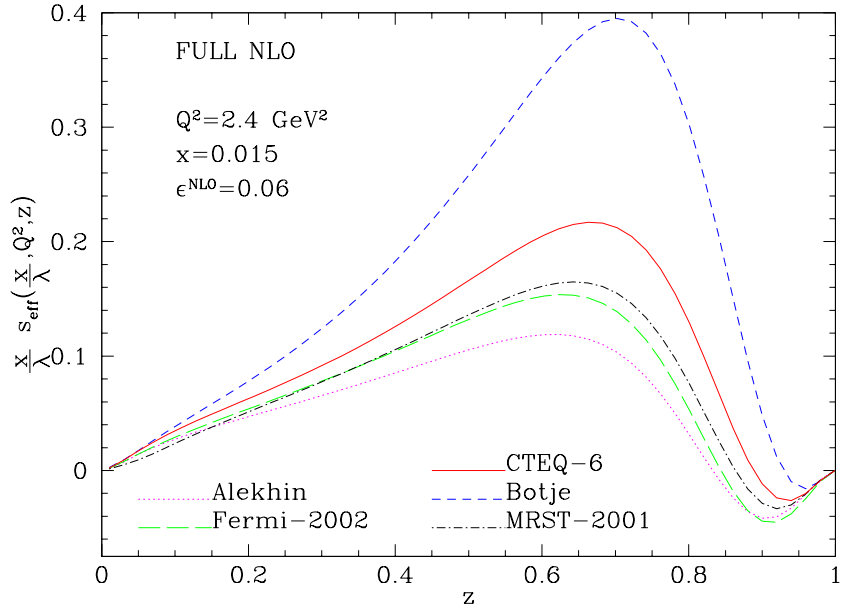}}
\caption{\footnotesize Comparison of the behaviour for
$\frac{x}{\lambda}s_{eff}\left(\frac{x}{\lambda},Q^2,z\right)$ obtained
for different PDF sets: CTEQ-6, MRST-2001, Fermi-2002, Alekhin, Botje.}\label{pdf_graph}
\end{figure}

\noindent
Finally we want to highlight the strong dependence on the PDF set:
in Fig.\ref{pdf_graph} the behaviour of $(x/\lambda) s_{eff}(x/\lambda,Q^2,z)$
is shown for PDF sets contained in the LHA-PDF package \cite{LHA}.

\newpage
\subsection{Dependence on Fragmentation Functions}\label{pgf_FF}

\begin{figure}[!h]
\centering{\includegraphics[angle=270,width=12cm]{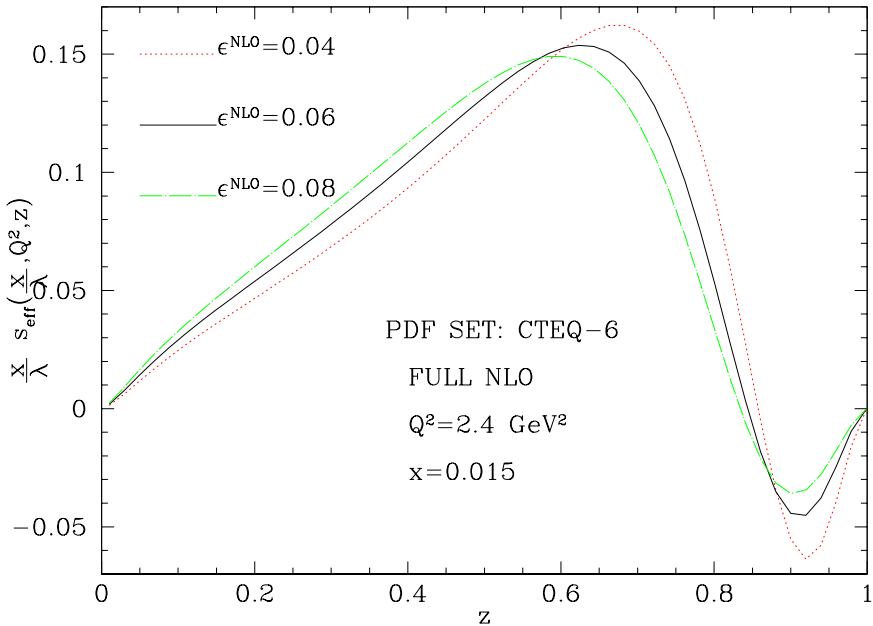}}
\caption{\footnotesize Dependence of
$\frac{x}{\lambda}s_{eff}(\frac{x}{\lambda},Q^2,z)$
on $\varepsilon_P$ parameter.
}\label{graph_ff}
\end{figure}

\noindent
In paragraph \ref{HP_correspondence} we have introduced in 
a generic way the Fragmentation Functions in order to link 
the partonic to the hadronic level, analogously to PDF;
with regard to massive quarks production in paragraph 
\ref{Q_channel} we have seen that FF are independent
of a factorization scale $\mu_F^{2}$.
Now we show some models.\\ 
Fragmentation Functions employed to reproduce numerical results exposed till now
is the Peterson one \cite{PETERSON}:
\begin{equation}\label{FF_peterson}
D(z,\varepsilon_P)=N\frac{1}{z\left[1-z^{-1}-\varepsilon_P(1-z)^{-1} \right]^{2}}
\end{equation}
Normalization factor $N$ is found imposing $\int_0^1dzD(z,\varepsilon_P)=1$
and obviously it depends on the choice of $\varepsilon_P$.
To perform an analysis at LO, the relevant value of $\varepsilon_P$ 
corresponds to $\varepsilon_P=0.20\pm 0.04$ (taken by a comparison 
with experimental data \cite{Data_1});
for an analysis at NLO the suggested value
corresponds to $\varepsilon_P=0.06\pm 0.03$
(from non-DIS data \cite{EPSILON_NLO},\cite{PDG_hern}).
Finally most recent investigations propose values of about
$\varepsilon_P^{NLO}\simeq 0.02\div 0.035$
\cite{Cacciari:1997du},\cite{Nason:1999zj}.
In Fig.\ref{graph_ff}, $(x/\lambda) s_{eff}(x/\lambda,Q^2,z)$
is displayed for $\varepsilon_P$ equal to $0.04$, $0.06$, $0.08$.

\noindent
A very widely employed FF in literature is the one of Collins-Spiller \cite{COLLINS-SPILLER}:
\begin{equation}
D(z,\varepsilon_{CS})=N\left[\frac{1-z}{z}+\frac{(2-z)\varepsilon_{CS}}{1-z} \right]
\frac{1+z^2}{\left[1-z^{-1}-\varepsilon_{CS}(1-z)^{-1} \right]^{2}}
\end{equation}
For a NLO analysis, in \cite{NOMAD} is reported an average value
$\varepsilon_{CS}=0.13\pm 0.08$.
\vspace{-0.5cm}
\begin{figure}[!h]
\centering{\includegraphics[angle=270,width=9.5cm]{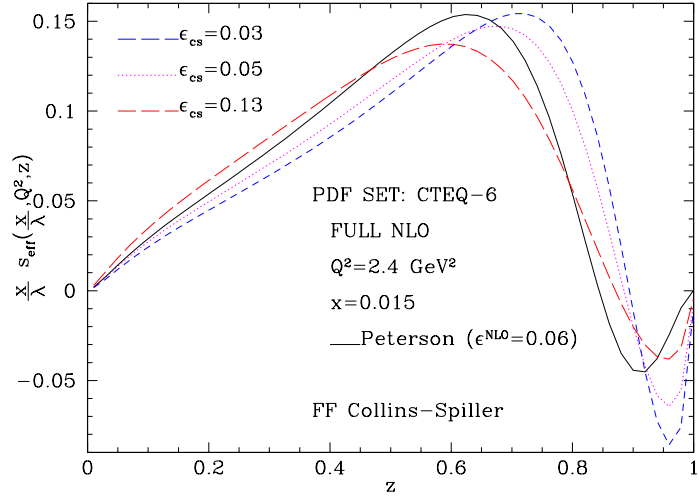}}
\caption{\footnotesize Behaviour of $\frac{x}{\lambda}s_{eff}(\frac{x}{\lambda},Q^2,z)$
obtained for a Collins-Spiller FF.}
\label{graph_ff_cs}
\end{figure}

\noindent
At last we take into account a Kartvelishvili FF \cite{KARTVELISHVILI}, $D(z,\alpha)=Nz^{\alpha}(1-z)$.
\vspace{-0.75cm}
\begin{figure}[!h]
\centering{
\includegraphics[angle=270,width=9.5cm]{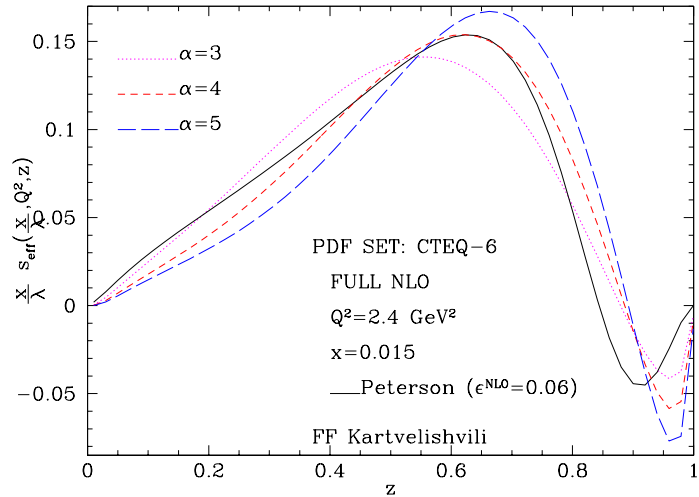}}
\caption{\footnotesize Behaviour of $\frac{x}{\lambda}s_{eff}(\frac{x}{\lambda},Q^2,z)$
obtained for a Kartvelishvili FF.}
\label{graph_ff_k}
\end{figure}

\clearpage
\newpage
\noindent
So far we have tacitly assumed that it is possible to describe
hadronization of the produced heavy quark simply introducing 
a Fragmentation Function and convolve with Semi-Exclusive Coefficient
Functions.
It is important to notice that actually this approach is not 
rigorously correct {\it a priori}.
First of all, as pointed out in paragraph \ref{FRAGMENTATION},
being massive the quark, it is arbitrary to choose a way to 
rescale partonic quantities (as energy, four-momentum, momentum, etc.)
to describe the corresponding hadronic quantities:
only when the transverse momentum is high, differences 
among the many approaches are negligible.
Secondarily when the transverse momenta are low, it is not
correct to assume that fragmentation is really independent 
of the remainder of the event;
in general it is not, because the single quark fragmenting 
must exchange colour charge with the rest of particles of the process
and only when the transverse momentum is high, these effects 
are suppressed.

\subsection{Dependence on Charm Quark Mass}

\begin{figure}[!h]
\centering{\includegraphics[angle=270,width=14cm]{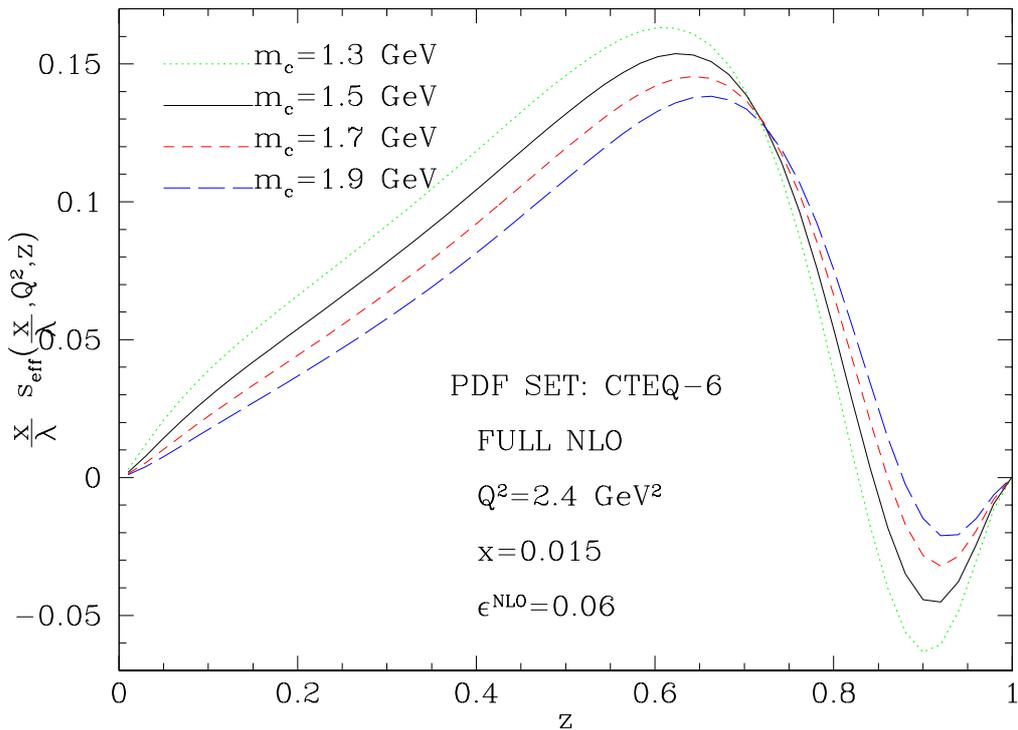}}
\caption{\footnotesize Dependence of $\frac{x}{\lambda}s_{eff}(\frac{x}{\lambda},Q^2,z)$
from the mass of Charm quark.}\label{graph_mass}
\end{figure}

\clearpage
\newpage
\subsection{Scale Dependence}

\begin{figure}[!h]
\centering{\includegraphics[angle=270,width=14cm]{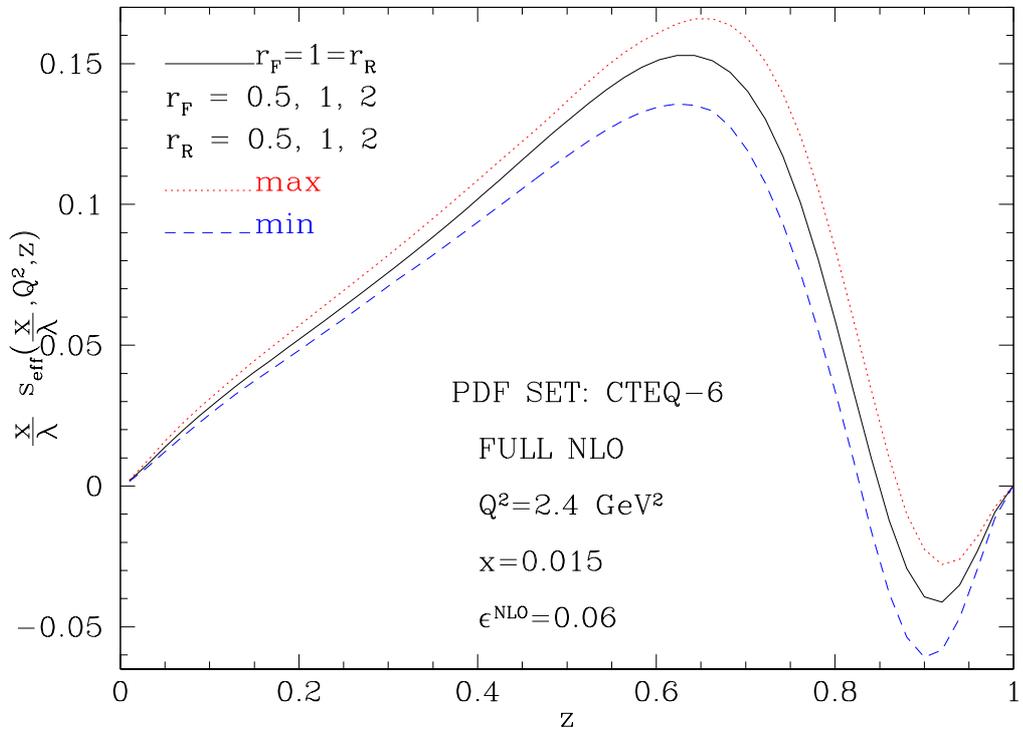}}
\caption{\footnotesize Dependence of $\frac{x}{\lambda}s_{eff}(\frac{x}{\lambda},Q^2,z)$
on the factorization ($\mu_F^2$) and renormalization ($\mu_R^2$) scales.
The continuous line is the result for $\mu_F^2=\mu_R^2=(Q^2+m_c^2)$
whereas the upper (lower) line is the envelopment given by maxima (minima)
obtained varying the parameters $r_{F,R}=\mu_{F,R}^2/(Q^2+m_c^2)$ in $\{0.5,1.0,2.0\}$,
excluding combinations $\{r_{F},r_R\}=\{0.5,2\}$ and $\{r_{F},r_R\}=\{2,0.5\}$.
}\label{graph_scale}
\end{figure}

\noindent
Until now factorization scale $\mu^2_F$ has been set $Q^2+m_c^2$ 
in PDF and Coefficient Functions;
the same has been done with regard to the renormalization scale $\mu_R^2$,
appearing as argument of the coupling constant $\alpha_s$.
In order to investigate the dependence of $(x/\lambda) s_{eff}(x/\lambda,Q^2,z)$
on these two scales, we introduce a compact notation rewriting
$\mu^2_{F,R}=(Q^2+m_c^2)r_{F,R}$.
In graph \ref{graph_scale} there are shown the results for $r_F=1=r_R$ and
the envelopment given by maxima and minima for $r_{F,R}\in \{0.5,1,2\}$ 
independently varying (excluding the cases $\{r_{F},r_R\}=\{0.5,2\}$ and
$\{r_{F},r_R\}=\{2,0.5\}$).

\chapter{Mellin Space}\label{chapter_numerical_tools}

{\it In this chapter we show that working in 
Mellin space allows to elementarily deal with
convolutions of Coefficient Functions, PDF, 
FF and with evolution of Parton Distribution Functions;
in particular the formalism is very advantageous
in order to include resummation.\\
The main difficulties to implement Mellin transforms 
for Coefficient Functions carried out in chapter
\ref{chapter_Coeff_Funct} have been pointed out.
}\\
\\
\\
\noindent
Main result of this work essentially consists
in the analytical calculation of functions $\mathscr{F}_k(\chi,y,z)$
that we found to have the form (eq.\ref{EXCLUSIVE_cross})
\begin{equation}\label{F_x_space}
\mathscr{F}_k(\chi,y,z)=
\sum_a\int_{\chi}^1\int_{\text{max}\{z,\zeta_{min}\}}^1
\!\!\!\!\!\!\!\!\!\!\!\!\!\!\!\!\!\!\!\!\!\!\!
\hat{\mathscr{F}}_k^{a}(\xi,y,\zeta,\mu^2)
\mathfrak{f}^{a}\left(\frac{\chi}{\xi},\mu^2\right)
\mathfrak{D}\left(\frac{z}{\zeta}\right)\frac{d\xi}{\xi}\frac{d\zeta}{\zeta}
\end{equation}
We define the Mellin Transform (MT) $h(N)$, being $N\in\mathbb{C}$,
for a function $h(t)$ with support in $t\in[0,1]$:
\begin{equation*}
h(N)\equiv\int_0^1dt t^{N-1} h(t)
\end{equation*}
Taking MT of \ref{F_x_space} with respect to $\chi$ and $z$
variables, neglecting for sake of simplicity the dependence 
of Coefficient Functions from $y$ and  $\mu^2$, we obtain
\begin{equation*}
\begin{split}
\mathscr{F}_k(N,M)&=\int_0^1d\chi \chi^{N-1}\int_0^1dz z^{M-1}
\mathscr{F}_k(\chi,z)\\
&=\sum_a\mathfrak{f}^{a}(N,\mu^2)\mathfrak{D}(M)
\int_0^1d\xi \xi^{N-1}\int_{\zeta_{min}}^1d\zeta \zeta^{M-1}
\hat{\mathscr{F}}_k^{a}(\xi,\zeta)
\end{split}
\end{equation*}
with $\zeta_{min}=(1-\lambda)\xi/(1-\lambda\xi)$; setting
\begin{equation}\label{DMT}
\mathscr{G}^{a}_k(N,M)\equiv
\int_0^1d\xi \xi^{N-1}
\int_{\zeta_{min}}^1d\zeta \zeta^{M-1}
\hat{\mathscr{F}}^{a}_k(\xi,\zeta)
\end{equation}
we can rewrite the equality in a factorised form 
\begin{equation}\label{F_m_space}
\mathscr{F}_k(N,M)=\sum_a\mathfrak{f}^{a}(N,\mu^2)\mathfrak{D}(M)
\mathscr{G}^{a}_k(N,M)
\end{equation}
\noindent
A very important reason to work in Mellin space
concerns resummation.
In chapter \ref{chapter_results} we have highlighted
that $\frac{x}{\lambda}s_{eff}(\frac{x}{\lambda},Q^2,z)$
(deduced just from perturbative NLO calculation)
has no quantitative meaning in the region $z\gtrsim 0.85$ :
in fact it is fundamental to consider also resummation 
in order to work out accurate and physically meaningful results.
Using MT formalism an observable $\sigma$ resummed up to
Next-to-Leading Log (NLL) + Next-to-Leading Order (NLO) 
is given by
\begin{equation}\label{RESUMM}
\sigma^{(Res)}(N)=\bigg[\sigma^{(NLL)}(N,\alpha_s)-
\sigma^{(NLL)}(N,\alpha_s)\bigg|_{\alpha_s}
\bigg]+\sigma^{(NLO)}(N,\alpha_s)
\end{equation}
where $\sigma^{(NLL)}(N,\alpha_s)|_{\alpha_s}$ represents the 
expansion of $\sigma^{(NLL)}(N,\alpha_s)$ truncated to 
$\alpha_s$ order.
Unlike NLO, theoretical calculation up to NLL is analytically 
feasible only in Mellin space:
consequently, inside such a space, the {\it matching} 
between NLO and NLL contributions is carried out in a natural 
and elementary way \cite{CATANI-TRENTADUE}.

\section{Analytical Treatment and Numerical Implementation} 

\begin{figure}[!h]
\centering
{\includegraphics[width=6cm]{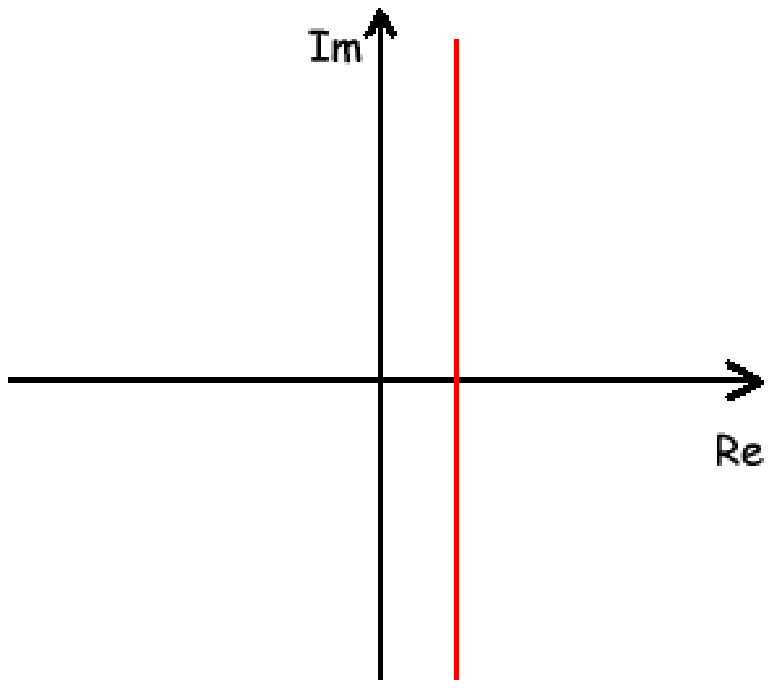}
\includegraphics[width=6cm]{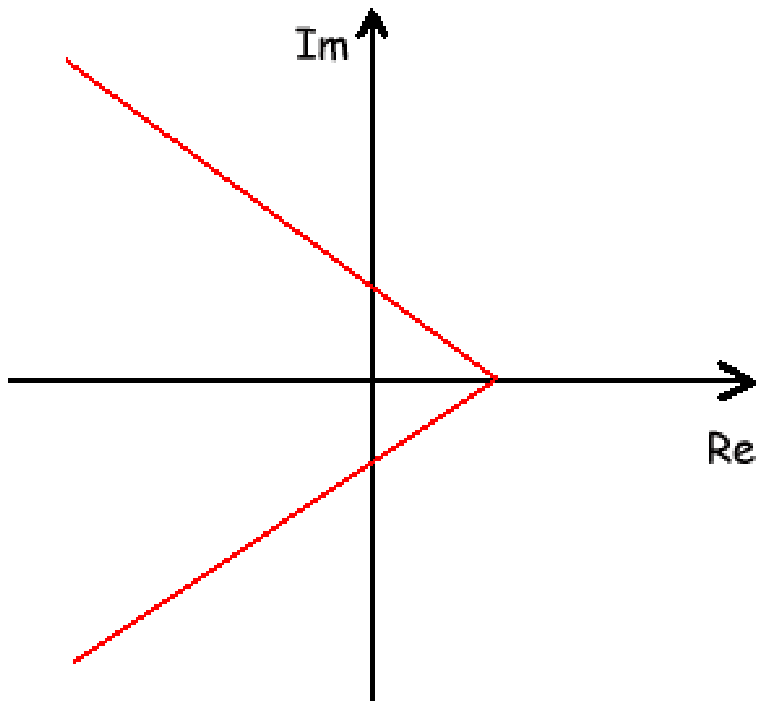}
}\caption{\footnotesize Some integration paths 
to evaluate an Inverse-Mellin-Transform.}\label{IMT_path}
\end{figure}

In Mellin space convolutions have therefore a simplified structure 
allowing to represent functions $\mathscr{F}_k$
as simple product of factors.
Nevertheless the experimental measurement for $\mathscr{F}_k$
is done through variables $\chi$, $z$ and not through 
{\it conjugate variables} $N$ and $M$,
then the Inverse-Mellin-Transform (IMT) of \ref{F_m_space}
has to be taken in order to compare with $\mathscr{F}_k(\chi,z)$.\\
IMT $h(t)$ for a function $h(N)$ is defined as
\begin{equation*}
h(t)\equiv\frac{1}{2\pi i}\int_{C}\!\!dN t^{-N} h(N)
\end{equation*}
where $C$ is a path passing on the right of all singularities
of the analytical continuation of $h(N)$ to the complex
plane.
So
\begin{equation}\label{X_factorization}
\mathscr{F}_k(\chi,z)=\frac{1}{(2\pi i)^2} \int_{C_1}\!\!dN\chi^{-N}
\int_{C_2}\!\!dMz^{-M} \mathscr{F}^{a}_k(N,M)
\end{equation}
At mathematical level, the most elementary path to evaluate the 
inverse-transform is the one illustrated on the left in Fig.\ref{IMT_path}; 
however it is not usable for numerical implementations 
(for usually encountered functions),
because there is not suppression of contributions
coming from the integrand when the imaginary part tends to infinity:
therefore it is not possible to achieve a (settled) convergence for the
result \cite{Kosower}.
In practice a contour like the one on the right of Fig.\ref{IMT_path} 
is employed, or its deformations to obtain results 
having a faster convergence.

\begin{figure}[!h]
\centering{\includegraphics[angle=270,width=14cm]{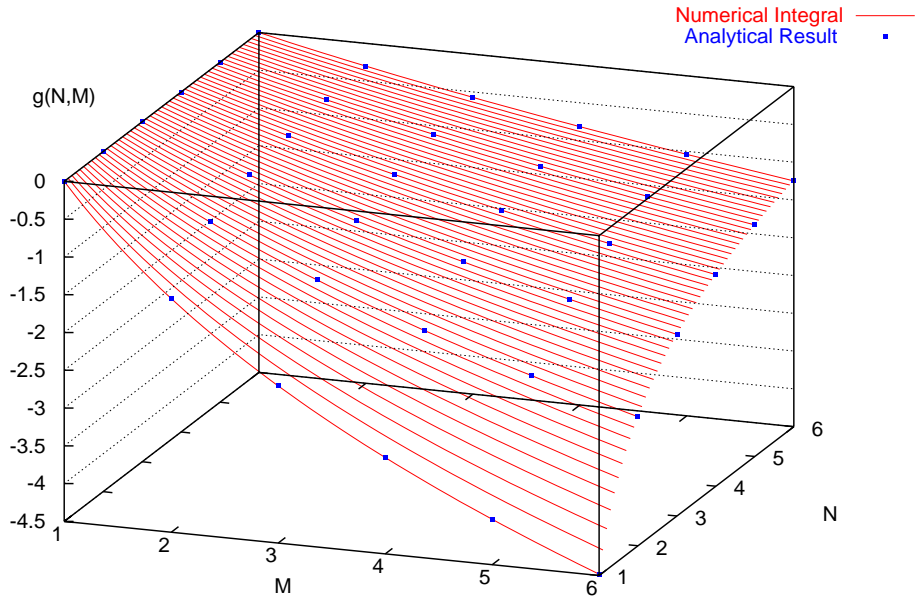}}
\caption{\footnotesize Expression of $g(N,M)$ in \ref{integral_result}
evaluated analytically \eqref{extend_result} and by numerical integration;
$\lambda=0.516$ ($Q^2=2.4\:GeV^2$, $m_c=1.5\:GeV$).
}\label{graph_gmn}
\end{figure}

\noindent
Notice that evaluating a function $g(N)$ along the first path, 
the values $\text{Re}(N)$ are definitely on the right of whatever real number initially fixed;
for the second path this is no more true and originates great issues about treatment 
of the double integral defined in \ref{DMT}.
In order to illustrate the problem in an incisive manner, an example is shown
without loss of generality.\\
Calculating the Mellin transform for $g(t)=t^\alpha$:
\begin{equation*}
g(N)=\int_0^1dtt^{N-1}t^\alpha=\frac{t^{N+\alpha}}{N+\alpha}\bigg|_0^1
\end{equation*}
In the region where $\text{Re}(N)>-\text{Re}(\alpha)$
the result is well defined and equal to $1/(N+\alpha)$;
however, when $\text{Re}(N)<-\text{Re}(\alpha)$, a divergence is found
evaluating integral in $0$.
In order to avoid this problem, MT is evaluated for
$\text{Re}(N)>-\text{Re}(\alpha)$ and the result is 
extended by {\bf analytical continuation} to the whole complex
plane, at the most excepting isolated poles on the real axis.
This explains why it is not possible {\it numerically} to carry out
Mellin transforms for contours not fully contained in the 
region on the right of a fixed real number.
Calculation in \ref{DMT} has to be analytically performed
and then the result extended to the complex plane, because
of requirements on the integration path.\\
We consider the double integral \ref{DMT} for the term 
\begin{equation}
g(\xi,\zeta)=\frac{1}{(1-\xi)_+}\frac{1}{(1-\zeta)_\oplus}
\end{equation}
contained in $\mathscr{G}_k^a(N,M)$ of \ref{DMT},
coming from Semi-Exclusive Coefficient Functions
in \ref{Q_coefficient_functions},
\begin{equation}\label{integral_result}
\begin{split}
g(N,M)=&\int_0^1d\xi \xi^{N-1}
\int_{\frac{(1-\lambda)\xi}{(1-\lambda\xi)}}^1d\zeta \zeta^{M-1}g(\xi,\zeta)\\
=&\int_0^1d\xi\frac{\xi^{N-1}}{(1-\xi)_+}
\int_{\frac{(1-\lambda)\xi}{(1-\lambda\xi)}}^1d\zeta
\frac{\zeta^{M-1}}{(1-\zeta)_\oplus}
\end{split}
\end{equation}
Making use of the expression
\begin{equation*}
\left(1-t^k\right)=(1-t)\sum_{j=1}^{k}t^{j-1}
\end{equation*}
a solution in compact form is found
\begin{equation}\label{compact_result}
g(N,M)=-\sum_{k=1}^{M-1}\frac{1}{k}
\sum_{j=1}^{k}\frac{{_2F}_1(1,N,N+j,\lambda)}{N+j-1}
\end{equation}
but it is difficult to analytically extend 
to the complex plane for $M$ index:
therefore such a result is not usable at the moment.\\
An alternative solution is given by
\begin{equation}\label{extend_result}
\begin{split}
g(N,M)=&S_{1}\left(M-1\right)\Psi\left(N\right)\\
&+\sum_{k=0}^{\infty} \sum_{j=0}^{\infty} \bigg[
\frac{\Gamma(M+k+j)}{\Gamma(M+k+1)}\Psi(M+N+k+j)(1-\lambda)^{M-1}\\
&-\frac{\Gamma(1+k+j)}{\Gamma(2+k)}\Psi(1+N+k+j)
\bigg]\frac{\lambda^{j}}{j!}(1-\lambda)^{k+1}
\end{split}
\end{equation}
where $\Psi(t)$ function is the logarithmic derivative
of $\Gamma(t)$ function
\begin{equation}
\Psi(t)=\frac{d}{dt}\log\Gamma(t)=\frac{\Gamma'(t)}{\Gamma(t)}
\end{equation}
whereas $S_1(t)$ is the analytical continuation of the
harmonic series to the whole complex plane
\begin{equation}
S_1(t)=\sum_{k=1}^{t}\frac{1}{k}=\Psi(t+1)+\gamma_e
\end{equation}
where $\gamma_e$ ($=0.57721...$) is the Euler constant. \\
Being $\int_{\zeta_{min}}^1 d\zeta h_{\oplus}(\zeta)=0$ 
because of {\it plus}-distribution properties,
it is verified from \ref{integral_result} that $g(N,1)= 0$; 
in fact expressions \ref{compact_result} and \ref{extend_result} 
satisfy this requirement. 
To check the correctness, these expressions have been 
evaluated for values of $N$ and $M$ integer and positive, 
varying $\lambda$, and they have been compared (Fig.\ref{graph_gmn}) 
to the analogous result numerically derived from \ref{integral_result}. \\

\noindent
Extension of result \ref{extend_result} to the complex plane is 
immediate because there are only isolated poles on the real axis:
such an expression is then usable with a generic contour 
to calculate an inverse-transform.
Numerical implementation of this contribution involves great
complications.
First of all the integration path must be good enough to 
guarantee stability and fast convergence for inverse-transform 
of $S_1$ and $\Psi$ functions. 
Furthermore the right evaluation of $g(N,M)$ for $N$ and $M$
with real and/or imaginary parts \textquotedblleft large\textquotedblright\:
in absolute value is hard;
in fact the terms of the sum are made of contribution tending 
to balance (first line in \ref{extend_result} with the second one)
and the nearly cancellation of large numbers implicates
a not trivial control of accuracy.
A method for fast converging of the 
Inverse-Mellin-Transform inside a restricted region of 
$N$ and $M$ is therefore needed or it is mandatory to use multiple 
\cite{CLN},\cite{GNU_MP} and adjustable precision in order to deal
with sums containing a high number of terms;
for the second case in particular, there is also the
problem that some coefficients can become very big
and go out the maximum range of a computer.\\
\\
It can be shown that through $g(N,M)$ and few other contributions
it is possible to write, in principle, the whole Coefficient Functions.
Terms proportional to $\delta(1-\zeta)$ contained in 
\ref{Q_coefficient_functions} and \ref{G_coefficient_functions}
give results that can be deduced from relations in
\cite{BLUMLEIN}.
For term
$$(1-\zeta)\left(\frac{1-\lambda\xi}{1-\xi}\right)^2
\left[\frac{1-\xi}{(1-\lambda\xi)^2}\right]_+$$
in \ref{Q_coefficient_functions} we have to explicitly calculate
the double integral \ref{DMT}.\\
By means of the identity (as distribution)
\begin{equation*}
1=\frac{(1-\xi)}{(1-\xi)_+}\frac{(1-\zeta)}{(1-\zeta)_{\oplus}}
\end{equation*}
any terms of remaining contributions in Coefficient Functions
can be rewritten in the form
\begin{equation*}
r(\xi,\zeta)=\sum_{a,b}C_{ab}\frac{\xi^a\zeta^b}{(1-\xi)_+(1-\zeta)_{\oplus}}
\end{equation*}
where $C_{ab}$ is an opportune numerical coefficient,
containing at the most powers of $\lambda$.
Using the properties of Mellin transforms,
if $h(N)$ is the MT of $h(t)$, then the one for
$t^{\alpha}h(t)$ is given by $h(N+\alpha)$, so
\begin{equation*}
r(N,M)=\sum_{a,b}C_{ab}g(N+a,M+b)
\end{equation*}
Theoretically it is possible to obtain the MT of the whole 
Coefficient Functions in this way. 
In practice however such an approach is opportune 
only if a fast and accurate routine is found
to calculate $g(N,M)$; otherwise it is preferable
to evaluate the double integral \ref{DMT}
for each single term contained in $\mathscr{G}_k^a(\xi,y,z)$,
running into the same problems encountered evaluating $g(N,M)$,
optimizing algorithms according to the  necessities.\\ 
\\
Complications in evaluating integrals \ref{DMT} originates 
from the fact that we are dealing with  
Coefficient Functions both differential and for massive quarks.
For the massless case ($\lambda=1$), expression \ref{DMT}
in fact reduces to
\begin{equation*}
\begin{split}
\mathscr{G}^{a}_k(N,M)=&
\int_0^1d\xi \xi^{N-1}
\int_{0}^1d\zeta \zeta^{M-1}
\hat{\mathscr{F}}^{a}_k(\xi,\zeta)\\
=&\sum_j\left[\int_0^1d\xi \xi^{N-1}
\mathscr{U}^{a}_{j,k}(\xi)\right]
\left[\int_{0}^1d\zeta \zeta^{M-1}
\mathscr{V}^{a}_{j,k}(\zeta)\right]
\end{split}
\end{equation*}
that is $\mathscr{G}^{a}_k(N,M)$ becomes a sum of terms 
{\bf factorised} as product of two {\it independent} Mellin transforms.\\
The inclusive case involves, by definition, only one 
variable and it is simply recovered setting $M=1$ in \ref{DMT};
$\mathscr{G}^{a}_k(N,1)$ is then turned out to be the Mellin
transform of expressions in paragraph \ref{C_F_inclusive}.
For both cases, thanks to MT of functions and most recurring 
distributions \cite{BLUMLEIN}, there are not particular issues.\\
For the situation differential and massive at the same time, 
the inner integral in \ref{DMT} depends on $\xi$ variable:
this is the root cause for the encountered complexity.

\section{PDF Treatment}

In the previous section we have pointed out the difficulties
of a numerical implementation to calculate  
$\mathscr{F}_k(\chi,z)$ structure functions 
using the Mellin space formalism.
The advantage of such an approach justifies the investment on investigation about
the described problems, even if effective 
tools to numerically calculate relation \ref{F_x_space} already exist.\\
Besides being the natural environment to handle resummation \ref{RESUMM},
working in Mellin space is convenient also to deal with PDF.

\subsection{Evolution Equation}
Dependence of PDF on the scale $\mu^2$ is regulated by 
the matrix evolution equation DGLAP (\cite{DISSERTORI}, \cite{ESW}), 
coupling $2n_f+1$ distribution functions of quarks, antiquarks and gluons
($n_f$ is the flavours number)
\begin{equation}\label{DGLAP}
\mu^2\frac{\partial}{\partial\mu^2}\left[
\begin{array}{c} q_i(x,\mu^2)\\ g(x,\mu^2)\end{array}
\right] = \frac{\alpha_s(\mu^2)}{2\pi}\sum_{q_j}\int_x^1\frac{d\xi}{\xi}\left[
\begin{array}{cc} P_{q_i q_j}(\xi) & P_{q_i g}(\xi)\\ P_{gq_j}(\xi) &
P_{gg}(\xi)\end{array}\right]\left[
\begin{array}{c} q_j(x/\xi,\mu^2)\\ g(x/\xi,\mu^2)\end{array}\right]
\end{equation}
where $q_i$ and $q_j$ can vary along all the freedom degrees of quarks.\\
We introduce combinations $q_{NS}$ of {\it non-singlet}:
\begin{equation}\label{NS_combination}
\begin{split}
V_i&=q_i^-\\
T_3&=u^+ - d^+\\
T_8&=u^+ + d^+ - 2s^+\\
T_{15}&=u^+ + d^+ + s^+ -3 c^+\\
T_{24}&=u^+ + d^+ + s^+ + c^+ - 4b^+\\
T_{35}&=u^+ + d^+ + s^+ + c^+ + b^+ - 5t^+
\end{split}
\end{equation}
with $q_i^{\pm}=q_i\pm\overline{q}_i$, where $q_i=u,d,c,s,t,b$ 
obviously are quark flavours;
for these combinations the matrix equation \ref{DGLAP} 
decouples in $2n_f-1$ independent scalar equations
\begin{equation}\label{not_singlet}
\begin{split}
&\mu^2\frac{\partial}{\partial\mu^2} q_{NS}^{V}(x,\mu^2)
 = \frac{\alpha_s(\mu^2)}{2\pi}\int_x^1\frac{d\xi}{\xi}
 P_{qq}^{-}(\xi) q_{NS}^{V}(x/\xi,\mu^2)\\
&\mu^2\frac{\partial}{\partial\mu^2} q_{NS}^{T}(x,\mu^2)
 = \frac{\alpha_s(\mu^2)}{2\pi}\int_x^1\frac{d\xi}{\xi}
 P_{qq}^{+}(\xi) q_{NS}^{T}(x/\xi,\mu^2)
\end{split}
\end{equation}
having defined $P^{\pm}=P^{NS}_{qq}\pm P^{NS}_{q\overline{q}}$,
where $P^{NS}$ have been introduced by decomposing
\begin{equation*}
P_{q_i q_k}=\delta_{ik}P^{NS}_{qq}+P^{S}_{qq}
\qquad
P_{q_i \overline{q}_k}=\delta_{ik}P^{NS}_{q\overline{q}}+P^{S}_{q\overline{q}}
\end{equation*}
We construct the only {\it singlet} distribution as
\begin{equation}\label{S_combination}
\Sigma=\sum_iq_i^+=\sum_i \left(q_i+\overline{q}_i\right)
\end{equation}
Its evolution is coupled to the distribution of the gluon
\begin{equation}\label{singlet}
\mu^2\frac{\partial}{\partial\mu^2}\left[
\begin{array}{c} \Sigma(x,\mu^2) \\ g(x,\mu^2)\end{array}
\right] = \frac{\alpha_s(\mu^2)}{2\pi}\int_x^1\frac{d\xi}{\xi}\left[
\begin{array}{cc} P_{qq}(\xi) & 2n_f P_{qg}(\xi)\\
                  P_{gq}(\xi) & P_{gg}(\xi)\end{array}\right]\left[
\begin{array}{c}\Sigma(x/\xi,\mu^2)\\ g(x/\xi,\mu^2)\end{array}\right]
\end{equation}
so now 
\begin{align*}
P_{qq}&\equiv P^++n_f\left( P^{S}_{qq}+ P^{S}_{q\overline{q}}\right)\\
P_{qg}&\equiv P_{q_i g}=P_{\overline{q}_i g}\\
P_{gq}&\equiv P_{g q_i}=P_{g \overline{q}_i}
\end{align*}
What has been worked out so far allows to simplify the 
matrix structure of evolution equations, but 
the base structure of such a kind of integral-differential
equations is left unchanged.\\
Applying MT with respect to $x$ variable for evolution
equations \ref{not_singlet} and \ref{singlet},
it follows
\begin{equation}
\begin{split}
&\mu^2\frac{\partial}{\partial\mu^2} q_{NS}^{V}(N,\mu^2)
 = \frac{\alpha_s(\mu^2)}{2\pi} \gamma_{qq}^{-}(N) q_{NS}^{V}(N,\mu^2)\\
&\mu^2\frac{\partial}{\partial\mu^2} q_{NS}^{T}(N,\mu^2)
 = \frac{\alpha_s(\mu^2)}{2\pi} \gamma_{qq}^{+}(N) q_{NS}^{T}(N,\mu^2)
\end{split}
\end{equation}
\begin{equation}
\mu^2\frac{\partial}{\partial\mu^2}\left[
\begin{array}{c} \Sigma(N,\mu^2) \\ g(N,\mu^2)\end{array}
\right] = \frac{\alpha_s(\mu^2)}{2\pi}\left[
\begin{array}{cc} \gamma_{qq}(N) & 2n_f \gamma_{qg}(N)\\
                  \gamma_{gq}(N) & \gamma_{gg}(N)\end{array}\right]\left[
\begin{array}{c}\Sigma(N,\mu^2)\\ g(N,\mu^2)\end{array}\right]
\end{equation}
where {\it anomalous dimensions} $\gamma_{ab}^{(\pm)}$ ($a,b\in \{q,g\}$) 
are given by
\begin{equation}
\gamma_{ab}^{(\pm)}(N,\alpha_s)=\int_0^1dx x^{N-1}P_{ab}^{(\pm)}(x,\alpha_s)
\end{equation}
We have highlighted the dependence of $P_{ab}$ (and therefore of 
$\gamma_{ab}$) also on $\alpha_s$ and consequently on the scale $\mu^2$
therein contained.
Under Mellin transform convolution integrals become multiplications,
so an analytical solution is easily carried out; 
for the case of non-singlet it follows (with opportune $\gamma(N,t)$)
\begin{equation}\label{M_evolution}
q_{NS}(N,\mu^2)=q_{NS}(N,\mu^2_0)\exp\left\{
\int_{\mu^2_0}^{\mu^2}\frac{dt}{t}\frac{\alpha_s(t)}{2\pi} \gamma(N,t)\right\}
\end{equation}
For the case of singlet $\Sigma$ and of $g$ it is possible to proceed 
analogously even if it is a little bit more complicated because
of the reciprocal dependence.
Inverting relations \ref{NS_combination} and \ref{S_combination}
distributions of $n_f$ quarks, $n_f$ antiquarks and gluon $g$
are obtained.
We want to point out that, because of \ref{M_evolution}, 
in Mellin space PDF have the form
\begin{equation}\label{Evolutor}
\begin{split}
&\mathfrak{f}^a(N,\mu^2)=\mathfrak{f}^a(N,\mu^2_0)
\mathscr{E}\left(\mu^2,\mu^2_0, N\right)\\
&\mathscr{E}\left(\mu^2,\mu^2_0, N\right)=
\exp\left\{
\int_{\mu^2_0}^{\mu^2}\frac{dt}{t}\frac{\alpha_s(t)}{2\pi}
\gamma(N,t)\right\}
\end{split}
\end{equation}
where $\mathfrak{f}^a(N,\mu^2_0)$ is the distribution function
of a parton inside a nucleon at the fixed scale $\mu^2_0$, 
whereas $\mathscr{E}$ is the operator of evolution from 
a scale $\mu^2_0$ to $\mu^2$ \cite{Kosower},
\cite{Evo_Eq_1},\cite{Evo_Eq_2}.

\newpage
\subsection{Modellization}

\begin{figure}[!h]
\centering{\includegraphics[angle=270,width=12.8cm]{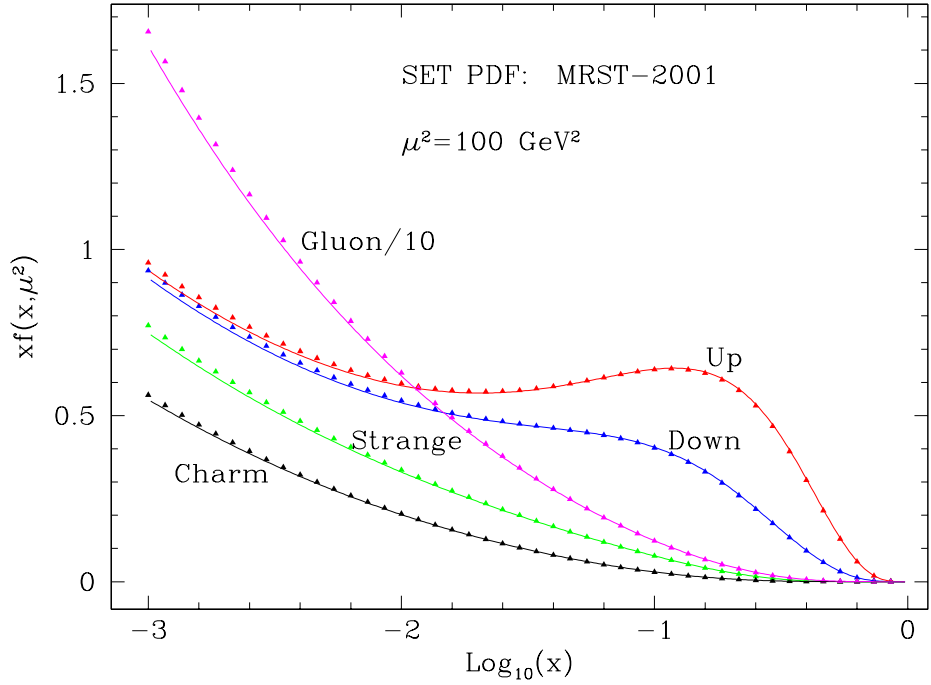}}
\centering{\includegraphics[angle=270,width=12.8cm]{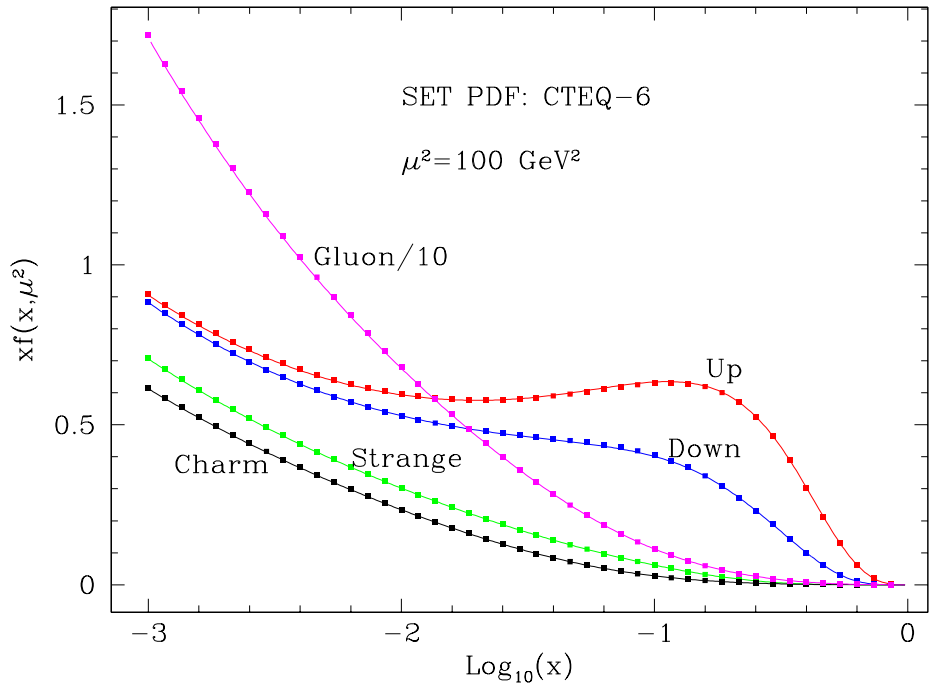}}
\caption{\footnotesize Behaviour of $xf(x,\mu^2)$ provided by 
LHA-PDF (points) and by our code (continuous line), 
for a scale $\mu^2=100\:GeV^2$
and for each parton (up, down, strange, charm, gluon).
At the top inside the figure, parametrisation
MRST-2001 is taken into account, whereas at the bottom, 
the CTEQ-6 one.}
\label{pdf_set}
\end{figure}

\noindent
As continuation of the results of the previous paragraph,
we have implemented a numerical code enabling to obtain 
values of PDF for an arbitrary scale ($\mu^2>\mu_0^2)$,
freely choosing a parametrisation for $f_a(N,\mu_0^2)$ 
(or equivalently for $f_a(\chi,\mu_0^2)$, taking then the MT);
to achieve this goal we partially made use of a code written by
Dr.~S.~Weinzierl \cite{FEPD}.\\
As preliminary test to verify the good working, 
we have reproduced the MRST-2001 PDF set calculating
Mellin transforms for parton distributions parametrised at
the scale $\mu_0^2=1\:GeV^2$, illustrated in \cite{MRST2001};
\begin{equation*}
\begin{split}
xu_V(x)=&0.158x^{0.25}(1-x)^{3.33}(1+5.61x^{0.5}+55.49x)\\
xd_V(x)=&0.040x^{0.27}(1-x)^{3.88}(1+52.73x^{0.5}+30.65x)\\
xS(x)=&0.222x^{-0.26}(1-x)^{7.10}(1+3.42x^{0.5}+10.30x)\\
xg(x)=&1.900x^{0.09}(1-x)^{3.70}(1+1.26x^{0.5}-1.43x)-0.21x^{-0.33}(1-x)^{10}
\end{split}
\end{equation*}
with a structure for light flavours of the sea given by
\begin{equation*}
\begin{split}
2\overline{u}&=0.4S-\Delta\qquad
2\overline{d}=0.4S+\Delta\qquad
2\overline{s}=0.2S\Delta\\
x\Delta&=x(\overline{d}-\overline{u})=
1.195x^{1.24}(1-x)^{9.10}(1+14.05x-45.52x^2)
\end{split}
\end{equation*}
We have compared our results at many scales $\mu^2$
to the ones provided by LHA-PDF \cite{LHA} library:
in Fig.\ref{pdf_set} we show a comparison for distributions
of some partons at $\mu^2=100\:GeV^2$ as scale.

\noindent
After that we have implemented also parametrisations for CTEQ6
\cite{CTEQ_6} and the ones given by ZEUS collaboration.\\
An interesting possibility consists in being able to define
also asymmetric distributions with regard to quark/antiquark 
pairs coming from gluon splitting of the sea;
in particular this holds for $s$ and $\overline{s}$,
up until now assumed equal by prescription in standard PDF sets.\\
Rewriting \ref{F_m_space} in the form
\begin{equation}
\mathscr{F}_k(N,M)=\sum_a\mathfrak{f}^{a}(N,\mu^2_0)\mathscr{E}(\mu_0^2,\mu^2)
\mathfrak{D}(M)
\mathscr{G}^{a}_k(N,M)
\end{equation}
we can also carry out $\mathscr{F}_k(\chi,z)$ numerically calculating
the Inverse-Mellin Transform with respect to $N$ and $M$:
through a double integral therefore is obtained the results for the
two convolutions of Coefficient Functions with FF and PDF, 
including their evolutions, without solving by means of 
numerical methods integral-differential equations \ref{not_singlet} 
and \ref{singlet}, furthermore depending on 
the chosen functional form for $f(\chi,\mu_0^2)$.

\chapter*{Conclusions and Outlook}
\addcontentsline{toc}{chapter}{Conclusions and Outlook}

In this research we have handled in a general way the
Deep Inelastic Scattering processes mediated by Charged Currents.
Following the classical approach to DIS, we have parametrised 
an exclusive hadronic cross section with multi-differential
structure functions ($d^jF_{k}/\prod_i^jd\alpha_i$), 
including in the analysis the mass of charged leptons.\\
In the parton model context, we have reconstructed the structure 
functions from the Partonic Distribution Functions (PDF)
and Fragmentation Functions (FF), and in the framework of 
Quantum Chromodynamics (QCD) we have carried out the semi-exclusive
Coefficient Functions up to Next-to-Leading Order for heavy quark production.\\
This result was already partially known \cite{GKRR},\cite{KRST};
we can therefore confirm it independently and now we have the full analytic control
of the model.

The obtained results have been numerically implemented using C/C++ language,
in order to carry out quantitative analysis for a more complete 
phenomenological study.
In the meantime, the theoretical uncertainties 
coming from physical inputs (heavy quark mass, Fragmentation Functions,
parton densities, ...) or from non-physical parameters (as 
factorization and renormalization scales) have been investigated. 

In this way we have then verified that the semi-exclusive result 
obtained through calculations at fixed order (NLO) is quantitatively 
inadequate in some kinematic regions, pointing out the need 
of considering also contributions originated by higher orders 
and then the necessity of all-orders-resummed calculation.

In order to simplify the inclusion of resummation in the analysis, 
we have considered the possibility to perform the convolution
among Partonic Densities, Coefficient Functions and Fragmentation
Functions directly in Mellin space. For this purpose we have written
a C/C++ code returning evolved Mellin moments of certain parametrisations
for modern PDF sets, as CTEQ6 and MRST2001.\\
The only open item is then the numerical treatment of analytical 
expressions extended to the whole complex plane for the
moments of the semi-exclusive Coefficient Functions.
Results for this ingredient seem to be hard to obtain numerically 
and their implementation has not yet been completed.

As first application of our results, we have considered 
the investigation of Strange quark distribution inside nucleons.
The CC DIS process with production of a hadron containing Charm 
is actually very useful to this purpose, because it allows to study
independently the distributions of quarks and antiquarks.
Therefore our semi-exclusive results permit to precisely reproduce 
experimentally accessible observables and to accurately measure 
partonic distribution of Strange quark and 
$s - \bar s$ asymmetry. 
 
Although the usual $x$-space convolution is satisfactory with 
regard to the described approach, in the future we will try to carry out 
an effective numerical implementation of Mellin transforms 
for the semi-exclusive Coefficient Functions.
This will allow to include in the (NLO) analysis the resummation 
up to Next-to-Leading Logarithm (NLL) in a natural way, so eventually 
we will have a theoretical result reliable and accurate 
in the whole kinematic region of interest.

A further development of the numerical implementation should be 
the inclusion of decay of charmed hadrons in muons, using a Montecarlo 
approach. 
This would enable to simulate in detail di-muonic events experimentally 
measured \eqref{pgf_exp} and therefore to describe the phenomenology 
in a more accurate and complete way.

{\appendix
\chapter{DIS - Details and Formalism}\label{DIS Formalism}
\section{Electroweak Leptonic Tensor}
\label{leptonic_tensor} \rm
The most generic electroweak leptonic vertex is (neglecting
overall coupling constants)
\begin{equation*}
\begin{split}
\langle \ell_{\it out},s| j_{L}^{\mu} |\ell_{\it in},r \rangle&=
\overline{\Psi}^{(s)}_{\ell{out}}(k_{out})\Gamma^{\mu}\Psi^{(r)}_{\ell{in}}(k_{in})\\
&=\overline{\Psi}^{(s)}_{\ell{out}}(k_{out})
\gamma^{\mu}\frac{(V-A\gamma_{5})}{2}
\Psi^{(r)}_{\ell{in}}(k_{in})
\end{split}
\end{equation*}
where $V$ and $A$ are the appropriate couplings of the leptonic 
current to the electroweak bosons.
\begin{figure}[!h]\label{leptonic_current}
\centering\epsfig{file=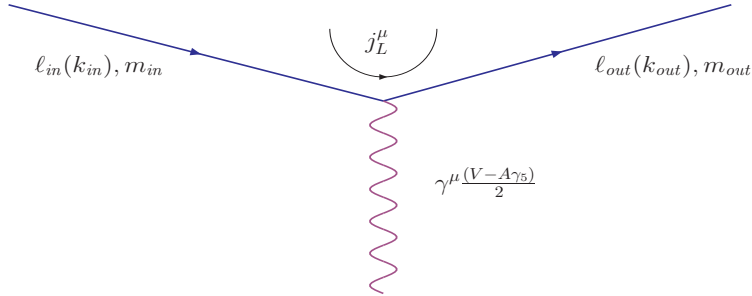,width=10cm,height=4cm}
\caption{ \footnotesize{Generic electroweak leptonic vertex.}}
\end{figure}

\noindent Keeping the squared modulus and summing over the 
polarization states ($s$,$r$) of the leptons, it is straightforward 
to obtain
\begin{equation}
\begin{split}
\label{generic_leptonic_tensor}
L^{\mu\nu}=&\left(V^{2}+A^{2}\right)
\left[k_{out}^{\mu}k_{in}^{\nu}+k_{in}^{\mu}k_{out}^{\nu}-(k_{in}k_{out})g^{\mu\nu}\right]\\
&+2iVA\epsilon^{\mu\alpha\nu\beta}k_{\alpha}^{in}k_{\beta}^{out}
+\left(V^{2}-A^{2}\right)m_{k_{in}}m_{k_{out}}g^{\mu\nu}
\end{split}
\end{equation}
the most generic leptonic tensor.

\newpage
\begin{itemize}
\item Electromagnetic case is carried out for $V=2$, $A=0$ and
obviously\\ 
$m_{k_{in}}=m_{k_{out}}\equiv m$:\\
\begin{equation}
\label{EM_tensor}
L^{\mu\nu}_{EM}=
4\left[k_{out}^{\mu}k_{in}^{\nu}+k_{in}^{\mu}k_{out}^{\nu}-(k_{in}k_{out})g^{\mu\nu}\right]
+4m^{2}g^{\mu\nu}
\end{equation}
\item Charged Electroweak case is carried out for $V=1$, $A=1$ and
the masses are allowed to be different (however their contributions disappear):\\
\begin{equation}
\label{EW_CC_tensor}
L^{\mu\nu}_{CC}=2\left[k_{out}^{\mu}k_{in}^{\nu}+k_{in}^{\mu}k_{out}^{\nu}-(k_{in}k_{out})g^{\mu\nu}\right]
+2i\epsilon^{\mu\alpha\nu\beta}k_{\alpha}^{in}k_{\beta}^{out}
\end{equation}
\item  Neutral Electroweak case is carried out for $V=g_{V}$,
$A=g_{A}$ and equal masses ($m_{k_{in}}=m_{k_{out}}\equiv m$):\\
\begin{equation}
\label{EW_NC_tensor}
\begin{split}
L^{\mu\nu}_{NC}=&\left(g_{V}^{2}+g_{A}^{2}\right)
\left[k_{out}^{\mu}k_{in}^{\nu}+k_{in}^{\mu}k_{out}^{\nu}-(k_{in}k_{out})g^{\mu\nu}\right]\\
&+2ig_{V}g_{A}\epsilon^{\mu\alpha\nu\beta}k_{\alpha}^{in}k_{\beta}^{out}
+\left(g_{V}^{2}-g_{A}^{2}\right)m^{2}g^{\mu\nu}
\end{split}
\end{equation}
where $g_{V}$ and $g_{A}$ are the (leptonic) fermion couplings 
to $Z^{0}$.
\end{itemize}
If the current is made of antiparticles rather than particles, 
then, instead of $j_{L}^{\mu}$, we must use 
\begin{equation*}
\Gamma^{\mu}\rightarrow C^{-1}j_{L}^{\mu}C=
-\gamma^{\mu}\frac{(V+A\gamma_{5})}{2}
\end{equation*}
In this way the result comes from \eqref{generic_leptonic_tensor} 
simply performing the change $A\rightarrow -A$ 
($C$ is the charge-conjugation operator; 
see \cite{HAMA} pages 298 and 303).\\
Notice that no averaging over the initial states (of spin) has been done.
\begin{figure}[!h]
\centering\epsfig{file=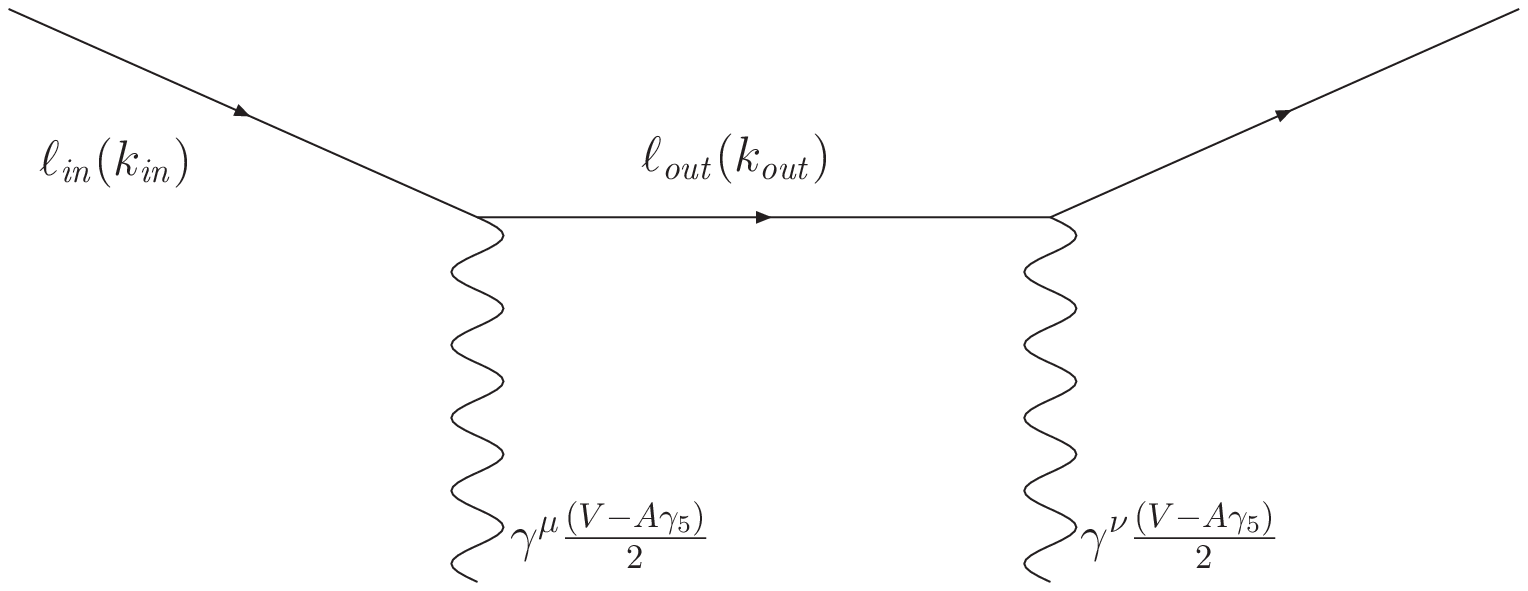,width=12 cm}
\caption{ \footnotesize{Leptonic Tensor.}}
\end{figure}

\newpage
\section{Hadronic Tensor}\label{generic_hadronic_tensor}

In \eqref{generic_cross_section} we found the term
\begin{equation*}
\begin{split}
\hat{W}^{\alpha\beta}(2\pi)^{4}\delta^{(4)}(P+& q-p_{X})\equiv
\frac{W^{\alpha\beta}}{4\pi}(2\pi)^{4}\delta^{(4)}(P+q-p_{X})\\
=&\frac{1}{4\pi}\sum_{\text{pol,X}}
\langle P| J_{H}^{\dagger\beta}(0)| X \rangle
\langle X| J_{H}^{\alpha}(0) |P \rangle
(2\pi)^{4}\delta^{(4)}(P+q-p_{X})
\end{split}
\end{equation*}
where polarisation states are understood and the tensor has not been 
averaged over initial spin states (we have done expressly in the 
main calculation);
rewriting $\delta$-Dirac as inverse Fourier transform
$$\delta^{(4)}(P+q-p_{X})=\frac{1}{(2\pi)^{4}}\int d^{4}te^{+i(P+q-p_{X})t}$$
then
\begin{equation*}
\begin{split}
H^{\alpha\beta}\equiv&\hat{W}^{\alpha\beta}(2\pi)^{4}\delta^{(4)}(P+ q-p_{X})\\
=&\frac{1}{4\pi}\int d^{4}te^{iqt}\sum_{\text{pol,X}}
\langle P|  e^{+iPt} J_{H}^{\dagger\beta}(0) e^{-ip_{X}t}| X \rangle
\langle X | J_{H}^{\alpha}(0)  |P \rangle\\
=&\frac{1}{4\pi}\int d^{4}te^{iqt}
\langle P| J_{H}^{\dagger\beta}(t)J_{H}^{\alpha}(0)|P \rangle\\
=&\frac{1}{4\pi}\int d^{4}te^{iqt}
\langle P|\left[ J_{H}^{\dagger\beta}(t),J_{H}^{\alpha}(0)\right]|P \rangle
\end{split}
\end{equation*}
that is $H^{\alpha\beta}$ can be expressed by means of the four-momenta 
of the particles of the process.
Physics of vertex involving the nucleon is supposed unknown and then
it is described in the most general way with a second rank 
{\it Lorentz-invariant} tensor.\\
\\
It is possible to construct such a tensor using only the momenta of 
incoming particles in the vertex (Fig.\ref{Generic_Vertex});
in fact the final state is fixed by four-momentum conservation ($p'= p+q$). 
\begin{equation}
\label{generic_tensor}
\begin{split}
H^{\alpha\beta}=&
-W_{1}g^{\alpha\beta}
+W_{2}p^{\alpha}p^{\beta}\\
&-iW_{3}\epsilon^{\alpha\beta}_{\:\:\:\:\rho\tau}p^{\rho}q^{\tau}
+W_{4}q^{\alpha}q^{\beta}\\
&+W_{5}(p^{\alpha}q^{\beta}+q^{\alpha}p^{\beta})
+iW_{6}(p^{\alpha}q^{\beta}-q^{\alpha}p^{\beta})
\end{split}
\end{equation}
\begin{figure}[!h]
\centering\epsfig{file=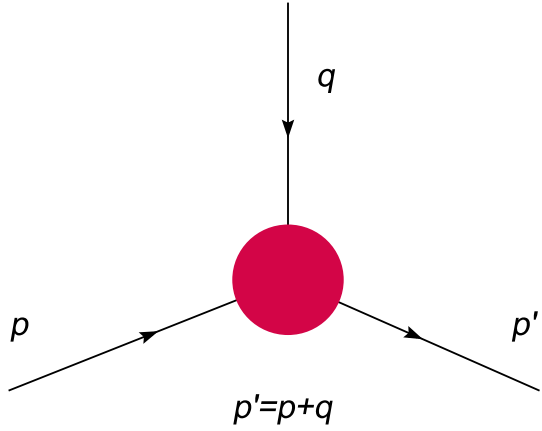} \caption{
\footnotesize{Hadronic Tensor: incoming and outgoing momenta.}}
\label{Generic_Vertex}
\end{figure}

\noindent 
(cf. \cite{ITZU}, page 532)
where $W_{k}$ depend on scalar quantities
made of $p$ and $ q$ (i.e. $W_{k}=W_{k}(p^2,pq,q^{2})$).
It is interesting to notice that there are not
$\gamma$-matrices: in fact we have already summed over spin
states (\cite{HAMA}, page 180).\\
If we suppose that interaction is purely electromagnetic, then
antisymmetric terms disappear (because of parity conservation) and the
current is preserved $(q_{\alpha}H^{\alpha\beta}=q_{\beta}H^{\alpha\beta}=0)$; 
in this case the tensor structure is simpler:
\begin{equation*}
H_{em}^{\alpha\beta}=
W_{1}\left(-g^{\alpha\beta}+\frac{q^{\alpha}q^{\beta}}{q^{2}}\right)
+W_{2}
\left(p^{\alpha}-\frac{pq}{q^{2}}q^{\alpha}\right)
\left(p^{\beta}-\frac{pq}{q^{2}}q^{\beta}\right)
\end{equation*}
For parity violating processes (i.e. weak) with current conservation
(i.e. for massless particles) then \cite{STRM}
\begin{equation*}
H_{ew}^{\alpha\beta}=
W_{1}\left(-g^{\alpha\beta}+\frac{q^{\alpha}q^{\beta}}{q^{2}}\right)
+W_{2} \left(p^{\alpha}-\frac{pq}{q^{2}}q^{\alpha}\right)
\left(p^{\beta}-\frac{pq}{q^{2}}q^{\beta}\right)
-iW_{3}\epsilon^{\alpha\beta}_{\:\:\:\:\rho\tau}p^{\rho}q^{\tau}
\end{equation*}
For the case of DIS with Charged Currents (parity violation), 
with massive final states ($m_{p'}\neq 0\Rightarrow$ not conserved 
current), then we must use the most generic 
expression \ref{generic_tensor}.
It is straightforward to demonstrate that term $W_{6}$ never
contributes when contracted even with the most generic electroweak
leptonic tensor \eqref{generic_leptonic_tensor}, because of symmetries 
and anti-symmetries of involved tensors, being $q^{\mu}=k_{\it in}^{\mu}-k_{\it out}^{\mu}$;
then the term $i(p^{\alpha}q^{\beta}-q^{\alpha}p^{\beta})W_{6}$
can be definitively neglected.\\
Eventually we can adopt as hadronic tensor 
\begin{equation}
\label{hadronic_tensor}
\boxed{
\begin{split}
H^{\alpha\beta}=&
-(2pq)H_{1}g^{\alpha\beta}
+4H_{2}p^{\alpha}p^{\beta}
-2iH_{3}\epsilon^{\alpha\beta}_{\:\:\:\:\rho\tau}p^{\rho}q^{\tau}\\
&+2H_{4}q^{\alpha}q^{\beta}
+2H_{5}(p^{\alpha}q^{\beta}+q^{\alpha}p^{\beta})
\end{split}
}
\end{equation}
that is, we have chosen a particular normalization for the coefficients.

\newpage
\section{Vector Bosons Propagator and Tensors Contraction} \label{boson_propagator} 
The most general expression for a massive ($M$) 1-spin boson propagator is
\begin{align}
B_{\rho\sigma}(\eta)=&-i\left(
\frac{g_{\rho\sigma}-q_{\rho}q_{\sigma}/M^{2}}{q^{2}-M^{2}+i\epsilon}+
\frac{q_{\rho}q_{\sigma}/M^{2}}{q^{2}-M^{2}/\eta+i\epsilon}
\right)\notag \\
=&-i\left[
\frac{g_{\rho\sigma}}{q^{2}-M^{2}+i\epsilon}-
\frac{(1-\eta^{-1})q_{\rho}q_{\sigma}}{(q^{2}-M^{2}+i\epsilon)(q^{2}-M^{2}/\eta+i\epsilon)}
\right]\label{propagator}
\end{align}
where $\eta\in [0,1]$ is the St\"ueckelberg gauge parameter
(\cite{ITZU}, page 698).\\
Massless case is well defined \eqref{propagator} and easily
recovered in the limit $M\rightarrow 0$.
\begin{figure}[!h]
\centering\epsfig{file=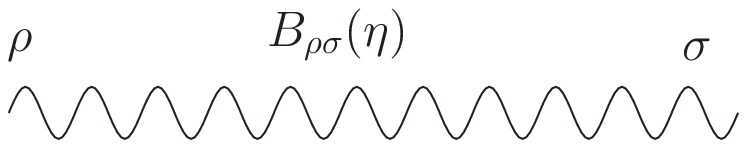,width=6 cm}
\caption{ \footnotesize{Boson Propagator.}}
\end{figure}
\\
We calculate contraction $L^{\mu\nu}T_{\mu\rho\nu\tau}H^{\rho\tau}$
for CC DIS, as defined respectively in \ref{squared_modulus}, \ref{EW_CC_tensor},
\ref{hadronic_tensor}.
\begin{equation}\label{final_contraction}
\begin{split}
L^{\mu\nu}\overline{B_{\mu\rho}}(\eta)B_{\nu\tau}(\eta)
H^{\rho\tau}=&
L^{\mu\nu}H_{\mu\nu}
+(1-\eta^{-1})^{2}
\frac{(q_{\mu}q_{\nu}L^{\mu\nu})(q_{\rho}q_{\tau}H^{\rho\tau})}{(q^2-M_{W}^{2}/\eta)^2}\\
&-(1-\eta^{-1})
\left[
\frac{(q_{\nu}L_{\rho}^{\:\:\nu})(q_{\tau}H^{\rho\tau})
+(q_{\mu}L^{\mu}_{\:\:\tau})(q_{\rho}H^{\rho\tau})}
{(q^2-M_{W}^{2}/\eta)}
\right]
\end{split}
\end{equation}
Notice that for the massless lepton case ($m_{\ell}=0$),
only the term $L^{\mu\nu}H_{\mu\nu}$ contributes because 
of $q_{\nu}L^{\mu\nu}=0=q_{\mu}L^{\mu\nu}$; 
also for the Feynman gauge case ($\eta=1$) only the first term 
contributes, nevertheless it still depends on the lepton mass.  \\
According to conventions of chapter \ref{chapter_DIS}
(see \ref{Kine_DIS}) 
\begin{equation}\label{L_H}
\begin{split}
L^{\mu\nu}&H_{\mu\nu}\!=8\frac{ME}{y}\bigg\{\!\bigg.
\left(\!xy^{2}+\!\frac{m_{\ell}^{2}y}{2ME}\right)H_{1}\upsilon
\!+\!\left[\left(1-\frac{m_{\ell}^{2}}{4E^{2}}\right)\!-\!\left(1+\frac{M}{2E}x\right)\!y
\right]2H_{2}\upsilon\\
&+\left[xy\left(1-\frac{y}{2}\right)-\frac{m_{\ell}^{2}y}{4ME}\right]2H_{3}\upsilon
+\left[\frac{m_{\ell}^{2}(m_{\ell}^{2}+Q^2)}{4M^2E^2}\right]H_{4}\upsilon
-\left[\frac{m_{\ell}^{2}}{ME}\right]H_{5}\upsilon
\bigg\}\bigg.
\end{split}
\end{equation}

\newpage
\noindent
We have computed the contribution of the remaining two terms in 
\ref{final_contraction}
and it is easy to show that the full result \eqref{unitary_chi}
is obtained from \ref{L_H} simply multiplying each 
coefficient of $H_k$ by $[1+\chi_{k}(\eta,x,y)]$, where
\begin{equation*}
\begin{split}
&\chi_{1}= -\frac{m_{\ell}^{2}(1-\eta^{-1})}{(Q^2+M^2_W/\eta)^2}
\frac{(Q^2\eta+Q^2+2M_{W}^{2})}{2\eta}    \\
&\chi_{2}=\frac{(\eta-1)}{(\eta Q^2+M^2_W)}
\frac{4m_{\ell}^{2}E^{2}y}
{\left[4(y-1)E^{2}+m_{\ell}^{2}+Q^{2}\right]}
\left[1\!-\frac{(\eta-1)}{4(\eta Q^2+M^2_W)}y(m_{\ell}^{2}+Q^2)  \right]\\
&\chi_{3}= 0    \\
&\chi_{4}= \frac{(1-\eta)Q^2}{(\eta Q^2+M^2_W)^2}(Q^2+2M_{W}^{2}+\eta Q^2)    \\
&\chi_{5}= -\frac{(\eta-1)}{(\eta Q^2+M^2_W)}\left[ Q^2+(Q^2+m_{\ell}^{2})\frac{y}{2}\right]
+\frac{(\eta-1)^2}{(\eta Q^2+M^2_W)^2}\left[ Q^2(Q^2+m_{\ell}^{2})\frac{y}{2}\right]
\end{split}
\end{equation*}
(notice that $Q^2$ depends on $y$ through $Q^2=2MExy$).\\
Factors $\chi_{j}$ are originated from terms $q_{\rho}q_{\tau}/M_{W}^{2}$
of $W$ boson propagator whilst terms containing $m_{\mu}$ in \ref{L_H} 
come directly from kinematics.\\
To carry out a calculation, it is necessary to choose a particular gauge:
for example $\eta=1$ corresponds to {\it Feynman} gauge, 
$\eta=\infty$ to {\it Lorentz} gauge, 
$\eta=0$ to {\it unitary} gauge and so on.
To avoid to introduce contributions of ghosts, we prefer to choose a 
physical gauge as the unitary one is (\cite{SIEGEL}, chapter VI, section B),
so finally we obtain 
\begin{equation}
\begin{split}\label{unitary_chi}
&\chi_{1}(\eta=0)= \frac{m_{\ell}^{2}(Q^2+2M_{W}^{2})}{2M_{W}^{4}}    \\
&\chi_{2}(\eta=0)= -\frac{m_{\ell}^{2}E^{2}y
\left[4M_{W}^{2}+y(Q^{2}+m_{\ell}^{2})\right]}
{M_{W}^{4}\left[4(y-1)E^{2}+m_{\ell}^{2}+Q^{2}\right]}    \\
&\chi_{3}(\eta=0)= 0    \\
&\chi_{4}(\eta=0)= \frac{Q^{2}(Q^2+2M_{W}^{2})}{M_{W}^{4}}    \\
&\chi_{5}(\eta=0)= \frac{Q^{2}}{M_{W}^{2}} +
\frac{(Q^{2}+M_{W}^{2})(Q^2+m_{\ell}^{2})}{2M_{W}^{4}}y
\end{split}
\end{equation}
This result agrees with the one reported in \cite{KRRE}.


\newpage
\section{Structure Functions}\label{contraction}
In this appendix, we briefly illustrate the result of the contraction 
between hadronic and leptonic tensors; furthermore we analyse the phase-space 
for the charged lepton in the final state and we give some considerations
about the general structure of functions $d^jF_{k}/\prod_i^j d\alpha_i$.\\
\\
Calculating contraction \ref{final_contraction} in unitary gauge 
($\chi_k=\chi_k(\eta\!=\!0)$, \ref{unitary_chi}), the full result is
\begin{equation}
\begin{split}\label{first}
L^{\mu\nu}&\overline{B_{\mu\rho}}(\eta=0)B_{\nu\tau}(\eta=0)H^{\rho\tau}=\\
&2\frac{ME}{y}
\bigg\{
\left[xy^{2}+\frac{m_{\ell}^{2}y}{2ME}\right](1+\chi_{1})H_{1}4\upsilon
+\left[\left(1-\frac{m_{\ell}^{2}}{4E^{2}}\right)-\left(1+\frac{M}{2E}x\right)y
\right]\\
&\cdot(1+\chi_{2})H_{2}8\upsilon
+\left[xy\left(1-\frac{y}{2}\right)-\frac{m_{\ell}^{2}y}{4ME}\right](1+\chi_{3})H_{3}8\upsilon\\
&+\left[\frac{m_{\ell}^{2}(m_{\ell}^{2}+Q^2)}{4M^2E^2}\right](1+\chi_{4})H_{4}4\upsilon
-\left[\frac{m_{\ell}^{2}}{ME}\right](1+\chi_{5})H_{5}4\upsilon
\bigg\}
\end{split}
\end{equation}
We introduce here the parametrisation $f_{k}=A_k H_{k}$,
with $A_{1,5}=4\upsilon$, $A_{2,3}=8\upsilon$ and $A_{4}=4x\upsilon$.
Obviously many others exist;
we have chosen it in order to obtain \ref{full_contraction}.\\
\\
\\
\underline{Charged Massive Lepton Phase Space}\\
\begin{equation*}
\begin{split}
d\hat{\Phi}=\frac{d^4k_{\it out}}{(2\pi)^3}
\delta_+\left(k_{\it out}^{2}-m_{\ell}^{2}\right)=&
\frac{d^3\underline{k}_{\it out}}{(2\pi)^3 2\sqrt{|\underline{k}_{\it out}|^{2}+m_{\ell}^{2}}}=
\frac{\kappa^{2}d\kappa d(1-\cos\theta) d\phi}{2(2\pi)^3 \sqrt{\kappa^2+m_{\ell}^{2}}} \\
\end{split}
\end{equation*}
using polar coordinates and with 
$\kappa\equiv |\underline{k}_{\it out}|$.\\
We identify $\theta$ as angle between the incoming lepton
and the outgoing one, that is between 
$\underline{k}_{\it in}$ and $\underline{k}_{\it out}$.\\
At this point we can choose a variables transformation 
\begin{equation*}
\begin{cases}
&x=\frac{Q^{2}}{2Pq}=\frac{2E\left(\sqrt{\kappa^{2}+m_{\ell}^{2}}-\kappa\cos\theta \right)-m_{\ell}^{2}}
{2M\left(E-\sqrt{\kappa^{2}+m_{\ell}^{2}}  \right)} \\
&y=\frac{Pq}{Pk_{\it in}}=1-\frac{\sqrt{\kappa^{2}+m_{\ell}^{2}}}{E}\\
\end{cases}
\end{equation*}
according to kinematics in \ref{Kine_DIS}.
It is then straightforward to calculate the Jacobian 
and to express muon phase-space in the form
\begin{equation}\label{muon_phase_space}
\boxed{d\hat{{\Phi}}=\frac{1}{(2\pi)^3}\frac{yME}{2}d\phi dxdy}
\end{equation}
It is interesting to notice that both massive and 
massless muon cases give the same result.
Phase-space $d\hat{\Phi}$ is then combined with \ref{first}
and re-inserted in the cross section to obtain
\ref{full_contraction} (and \ref{partonic_full_cross_section} 
with the appropriate partonic quantities).
Obviously other choices are allowed so,
the most general cross section can be written 
\begin{equation*}
d^{2}\sigma=\frac{G_{F}^{2}ME}{\pi}
\frac{M_{W}^{4}}{\left(Q^{2}(\Upsilon,\Sigma)+M_{W}^{2}\right)^{2}}
\left\{
\sum_{k=1}^{5}
c_{k}(\Upsilon,\Sigma)f_{k}(\Upsilon,\Sigma)d\Phi_{X}\right\}
J_{\Upsilon\Sigma}{d\Upsilon d\Sigma}
\end{equation*}
with opportune $\Upsilon$, $\Sigma$, $c_{k}$, $f_{k}$, $J_{\Upsilon\Sigma}$
(Jacobian of transformation), whilst $d\Phi_{X}$ is the phase-space for 
hadronic final state $X$.\\
If we choose to be completely inclusive over such a state, 
we can then identify functions $F_{k}$ with $\int d\Phi_{X} f_{k}$;
else rewriting $X$ phase-space as 
\begin{equation}\label{Phi_X_fact}
d\Phi_{X}=d\Phi_{X}\prod_{i}^{j}d\alpha_{i}\delta(\alpha_{i}-g_{i}(\Phi_X))=
d\hat{\Phi}_{X}\prod_{i}^{j}d\alpha_{i}
\end{equation}
where $g_i$ functions define $\alpha_i$ variables (observables),
it is then natural to identify 
\begin{equation}\label{F_differential}
\frac{d^{j}F_{k}}{\prod_{i}^{j}d\alpha_{i}}\equiv\int d\hat{\Phi}_{X}f_{k}
\end{equation}
obtaining a more exclusive result.

\newpage
\section{Projectors $P_{k}^{\mu\nu}$}\label{appendix_Projectors}
In order to work in the context of dimensional regularization, the 
expressions of projectors $P_{k}^{\mu\nu}$ in a $D$-dimensional 
space-time are given by 
 \begin{align}
 P_{1}^{\mu\nu}=&
 -(2pq)^{2}g^{\mu\nu}+4Q^{2}p^{\mu}p^{\nu}
 +2(2pq)(p^{\mu}q^{\nu}+q^{\mu}p^{\nu})&\notag\\
 P_{2}^{\mu\nu}=&
 -Q^{2}(2pq)^{2}g^{\mu\nu}+4(D-1)Q^{4}p^{\mu}p^{\nu}\notag\\
 &+2Q^{2}(D-1)(2pq)(p^{\mu}q^{\nu}+q^{\mu}p^{\nu})
 +(D-2)(2pq)^{2}q^{\mu}q^{\nu}\notag\\
 P_{3}^{\mu\nu}=&-i\epsilon^{\mu\nu}_{\:\:\:\:\sigma\tau}
 p^{\sigma}q^{\tau}\label{Projectors}\\
 P_{4}^{\mu\nu}=&p^{\mu}p^{\nu}\notag\\
 P_{5}^{\mu\nu}=&
 -(2pq)^{2}g^{\mu\nu}+4(D-1)Q^{2}p^{\mu}p^{\nu}
 +D(2pq)(p^{\mu}q^{\nu}+q^{\mu}p^{\nu})\notag
 \end{align}
When $D=4+2\varepsilon$ their contractions with the partonic
tensor $h_{\mu\nu}^a$ \eqref{partonic_tensor} provide
\begin{equation*}
P_{k}^{\mu\nu}h_{\mu\nu}^a=p_k(\varepsilon,2pq)h_{k}^a
\end{equation*}
where $p_k$ coefficients are 
\begin{equation}\label{Projectors_h_j}
\begin{split}
&p_1(\varepsilon,2pq)=2(1+\varepsilon)(2pq)^{3}\\
&p_2(\varepsilon,2pq)=2(1+\varepsilon)(2pq)^{4}\\
&p_3(\varepsilon,2pq)=(2pq)^{2}\\
&p_4(\varepsilon,2pq)=\frac{(2pq)^{2}}{2}\\
&p_5(\varepsilon,2pq)=2(1+\varepsilon)(2pq)^{3}
\end{split}
\end{equation}
It is important to remember that $q$ is the four-momentum of
$W$ boson exchanged, whilst $p$ is the one of the initial 
state parton.\\
A similar formalism is used in \cite{GOTT}.

\newpage
\section{Partonic Phase-Space}\label{partonic_phase_space}

The Phase-Space (PS) for two bodies, suitable to describe a final state containing 
a massive particle (heavy quark, $p'$) and one massless (quark or gluon, $k'$),
has the form
\begin{equation}\label{phase_space_NLO}
(PS)_{[2]}=\frac{d^Dp'}{(2\pi)^{D-1}} \delta_{+}\left(p'^{2}-m^{2}\right)
\frac{d^Dk'}{(2\pi)^{D-1}}\delta_{+}\left(k'^{2}\right)
(2\pi)^{D}\delta^{(D)}\left(k'+p'-q-p\right)
\end{equation}
expressed in a $D$-dimensional space-time.
Degrees of freedom enumeration:
2 x 4 coordinates for momenta, minus 2 conditions for mass-shell,
minus 4 because of four-momentum conservation,
minus 1 rotational symmetry axis around the 
centre of mass \cite{KMO}; at last only one independent 
not-trivial variable remains.\\
\\
We are going to carry out an expression depending on the only one degree of freedom.\\
Integrating $\delta^{(D)}$ through $d^Dk'$, we obtain $k'=p+q-p'$ and PS becomes
\begin{equation*}
\int \!d^Dk' (PS)_{[2]}=\frac{d^Dp'}{(2\pi)^{D-2}} \delta_{+}\left(p'^{2}-m^{2}\right)
\delta_{+}\left((p+q-p')^{2}\right)
\end{equation*}
Rewriting $d^Dp'=dp_0\!'d^{D-1}\underline{p}'$ and integrating first $\delta$-Dirac,
we recover on-mass-shell relation $p_{0}'\!=\sqrt{{|\underline{p}'|}^2+m^2}$.
Performing a variables change introducing (hyper-)spherical coordinates,
then the previous expression becomes
\begin{equation*}
\int d^Dk' dp_0\!'(PS)_{[2]}=\!
\frac{{|\underline{p}'|}^{D-2}}{2\sqrt{{|\underline{p}'|}^2+m^2}}
d{|\underline{p}'|}d\theta \left[\sin\theta\right] ^{D-3} \frac{d\Omega_{D-3}}{(2\pi)^{D-2}}
\delta_{+}\left((p+q-p')^{2}\right)
\end{equation*}
$\theta$ angle has not been defined because we have not yet
chosen a framework;
in literature $\theta$ is often chosen as angle between
the massive quark momentum and the straight line given by 
$\underline{p}+\underline{q}=0$ 
(i.e. working in the system reference of the centre of mass).\\
Introducing $\hat{w}\equiv\frac{(1+\cos\theta)}{2}\;\in[0,1]$,
$\hat{s}\equiv (p+q)^2$
and integrating $\delta$-Dirac with $d{|\underline{p}'|}$
(so $|\underline{p}'|=\frac{\hat{s}-m^2}{2\sqrt{\hat{s}}}$),
a compact result is obtained (remember $D=4+2\varepsilon$)
\begin{equation}\label{2body_PS}
\begin{split}
d\varphi_X^{NLO}\equiv
&\int d^Dk' dp_0\!'d{|\underline{p}'|}d\Omega_{D-3}(PS)_{[2]}
=d\hat{w}\left[\int d\hat{\varphi}_X^{NLO}\right] \\
=&\frac{1}{8\pi}\frac{1}{\Gamma(1+\varepsilon)}
\left(\frac{\hat{s}-m^2}{\hat{s}}  \right)
\left[\frac{(\hat{s}-m^2)^2}{4\pi\hat{s}}\right]^{\varepsilon}
\hat{w}^{\varepsilon}(1-\hat{w})^{\varepsilon} d\hat{w}\\
\end{split}
\end{equation}
where we have already integrated the factor $d\Omega_{D-3}$ according with G.11 
in \cite{STRM}.
Through \ref{2body_PS} we emphasize the dependence only on the not-trivial 
variable $\hat{w}$;
such a result is in agreement with eq.4 in \cite{KMO}.
\\
At this point one can integrate modulus-squared amplitudes over 
the whole PS to carry out an inclusive result either to remain differential
in order to convolve with Fragmentation Functions.
In the latter case, it is fundamental to notice that $\hat{w}$ is not well defined:
in fact it is not Lorentz invariant and furthermore it is not a 
{\it fragmentation variable}, that it is not possible to join in a natural way 
the final state parton momentum and/or energy to the corresponding observed hadron
(par. \ref{FRAGMENTATION}).
At Born level it is clear that $\hat{w}$ is not well defined:
in the reference system where $\underline{p}'=\underline{p}+\underline{q}=0$,
the massive quark produced is at rest and then it makes not sense
to define a scattering angle with respect to whatever direction.
Also at NLO there are difficulties in defining $\hat{w}$ \eqref{anomalus_pole}.
So, in order to avoid these drawbacks, a further variable change has to be done
\begin{equation}\label{zeta_definition}
\zeta\equiv\frac{p'\cdot p}{q\cdot p}
\end{equation}
as in \cite{GKRR}, where $p$ is the momentum of initial state parton
and $q$ the one of $W$ boson.
The link between $\zeta$ and $\hat{w}$ is given by
\begin{equation}
\zeta=\left[\frac{\hat{w}(1-\xi)+(1-\lambda)\xi}{(1-\lambda\xi)}\right]
\end{equation}
that becomes an identity in the massless limit ($m=0\rightarrow\lambda=1$).\\
Working with $\zeta$ is mathematically correct, however it is also possible to use $\hat{w}$ in a distributional sense, prescribing
\begin{equation}\label{t_hat_distribution}
f(\hat{w},\underline{p}'=0)\equiv \delta(1-\hat{w})
\left[\int_0^1f(t,\underline{p}')dt\right]_{\underline{p}'=0}
\end{equation}
where the artificial value $\hat{w}=1$ has been set in order that such a 
contribution is always included in convolutions; nevertheless 
\ref{t_hat_distribution} is not a mere unjustified prescription, 
but it follows relating the results obtained using $\zeta$ to those 
ones with $\hat{w}$ and using eq.18 in \cite{KRST}.
This last remark is very important and it solves the issue about
the appearance of an {\it anomalous} pole in \ref{anomalus_pole}. \\

\newpage
\noindent
Finally we illustrate PS for a single massive particle in the final state
\begin{equation}\label{phase_space_LO}
(PS)_{[1]}=\frac{d^4p'}{(2\pi)^{3}} \delta_{+}\left(p'^{2}-m^{2}\right)
(2\pi)^{4}\delta^{(4)}\left(p'-q-p\right)
\end{equation}
We introduce the definition of $\zeta$ with the prescription
$\delta\left(\zeta-\frac{p'\cdot p}{q\cdot p}\right)d\zeta$
and we decompose the space-phase in this way
\begin{equation}\label{ps_lo_fact}
\begin{split}
(PS)_{[1]}&=\left[
\frac{d^4p'}{(2\pi)^{3}} \delta_{+}\left(p'^{2}-m^{2}\right)
(2\pi)^{4}\delta^{(4)}\left(p'-q-p\right)
\delta\left(\zeta-\frac{p'\cdot p}{q\cdot p}\right)
\right]d\zeta\\
&\equiv d\hat{\varphi}_X^{LO}d\zeta
\end{split}
\end{equation}
Then easily 
\begin{equation}
\begin{split}
d\varphi_X^{LO}\equiv&
\left[\int d\hat{\varphi}_X^{LO}\right]d\zeta
=2\pi\frac{\lambda}{Q^2} \delta\left(1-\xi\right)
\delta\left(1-\zeta\right)d\zeta
\end{split}
\end{equation}
being
$$\xi=\frac{\hat{x}}{\lambda}\:\:\:\:\:\:\lambda=\frac{Q^2}{Q^2+m^2}$$
where $\hat{x}$ has been previously defined in \ref{parton_variables}
and $\lambda$ introduced in \ref{lambda_definition}.\\
Eventually the space-phase can be expressed through $\hat{w}$ variable as
\begin{equation*}
d\varphi_X^{LO}=\left[\int d\hat{\varphi}_X^{LO}(\hat{w})\right]d\hat{w}
\end{equation*}
with
\begin{equation*}
\int d\hat{\varphi}_X^{LO}(\hat{w})=
2\pi\frac{\lambda}{Q^2} \delta\left(1-\xi\right)
\delta\left(1-\hat{w}\right)
\end{equation*}

\newpage
\section{Light-Cone Variables}\label{light_cone}

Light-Cone coordinates are defined starting from a change of 
variables from the usual $(0,x,y,z)$ coordinates.\\
Given a vector $V^{\mu}$, its light-cone components are defined as
\begin{equation*}
V^{\pm}=\frac{V_0\pm V_z}{\sqrt{2}}\qquad \underline{V}^{\perp}=(V_x,V_y)
\end{equation*}
then it is possible to write $V^{\mu}=(V^+,V^-,V^{\perp})$.\\
It can be easily verified that Lorentz invariant scalar products 
have the form
\begin{equation*}
\begin{split}
V\cdot W&=V^+W^-+V^-W^+-\underline{V}^{\perp}\cdot\underline{W}^{\perp}\\
V\cdot V&=2V^+V^--|\underline{V}^{\perp}|^2
\end{split}
\end{equation*}
This set of coordinates has the useful property to transform very 
simply under boosts along the $z$-axis; when a vector is boosted
along this direction, light-cone coordinates show what are the large 
and small component of the vector (momentum).\\
In fact, boosting the coordinates along the $z$ axis, a new vector $V^{'\mu}$ 
is obtained: expressing it through ordinary $(0,x,y,z)$ components 
\begin{equation*}
V^{'}_{0}=\frac{V_0 + vV_z}{\sqrt{1-v^2}}\qquad
V^{'}_{z}=\frac{vV_0 + V_z}{\sqrt{1-v^2}}\qquad
V^{'}_{x}=V_{x}\qquad
V^{'}_{y}=V_{y}
\end{equation*}
whereas using light-cone coordinates
\begin{equation*}
V^{'\pm}=V^{\pm}e^{\pm\psi}\qquad
\underline{V}^{'\perp}=\underline{V}^{\perp}
\end{equation*}
where $v$ is the velocity defining the boost and the hyperbolic angle $\psi$ 
is set to $\frac{1}{2}\log\left(\frac{1+v}{1-v}\right)$, so $v=\tanh\psi$;
notice that if two boosts of parameters $\psi_1$ and $\psi_2$ are applied,
the result is an overall boost $\psi_1+\psi_2$.\\
\\
Now we apply the previous results to DIS:  
we fix $P$ ($P^2=M^2$) and $q$ ($-q^2=Q^2$) as vector-basis.
Using light-cone coordinates then
\begin{equation*}
P^{\mu}=\left(P^+,\frac{M^2}{2P^+},\underline{0} \right) \qquad
q^{\mu}=\left(-\eta P^+,\frac{Q^2}{2\eta P^+},\underline{0} \right)
\end{equation*}
where $\eta$ is defined through the implicit equation 
$$2qP=\frac{Q^2}{\eta}-\eta M^2$$
giving \ref{natch_var};
choosing $\hat{p}^{\perp}=0$, we define a class of reference systems 
so-called {\it collinear} 
\cite{Tung_Aivazis_Olness}.
In general a vector $\hat{p}$ can be decomposed in accordance with
\begin{equation*}
\hat{p}^{\mu}=C_P P^{\mu}+C_q q^{\mu}
\end{equation*}
Setting the constrains $\hat{p}^{2}=0$ and $\hat{p}^{+}=\xi P^+$,
then it is straightforward to carry out
$$ C_P=\frac{Q^2}{Q^2+\eta^2M^2}\xi \qquad C_q=-\frac{M^2\eta}{Q^2+\eta^2M^2}\xi $$
and finally
\begin{equation*}
\hat{p}^{\mu}=\left( \xi P^+,0,\underline{0}\right)
\end{equation*}
It is manifest that $\hat{p}^-\neq\xi P^-$.\\
Our last remark is that rewriting $P=(\sqrt{p^2+m^2},p_x,p_y,p)$ 
by means of light-cone coordinates
\begin{equation*}
P^{\pm}=\frac{p\pm\sqrt{p^2+m^2}}{\sqrt{2}}
\end{equation*}
then $P^+>P^-$ is always satisfied
(from this the definition of $P^+$ 
as \textquotedblleft large\textquotedblright\: component
and of $P^-$ as \textquotedblleft small\textquotedblright ).

\chapter{Mathematical Tools}
\section{Clifford Algebra}\label{Clifford}
Rules for the Dirac algebra in a $D$-dimensional space-time are

\begin{equation}
\begin{split}
\gamma_{\mu}\slh{a}\gamma^{\mu}&=-(D-2)\slh{a}\\
\gamma_{\mu}\slh{a}\slh{b}\gamma^{\mu}&=+4ab+(D-4)\slh{a}\slh{b}\\
\gamma_{\mu}\slh{a}\slh{b}\slh{c}\gamma^{\mu}&=-2\slh{c}\slh{b}\slh{a}-(D-4)\slh{a}\slh{b}\slh{c}\\
\end{split}
\end{equation}

\section{Distributions and Expansions}\label{DISTRIBUTIONS}
Some of the following expansions can be found 
in \cite{GOTT} and \cite{KRST}.

\begin{equation*}
\begin{split}
&\hat{w}^{\varepsilon}(1-\hat{w})^{\varepsilon}=1+\varepsilon\log \hat{w} +\varepsilon \log(1-\hat{w})+O(\varepsilon)\\
&\hat{w}^{\varepsilon}(1-\hat{w})^{1+\varepsilon}=(1-\hat{w})\left(1+\varepsilon\log \hat{w} +\varepsilon
\log(1-\hat{w})\right)+O(\varepsilon)\\
&\hat{w}^{\varepsilon}(1-\hat{w})^{-1+\varepsilon}=\frac{\delta(1-\hat{w})}{\varepsilon}+\frac{1}{(1-\hat{w})_{+}}+\varepsilon\frac{\log
\hat{w}}{1-\hat{w}} +\varepsilon \left[\frac{\log(1-\hat{w})}{1-\hat{w}}\right]_{+}+O(\varepsilon^{2})\\
&\hat{w}^{1+\varepsilon}(1-\hat{w})^{-1+\varepsilon}=\frac{\delta(1-\hat{w})}{\varepsilon}+\frac{\hat{w}}{(1-\hat{w})_{+}}+
\varepsilon\frac{\hat{w}\log \hat{w}}{1-\hat{w}} +\varepsilon \hat{w}\left[\frac{\log(1-\hat{w})}{1-\hat{w}}\right]_{+}
\!\!+O(\varepsilon^{2})
\end{split}
\end{equation*}

\begin{equation*}
\begin{split}
\xi^{-\varepsilon}(1-\xi)^{-1+2\varepsilon}&(1-\lambda \xi)^{-\varepsilon}=
\delta(1-\xi)\left[\frac{1}{\varepsilon}-\frac{(1-\lambda)^{\varepsilon}}{2\varepsilon}
-\varepsilon (1-\lambda)^{\varepsilon}\text{Li}_{2}(\lambda)\right]\\
&+\left[\frac{1}{1-\xi}\right]_{+}
-\varepsilon\frac{\log \xi}{1-\xi}+\varepsilon
\left[\frac{2\log(1-\xi)-\log(1-\lambda \xi)}{1-\xi}\right]_{+}
+O(\varepsilon^{2})
\end{split}
\end{equation*}

\begin{equation*}
(1-\xi)^{2\varepsilon}(1-\lambda \xi)^{-1-\varepsilon}=
\delta(1-\xi)\left[\frac{1}{\varepsilon}-\frac{(1-\lambda)^{\varepsilon}}{\varepsilon}
-K_{A}\right]
+\left[\frac{1}{1-\lambda \xi}\right]_{+}+O(\varepsilon)
\end{equation*}

\begin{equation*}
\begin{split}
(1-\xi)^{1+2\varepsilon}(1-\lambda \xi)^{-2-\varepsilon}=
&\delta(1-\xi)\left[\frac{1}{\varepsilon}-\frac{(1-\lambda)^{\varepsilon}}{\varepsilon}
-\frac{1+\lambda}{\lambda}K_{A}-\frac{(1-\lambda)^{\varepsilon}}{\lambda}\right]\\
&+\left[\frac{1-\xi}{(1-\lambda \xi)^2}\right]_{+}+O(\varepsilon)
\end{split}
\end{equation*}

\begin{equation*}
\xi^{-\varepsilon}(1-\xi)^{1+2\varepsilon}(1-\lambda \xi)^{-\varepsilon}=
(1-\xi)\left[1+\varepsilon\log\left( \frac{(1-\xi)^2}{(1-\lambda \xi)\xi}\right)\right]
+O(\varepsilon^2)
\end{equation*}

\noindent 
Plus distributions are defined as

$$\int_0^1d\xi f(\xi)\left[g(\xi)\right]_+\equiv
  \int_0^1d\xi\left[f(\xi)-f(1)\right]g(\xi)$$

$$\int_{\zeta_{min}}^1d\hat{\zeta}f(\hat{\zeta})\left[g(\hat{\zeta})\right]_{\oplus}\equiv
  \int_{\zeta_{min}}^1d\hat{\zeta}\left[f(\hat{\zeta})-f(1)\right]g(\hat{\zeta})$$
}

\noindent
{\Large \bf Acknowledgements}\\
\\
\noindent
First of all, I would like to thank Dr. Matteo Cacciari
for his willingness to follow my research: 
we have shared innumerable difficulties
and surprises met during this work.\\
\\
I am grateful to Federico Ceccopieri and 
Alberto Guffanti for many constructive discussions 
about QCD during the whole working period; 
Alberto in particular for the constant and indispensable 
consulting about information technology.\\ 
Among professors and researchers of the University of Parma
I would especially like to thank prof. M.Bonini, prof. M.Casartelli, 
prof. E.Onofri, Dr. F.Di Renzo and Dr. L.Griguolo.
I am grateful to prof. L.Trentadue for the helpful discussions.\\ 
I wish to acknowledge Dr. S.Weinzierl for his ever fast 
and efficient answers to my questions.\\
I am grateful to Dr. S.Catani for his helpfulness and patience 
to explain the Dipole Technique applied to NLO cross sections calculation.\\
I would like to thank Dr. S.O.Moch and prof. J.Bl\"umlein
for the hospitality at DESY-Zeuthen and the helpful
discussions on the very special subject of Mellin Transforms 
that we have met.\\
I would like to thank very much Dr. S.Kretzer for his courtesy
and support.\\
\\
\\
These are simply the main contributions.\\
I hope I have not forgotten anyone.\\
Actually there are a lot of people not mentioned here  
to whom I am grateful.\\
\\
Thank you.

\end{document}